\documentclass[twocolumn]{aastex631}
\usepackage{amsmath,amssymb,amstext}
\usepackage{gensymb}
\usepackage{graphicx}

\usepackage{xspace}

\graphicspath{{./}{figures/}}

\newcommand{\hst}{{\textit{HST}}\xspace}
\newcommand{\jwst}{{\textit{JWST}}\xspace}

\newcommand{\planetname}{{GJ~1132~b}\xspace}

\newcommand{\startemp}{{3261\,K\xspace}}

\newcommand{\eureka}{\texttt{Eureka!}\xspace}
\newcommand{\firefly}{\texttt{FIREFLy}\xspace}
\newcommand{\exotic}{\texttt{ExoTiC-JEDI}\xspace}

\newcommand{\POSEIDON}{\texttt{POSEIDON}\xspace}

\defcitealias{Lustig-Yaeger2023}{Lustig-Yaeger \& Fu et al.}
\defcitealias{Moran2023}{Moran \& Stevenson et al.}
\defcitealias{May2023}{May \& MacDonald et al.}

\shorttitle{GJ 1132 b Reveals its Secrets (Almost)}
\shortauthors{Bennett et al.}

\begin{document}

\title{Additional JWST/NIRSpec Transits of the Rocky M Dwarf Exoplanet GJ~1132~b Reveal a \\ Featureless Spectrum}

\author[0000-0002-9030-0132]{Katherine A. Bennett}
\affiliation{Department of Earth \& Planetary Sciences, Johns Hopkins University, Baltimore, MD 21218, USA}

\author[0000-0003-4816-3469]{Ryan J. MacDonald}
\affiliation{Department of Astronomy, University of Michigan, Ann Arbor, MI 48109, USA}
\affiliation{NHFP Sagan Fellow}

\author[0000-0002-1046-025X]{Sarah Peacock}
\affiliation{University of Maryland, Baltimore County, MD 21250, USA}
\affiliation{NASA Goddard Space Flight Center, Greenbelt, MD 20771, USA}
\affiliation{Consortium on Habitability and Atmospheres of M-dwarf Planets (CHAMPs), Laurel, MD, USA}

\author[0000-0002-9032-8530]{Junellie Perez}
\affiliation{Department of Earth \& Planetary Sciences, Johns Hopkins University, Baltimore, MD 21218, USA}
\affiliation{Consortium on Habitability and Atmospheres of M-dwarf Planets (CHAMPs), Laurel, MD, USA}

\author[0000-0002-2739-1465]{E. M. May}
\affil{Consortium on Habitability and Atmospheres of M-dwarf Planets (CHAMPs), Laurel, MD, USA}
\affiliation{Johns Hopkins Applied Physics Laboratory, 
Laurel, MD, 20723, USA}

\author[0000-0002-6721-3284]{Sarah E. Moran}
\affiliation{NASA Goddard Space Flight Center, Greenbelt, MD 20771, USA}
\affiliation{NHFP Sagan Fellow}

\author[0000-0001-8703-7751]{Lili Alderson}
\affiliation{Department of Astronomy, Cornell University, Ithaca, NY 14853, USA}
\affil{School of Physics, HH Wills Physics Laboratory, University of Bristol, Bristol, BS8 1TL, UK}

\author[0000-0002-0746-1980]{Jacob Lustig-Yaeger}
\affil{Consortium on Habitability and Atmospheres of M-dwarf Planets (CHAMPs), Laurel, MD, USA}
\affiliation{Johns Hopkins Applied Physics Laboratory, 
Laurel, MD, 20723, USA}

\author[0000-0003-4328-3867]{Hannah R. Wakeford}
\affil{School of Physics, HH Wills Physics Laboratory, University of Bristol, Bristol, BS8 1TL, UK}

\author[0000-0001-6050-7645]{David K. Sing}
\affiliation{Department of Earth \& Planetary Sciences, Johns Hopkins University, Baltimore, MD 21218, USA}
\affiliation{Department of Physics \& Astronomy, Johns Hopkins University, Baltimore, MD 21218, USA}

\author[0000-0002-7352-7941]{Kevin B. Stevenson}
\affil{Consortium on Habitability and Atmospheres of M-dwarf Planets (CHAMPs), Laurel, MD, USA}
\affiliation{Johns Hopkins Applied Physics Laboratory, 
Laurel, MD, 20723, USA}

\author[0000-0003-1240-6844]{Natasha E. Batalha}
\affiliation{NASA Ames Research Center, Moffett Field, CA 94035, USA}

\author[0000-0003-3204-8183]{Mercedes L\'opez-Morales}
\affiliation{Space Telescope Science Institute, Baltimore, MD 21218, USA}

\author[0000-0003-4157-832X]{Munazza K. Alam}
\affiliation{Space Telescope Science Institute, Baltimore, MD 21218, USA}

\author[0000-0003-3667-8633]{Joshua D. Lothringer}
\affiliation{Space Telescope Science Institute, Baltimore, MD 21218, USA}

\author[0000-0002-3263-2251]{Guangwei Fu}
\affiliation{Department of Physics \& Astronomy, Johns Hopkins University, Baltimore, MD 21218, USA}

\author[0000-0002-4207-6615]{James Kirk}
\affiliation{Department of Physics, Imperial College London, SW7 2AZ, London, UK}

\author[0000-0003-3305-6281]{Jeff A. Valenti}
\affiliation{Space Telescope Science Institute, Baltimore, MD 21218, USA}

\author[0000-0002-4321-4581]{L. C. Mayorga}
\affil{Consortium on Habitability and Atmospheres of M-dwarf Planets (CHAMPs), Laurel, MD, USA}
\affiliation{Johns Hopkins Applied Physics Laboratory, 
Laurel, MD, 20723, USA}

\author[0000-0001-7393-2368]{Kristin S. Sotzen}
\affil{Consortium on Habitability and Atmospheres of M-dwarf Planets (CHAMPs), Laurel, MD, USA}
\affiliation{Johns Hopkins Applied Physics Laboratory, 
Laurel, MD, 20723, USA}

\begin{abstract}

As an archetypal M-dwarf rocky exoplanet, GJ~1132~b has a varied history of atmospheric measurements. At 1.13~$\rm R_{\oplus}$, 1.66~$\rm M_{\oplus}$, and 580~K, it orbits a bright, slowly rotating M dwarf in a 1.6-day period, making it a prime target for characterization. In this study, we combine two JWST NIRSpec/G395H transits previously reported by \citetalias{May2023} (\citeyear{May2023}) with two new NIRSpec/G395M transits to constrain the presence of an atmosphere. This marks the first time the G395H and G395M modes have been combined for a single target, and we report no difference in the quality of data between the two modes. For rocky M-dwarf studies, G395H may still be preferred if stacking transits to utilize the high-resolution flux-calibrated stellar spectra and assess evolving stellar heterogeneity. GJ~1132~b's co-added transmission spectrum is best-fit with a flat line. A thin steam atmosphere is also consistent with the data, but this interpretation is driven almost entirely by the first transit, which suggests an increase in cool spot coverage-fraction derived from the flux-calibrated stellar spectra. This demonstrates the importance of always considering stellar heterogeneity evolution in multi-visit transits, and also the importance of a ``leave-one-transit-out" approach in modeling efforts of co-added transits. We combine these results with MIRI/LRS emission data \citep{Xue2024} to show that together, transmission and emission are consistent with only the thinnest of atmospheres. Given GJ~1132~b's age and distance from the star, a thin atmosphere is not likely stable. Therefore, the simplest explanation is that GJ~1132~b is indeed a bare rock.

\end{abstract}


\section{Introduction} \label{sec:intro}

Understanding whether rocky planets orbiting M dwarfs typically host atmospheres continues to be one of the most pressing questions in exoplanetary astronomy today. As the first step toward understanding how many nearby planets may be habitable, it is a critical question that nonetheless is proving a massive challenge to answer. Broadly, the community seeks to understand whether the cosmic shoreline, the hypothetical line dividing planets with and without atmospheres \citep{Zahnle2017}, exists outside our own Solar System. The shoreline is hypothesized to be due to some combination of planetary size and stellar irradiation. Planet size is important because larger planets have a higher escape velocity and thus can retain atmospheric particles more easily, whereas stellar irradiation matters because planets farther from their host stars experience lower rates of high energy radiation over their lifetimes and thus do not undergo complete atmospheric erosion. The shoreline may be different for close-in M dwarfs, however, because they experience higher levels of irradiation, when compared to Earth, Venus and Mars.

M dwarfs live and die slowly, meaning they experience a much more prolonged period of rapid rotation and high activity relative to other stars as they contract onto the main sequence (e.g., \citealt{Preibisch2005, Shkolnik2014}) and evolve (e.g., \citealt{Pizzolato2003, Pass2024}). Additionally (and relatedly), M-dwarfs experience higher rates of stellar flares (e.g., \citealt{Audard2000, Davenport2016}). Both of these factors may contribute to atmospheric mass loss and possibly complete atmospheric erosion early on in the lifetime of close-in M-dwarf rocky exoplanets (e.g., \citealt{Khodachenko2007, Tian2009, Owen2012, Peacock2019, vanLooveren2024}). There are still significant unknowns around this question due to uncertainties in high-energy stellar spectral energy distributions \citep{France2016}, especially the ratios between the extreme- far- and near-ultraviolet fluxes, which affect atmospheric chemistry \citep{Segura2005, Rugheimer2015}, precise flare rates \citep{France2020} and broadband flare energy distributions \citep{Brasseur2023, Paudel2024, Burton2025}, the mechanisms of atmospheric loss (nonthermal versus thermal processes; \citealt{Lammer2008, Owen2019}), and the variety of possible interior and atmospheric compositions (as well as outgassing mechanisms) dictated by initial planet formation \citep{Elkins2008, Kite2009}.

Nonetheless, M dwarfs offer the best opportunities right now to search for atmospheres on rocky exoplanets due to the optimally large planet-to-star radius ratio for planets around these small stars. This has led the community on a massive effort to detect atmospheres from the ground and space. With \jwst, the community has already looked at $\sim$20 targets $\rm <2\;R_{\oplus}$ in transmission, emission, or as a phase curve: e.g., LHS~475b (\citetalias{Lustig-Yaeger2023} \citeyear{Lustig-Yaeger2023}), GJ~486~b (\citetalias{Moran2023} \citeyear{Moran2023}, \citealt{WeinerMansfield2024}), TRAPPIST-1b \citep{Greene2023, Ducrot2024}, TRAPPIST-1c \citep{Zieba2023, Radica2025}, \planetname (\citetalias{May2023} \citeyear{May2023}, \citealt{Xue2024}), GJ~341~b \citep{Kirk2024}, GJ~367~b \citep{Zhang2024}, TOI-836~b \citep{Alderson2024}, 55~Cnc~e \citep{Hu2024}, LHS~1140~b \citep{Cadieux2024, Damiano2024},  L98-59~d \citep{Gressier2024}, L98-59~c \citep{Scarsdale2024}, L168-9~b \citep{Alam2025}, L98-59~b \citep{BelloArufe2025}, TOI-776~b \citep{Alderson2025}, LHS~1478~b \citep{August2025}, LTT~1445A~b \citep{Wachiraphan2025}, TOI-1685b \citep{Luque2024}, TOI-1468~b \citep{Meier2025}, and GJ~357~b \citep{Redai2025, Taylor2025b}. Nearly all of these have been inconclusive or hinted at no atmosphere. Only two planets - L98-59~d \citep{Gressier2024, Banerjee2024} and 55~Cnc~e \citep{Hu2024} - have thus far hinted at signs of secondary atmospheres, but these findings have yet to be confirmed. 

The focus of this study is the M dwarf rocky exoplanet GJ 1132 b. Discovered in 2015 by \cite{Berta2015}, \planetname is a $\rm 1.13\pm0.056\;R_{\oplus}$ \citep{Dittman2017}, $\rm 1.66\pm0.23\;M_{\oplus}$ \citep{Bonfils2018} planet in a 1.6-day orbital period around a $\rm 0.21\;R_{\odot}$, $\rm 0.18\;M_{\odot}$ 3270~K M~dwarf. Assuming $A_b=0$, the planet has $T_{\rm eq}\approx580$~K. The planet's mass and radius give it a bulk density of $\rm 6.3\pm1.3\;g\;cm^{-3}$, consistent with an Earth-like composition. Its star has a slow rotation period of $122$ days \citep{Cloutier2017, Bonfils2018} and is thought to be at least 5 Gyr old based on galactic kinematics, lack of H$\alpha$ emission, and comparison with Barnard's Star and Proxima Centauri \citep{Berta2015}. This suggests the star is currently not highly active, though it is important to note that \planetname likely still experienced prolonged high stellar activity in its past, based on the stellar spindown rates with age for M dwarfs (e.g., \citealt{Netwon2016, Pass2022, Pass2024}). From the estimated mass-dependent timescale of the saturated regime of XUV stellar radiation from \cite{Pass2024} (see their Equation 2), GJ~1132 could have remained in a heightened activity state for as long as $\sim3$~Gyr. Additionally, even older M dwarfs have well-documented frequent flare activity (e.g., \citealt{Vida2017}). 

\planetname has two possible siblings: GJ~1132~c, a $\rm 2.64\pm0.44\; M_{\oplus}$ minimum mass candidate at an 8.93-day orbital period (placing it in the habitable zone of GJ~1132) and another tentative planet with minimum mass $\rm 8.4^{+1.7}_{-2.5}\; M_{\oplus}$ with a 177 day orbital period \citep{Bonfils2018}. However, this latter signal is just as likely to be due to stellar contamination. The GJ~1132 system is just 12.6~pc away \citep{Xue2024}, and was the nearest transiting rocky exoplanet detected at the time of discovery in 2015, leaving it as one of the more promising targets for atmospheric characterization.

Indeed, \planetname has a long history of atmospheric detection claims and refutations. The first atmospheric investigation was by \cite{Southworth2017}, who observed nine transits in the \textit{griz} and \textit{JHK} bandpasses using the MPG 2.2~m telescope at ESO La Silla in Chile. They claimed an atmospheric detection based on increased transit depths measured in the \textit{z} and \textit{K} bands, which they found hinted at $\rm H_2O$ or $\rm CH_4$. Follow-up by \cite{Diamond-Lowe2018}, however, found a featureless spectrum from $\rm 0.7-1\;\mu m$ (covering the $z$ band) using five transits with the Magellan II Telescope at Las Campanas Observatory, consistent with a high mean molecular weight atmosphere or bare rock.

The story continued with space-based observations of \planetname. \cite{Waalkes2019} reported a non-detection of Ly$\alpha$ using \hst STIS/G140M and placed a $2\sigma$ upper limit on the effective size of an exosphere at 7.3 $\rm R_P$, which translates to an upper limit on the hydrogen mass-loss rate of $\rm 0.8-8\times10^8\;g\;s^{-1}$ (or $\rm 465-4650$ Earth atmospheric masses per Gyr). Meanwhile, \cite{Swain2021} found evidence for an $\rm H_2$-dominated atmosphere with HCN and $\rm CH_4$ using five transits with \hst WFC3/G141. They postulated this atmosphere originated via mantle $\rm H_2$ degassing following the loss of a primary atmosphere. However, \cite{Mugnai2021} and \cite{Libby-Roberts2022} both analyzed the same dataset and found a featureless spectrum, suggesting the \cite{Swain2021} results were due to differences in data reduction. Both \cite{Mugnai2021} and \cite{Libby-Roberts2022} found their data to be consistent with a high mean-molecular weight atmosphere, which could be thick or tenuous, or a bare rock. They also stated that their data are consistent with a cloudy primary atmosphere but deemed this scenario unlikely, as XUV irradiation likely long ago evaporated away any primordial $\rm H_2/He$ atmosphere on this planet. 

As part of GO Program 1981, our team observed \planetname in transmission with \jwst NIRSpec/G395H in 2023 (\citetalias{May2023} \citeyear{May2023}). We found two transits that told different stories: the first was consistent with either an $\rm H_2O$-dominated atmosphere with $\rm \sim1\%\;CH_4$ or stellar contamination, while the second was featureless. We investigated stellar and planetary atmospheric variability as well as instrumental systematics as drivers of this difference, but concluded that it was likely an unlucky random noise instance. We do note that our analysis found that if the low mean-molecular weight atmosphere detected by \cite{Swain2021} were real, there would have been large $\rm CH_4$ features (on the order of 400 ppm) detectable in the NIRSpec/G395H bandpass, but we did not observe this.

\planetname was also observed by \jwst in emission with GTO Program 1274 using MIRI/LRS \citep{Xue2024}. One eclipse revealed a dayside temperature $1\sigma$ below that expected for a bare rock with zero albedo and no heat redistribution, with the emission spectrum consistent with a featureless blackbody or a range of atmospheric compositions with $\leq10$ bar. 
Atmospheres containing at least 1\% CO$_2$, however, would need to be less than 10 mbar to be consistent with the data within $2\sigma$, given the large CO$_2$ absorption that would otherwise be present in the MIRI/LRS range.

A more recent ground-based study using CRIRES+ on the Very Large Telescope \citep{Palle2025} used the cross-correlation technique to search for HCN, $\rm CH_4$, and $\rm H_2O$ in \planetname's atmosphere, using two models: one based on the atmosphere put forth in \cite{Swain2021}, and the other based on the $\rm H_2O$-dominated atmosphere postulated by \citetalias{May2023} (\citeyear{May2023}). None of the molecules were detected, and they largely rule out the atmospheric scenario from \cite{Swain2021}, but were unable to reach sufficient sensitivity to support or refute the findings from \citetalias{May2023} (\citeyear{May2023}). Additionally, they did not detect 1083 nm He~I and placed a $3\sigma$ upper limit on the mass-loss rate of $1.63\times10^5$ Earth atmospheric masses per Gyr.

Taking all the atmospheric constraints to-date together, it is clear that that \planetname lacks a primordial $\rm H_2/He$ atmosphere, but it remains possible for the planet to host a thin secondary atmosphere. To follow-up on our first two transit observations and determine whether or not \planetname hosts a thin atmosphere, we observe two additional \planetname transits, this time with NIRSpec/G395M instead of G395H in order to avoid the G395H detector gap. In this paper, we report on all four visits together and what they tell us, in conjunction with the \jwst emission results, about the presence of an atmosphere on \planetname. In Sections \ref{sec:observations} and \ref{sec:reductions}, we describe our observations and data reduction process, respectively. We interpret the resulting spectrum in Section \ref{sec:analysis} and discuss both the technical and astrophysical implications of our findings in Section \ref{sec:discussion}.

\section{Observations} \label{sec:observations}

We observed four transits of \planetname with \jwst as part of Cycle 1 GO Program 1981 (P.I.s K. Stevenson and J. Lustig-Yaeger), all using the NIRSpec Bright-Object Time Series (BOTS) mode \citep{Espinoza2023}. For the first two transits (Visits 1 and 2, hereafter), we utilized NIRSpec/G395H, as reported by \citetalias{May2023} (\citeyear{May2023}). For the latter two transits (Visits 3 and 4, hereafter), we utilized the NIRSpec/G395M observing mode in order to diagnose whether or not the gap between the data from the NRS1/NRS2 detectors in our G395H observations confounded our results. Specifically, we did not know whether the potential water slope seen our first transit was due to an undiagnosed offset between the NRS1/NRS2 detectors. G395M exclusively uses the NRS1 detector and thus has no gap in the data, though it covers almost precisely the same wavelength range as G395H ($\sim\rm 2.9-5.1\;\mu m$), albeit at a lower resolving power ($R\sim1,000$ for G395M compared to $R\sim2,700$ for G395H). 

Our G395H observations occurred on 2023~February~25 and 2023~March~05, a little over a week apart, while our G395M observations occurred on 2024~February~08 and 2024~June~05. We acquired the target using the SUB32 subarray with the F140X filter. For the science exposures, we used the SUB2048 subarray with the NRSRAPID readout mode. Visits 1 and 2 (using G395H) used 14 groups per integration, with 814 integrations per visit (each lasting 4.62 hours including overhead). Visits 3 and 4 (using G395M) used 6 groups per integration, with 1732 integrations (4.60 hours) for Visit 3 and 1932 integrations (4.95 hours) for Visit 4. 

\section{Data Reduction} \label{sec:reductions}

As in \citetalias{May2023} (\citeyear{May2023}), we reduce the data with three independent reduction pipelines: \firefly \citep{Rustamkulov2022, Rustamkulov2023}, \eureka \citep{Bell2022}, and \exotic \citep{Alderson2022}. Below we detail the reduction process for each pipeline. The weighted mean (at $R\sim100$) of all four visits for each pipeline is shown in \autoref{fig:spectrum_reduction_compare}, and the fit orbital parameters for each reduction are given in \autoref{tab:orbital_parameters}. 

\subsection{\firefly} \label{sec:firefly_reduction}

\firefly begins with the \texttt{uncal} fits files and reduces the data through the spectroscopic light curve fitting. In addition to reducing the G395M data, we completely re-reduced the G395H data, in order to make sure all four visits were reduced with the same calibration files from the \texttt{jwst} pipeline (version 1.13.4 with CRDS context (pmap) 1241) and with the same version of \firefly.

\firefly first utilizes Stages 1 (for group-level corrections) and 2 (for integration-level corrections) of the \texttt{jwst} reduction pipeline. As in previous reductions, the only changes we make in Stage 1 are to apply a group-level $1/f$ subtraction, and to skip the dark current step and the jump step. In Stage 1, we use the default \texttt{jwst} superbias file, as we found in \citetalias{May2023} (\citeyear{May2023}) that at this time, the most up-to-date superbias file does a better job at removing the bias than our custom bias subtraction described in \citetalias{Moran2023} (\citeyear{Moran2023}). Stage 1 ends with ramp-fitting and the gain-scale step, after which in Stage 2 we only utilize the ``assign WCS" step - we skip flat-fielding - and proceed to \firefly's custom stages.

For the stellar extraction, we first apply a cosmic ray correction using \texttt{lascosmic} \citep{vanDokkum2001}. Additionally, we apply a second $1/f$ subtraction at the integration-level and measure the x-position and y-position detector shifts over the course of the observation. We extract the 1D stellar spectra by using a 4th order polynomial to compute the trace, and extracting the flux around an aperture full-width of $\sim5.3$ pixels across G395H and G395M. We determine the precise pixel width of the trace by finding the width that minimizes the out-of-transit scatter in the white light curve. This amounts to including all flux within 3.8$\sigma$ of the center of the PSF per column.

Before fitting the light curves, we trim the first 100 integrations for all four visits. We fit all observations separately, including the two different detectors for the G395H observations. We assume a circular orbit and fix the period to 1.628931 days \citep{Bonfils2018} while fitting for the semi-major axis in units of stellar radii ($a/R_s$), the impact parameter ($b$), the center-of-transit time ($T_0$), the flux, and the transit depth ($(R_p/R_s)^2$). We fit for limb darkening using the quadratic law. Initially, we used the $q_1/q_2$ Kipping parameterization \citep{Kipping2013} but found that $q_2$ was consistently not fit well as it trended toward zero. Instead, we shifted to using the classic $u_1/u_2$ parameterization \citep{Kopal1950}. While we find in this case that $u_1$ still trended toward zero (which is consistent with $q_2$ trending toward zero, since $q_2=0.5u_1(u_1+u_2)^{-1}$), the sigma differences between the individually fitted $a/R_s$ and $b$ decreases when switching from the $q_1/q_2$ to $u_1/u_2$ parameterization, and so we elect to stick with the latter. We set $u_1=0$ and fit for $u_2$. Once we fit all six white light curves individually and confirm that the best-fit system parameters are within 1$\sigma$ of each other, we calculate the weighted mean of the fitted $a/R_s$, $b$, and $u_2$ values. We then go back and refit the white light curves fixing $a/R_s$ and $b$ to their weighted means from all six light curves, and fixing $u_2$ to its weighted mean within each respective observing mode.

\begin{figure*}
\figurenum{1}
    \centering
    \hspace*{-1cm}
    \includegraphics[width=1.15\linewidth]{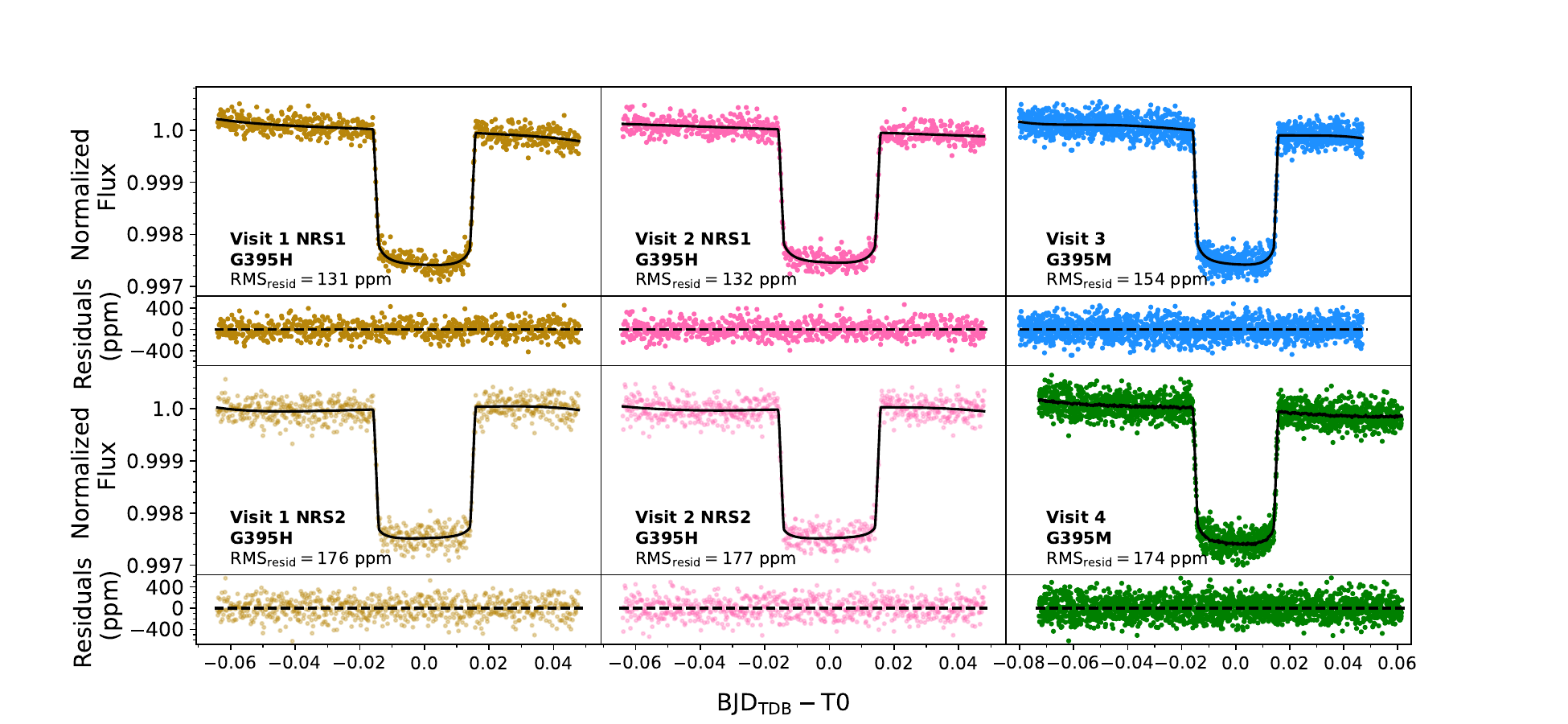}
    \caption{White light curve data (colored points) and best-fit models (black lines) for all six light curves using the \firefly reduction. Residuals are shown below each respective light curve, with the mean rms of the residuals ($\rm RMS_{resid}$) for each light curve given in the label. The complete figure set (one figure per reduction, three images total) is available in the online version of this paper.}
    \label{fig:wlcs}
\end{figure*}

To determine the most optimal systematics model in the light curve fits, we examine all combinations of the possible detrending vectors (8191 total possible combinations), which includes a polynomial in time (up to order six), as well as the x- and y- position shifts, their squares (e.g., $x^2$ where x is the subpixel position in the spectral direction) and their product ($x\times y$). We determine the most optimal combination of detrending vectors by determining which combination has the lowest Akaike Information Criterion (AIC) value. When fitting the white light curves, if a given fitting parameter is not statistically significant (to $\ge 3\sigma$), we then manually exclude it, except in the case where excluding it increases the red noise in the residuals. We found this to be the best method to determine what systematics model to use. If we simply relied on the model with the lowest Bayesian Information Criterion (BIC), there was almost always significant red noise in the residuals. We used a linear polynomial in time for Visit 2 NRS1, a 3rd-order polynomial for Visit 2 NRS2, and a 5th-order polynomial in time for Visit 1 NRS1 and NRS2 and Visit 3. For Visit 4, we used a 4th-order polynomial plus the y-position shift. To actually fit the white light curves, we use \texttt{emcee} \citep{emcee2013} with 1000 steps and 200 burn-ins. We show the individually-fitted white light curves and their residuals in \autoref{fig:wlcs}. Comparing the residuals on the NRS1 detector, the RMS of the G395M residuals are generally larger than those of G395H NRS1 due to the lower number of groups per integration in these visits.

\figsetstart
\figsetnum{1}
\figsettitle{White Light Curves of GJ~1132~b.}

\figsetgrpstart
\figsetgrpnum{1.1}
\figsetgrptitle{wlc_firefly}
\figsetplot{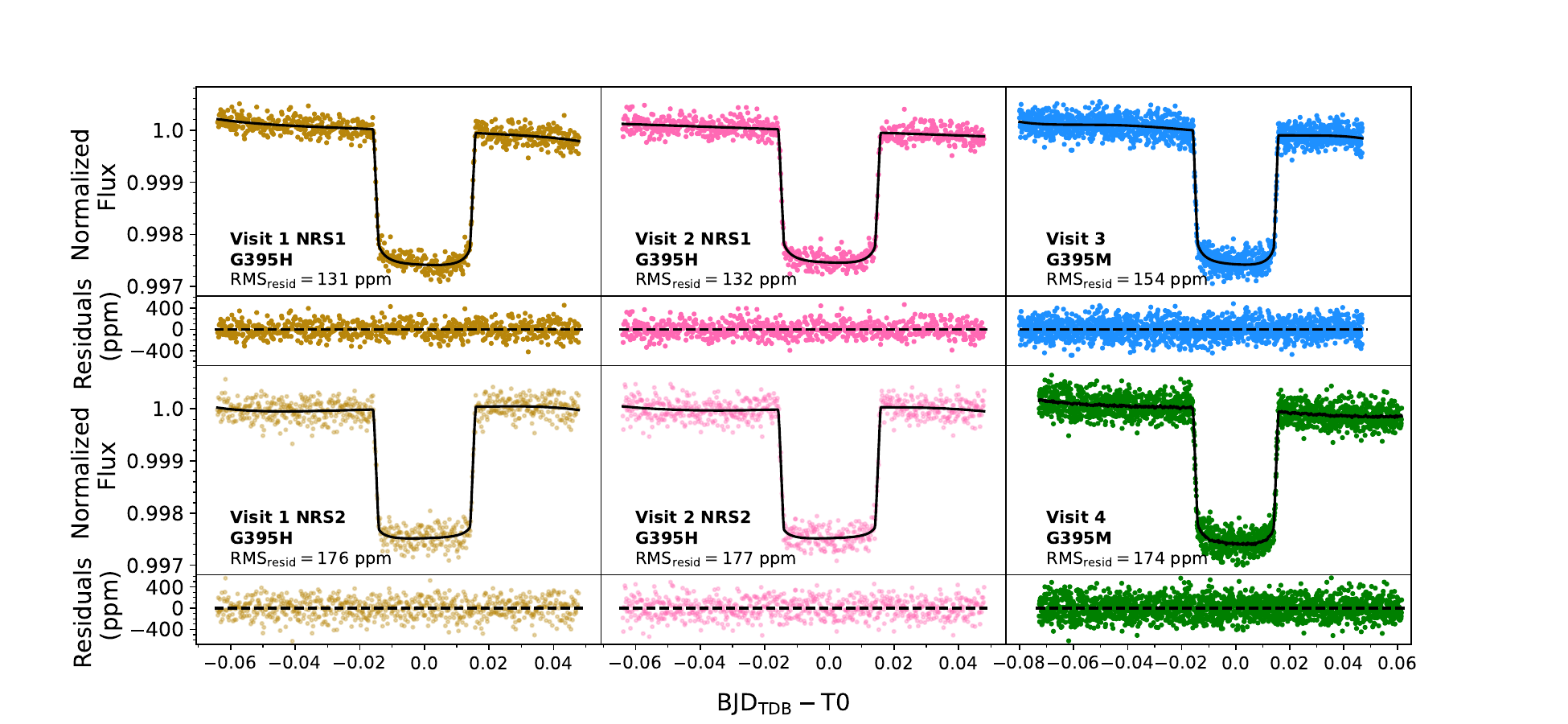}
\figsetgrpnote{White light curve data (colored points) and best-fit models (black lines) for all six light curves. Residuals are shown below each respective light curve, with the mean rms of the residuals for each light curve given in the label.}
\figsetgrpend

\figsetgrpstart
\figsetgrpnum{1.2}
\figsetgrptitle{wlc_eureka}
\figsetplot{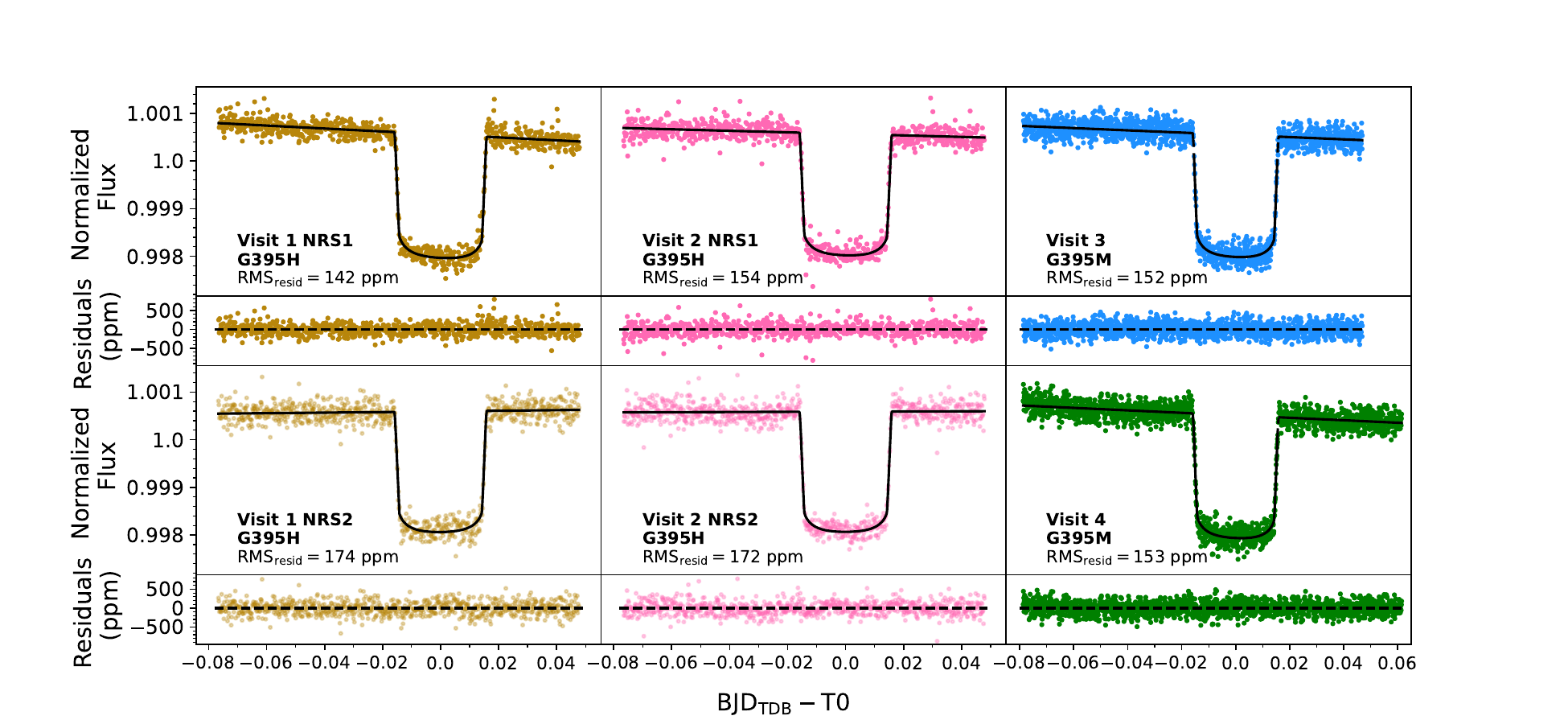}
\figsetgrpnote{White light curve data (colored points) and best-fit models (black lines) for all six light curves. Residuals are shown below each respective light curve, with the mean rms of the residuals for each light curve given in the label.}
\figsetgrpend

\figsetgrpstart
\figsetgrpnum{1.3}
\figsetgrptitle{wlc_exotic}
\figsetplot{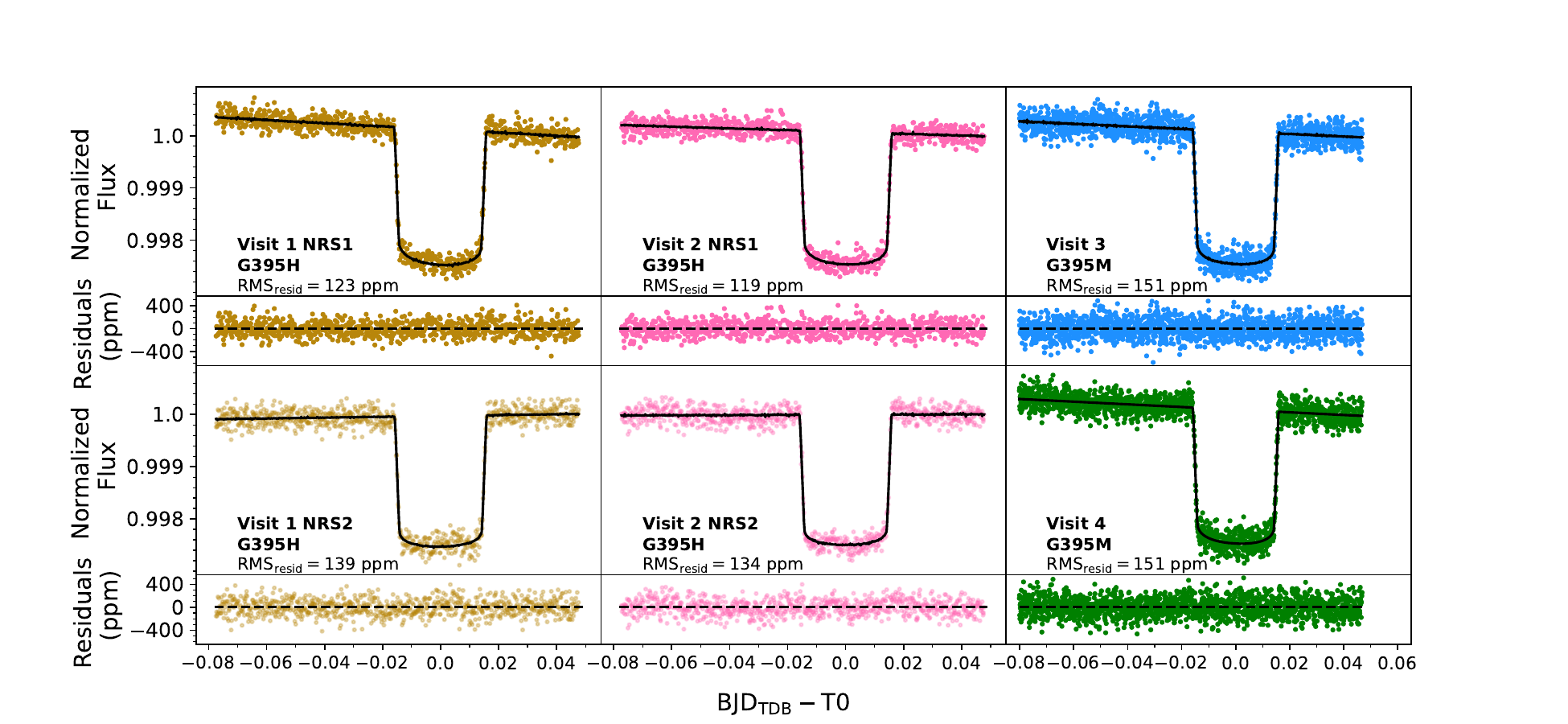}
\figsetgrpnote{White light curve data (colored points) and best-fit models (black lines) for all six light curves. Residuals are shown below each respective light curve, with the mean rms of the residuals for each light curve given in the label.}
\figsetgrpend

\figsetend

\figsetstart
\figsetnum{2}
\figsettitle{Spectroscopic Light Curves of GJ~1132~b.}

\figsetgrpstart
\figsetgrpnum{2.1}
\figsetgrptitle{slc_firefly}
\figsetplot{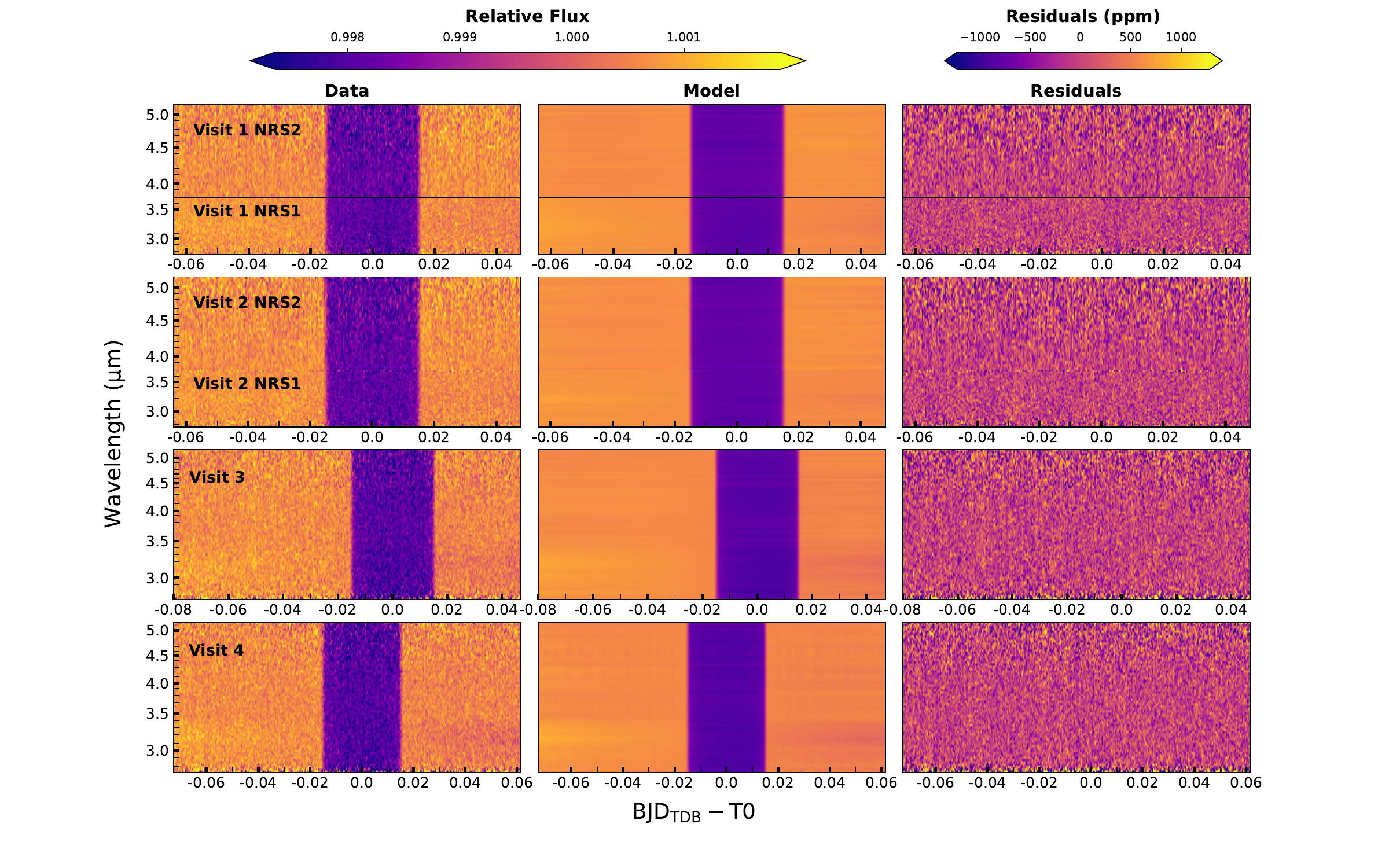}
\figsetgrpnote{Spectroscopic light curve data, model, and residuals for all six light curves. Each row reflects a different observation.}
\figsetgrpend

\figsetgrpstart
\figsetgrpnum{2.2}
\figsetgrptitle{slc_eureka}
\figsetplot{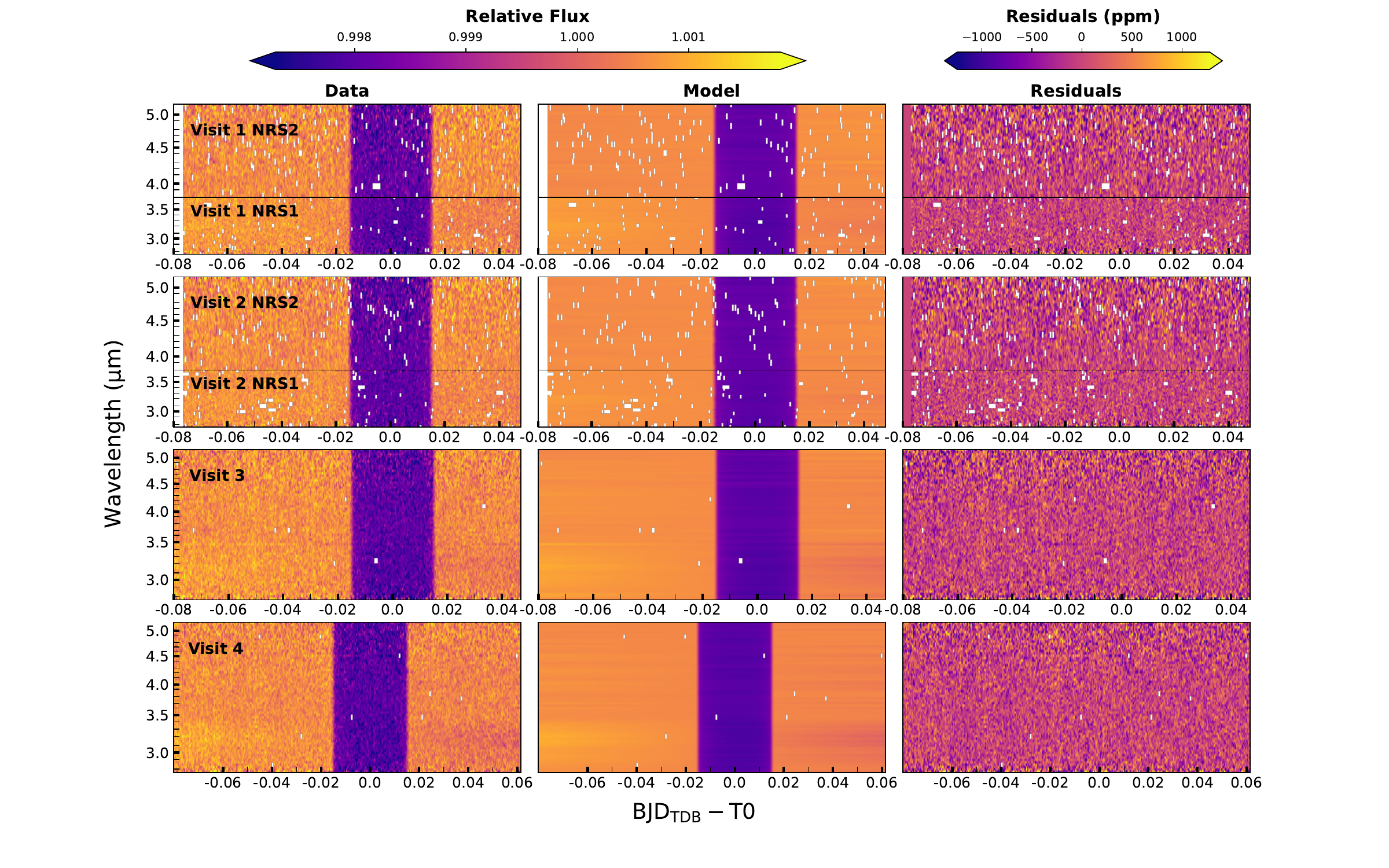}
\figsetgrpnote{Spectroscopic light curve data, model, and residuals for all six light curves. Each row reflects a different observation.}
\figsetgrpend

\figsetgrpstart
\figsetgrpnum{2.3}
\figsetgrptitle{slc_exotic}
\figsetplot{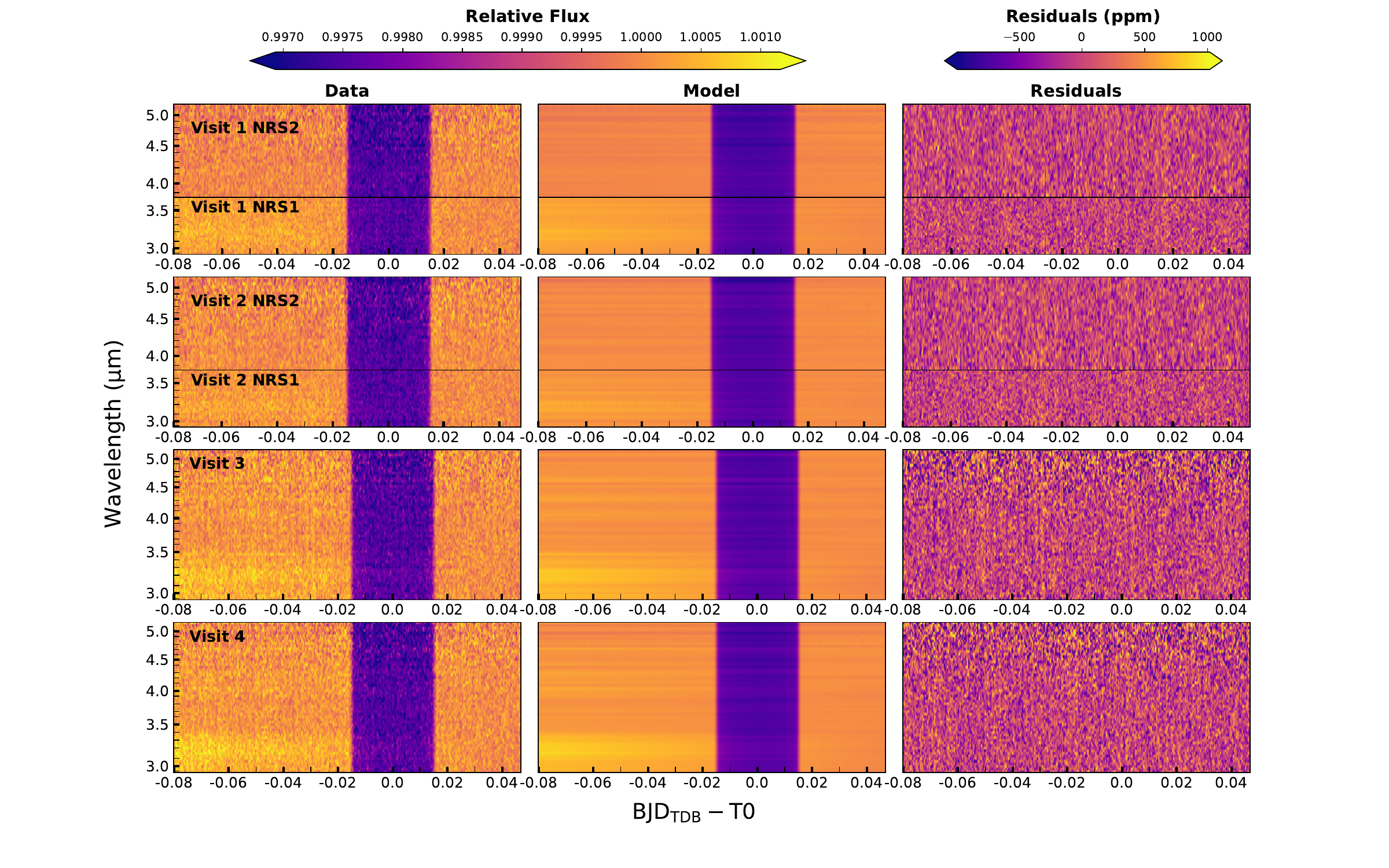}
\figsetgrpnote{Spectroscopic light curve data, model, and residuals for all six light curves. Each row reflects a different observation.}
\figsetgrpend

\figsetend

\begin{figure*}
\figurenum{2}
    \centering
    \hspace*{-1.9cm}
    \includegraphics[width=1.2\linewidth]{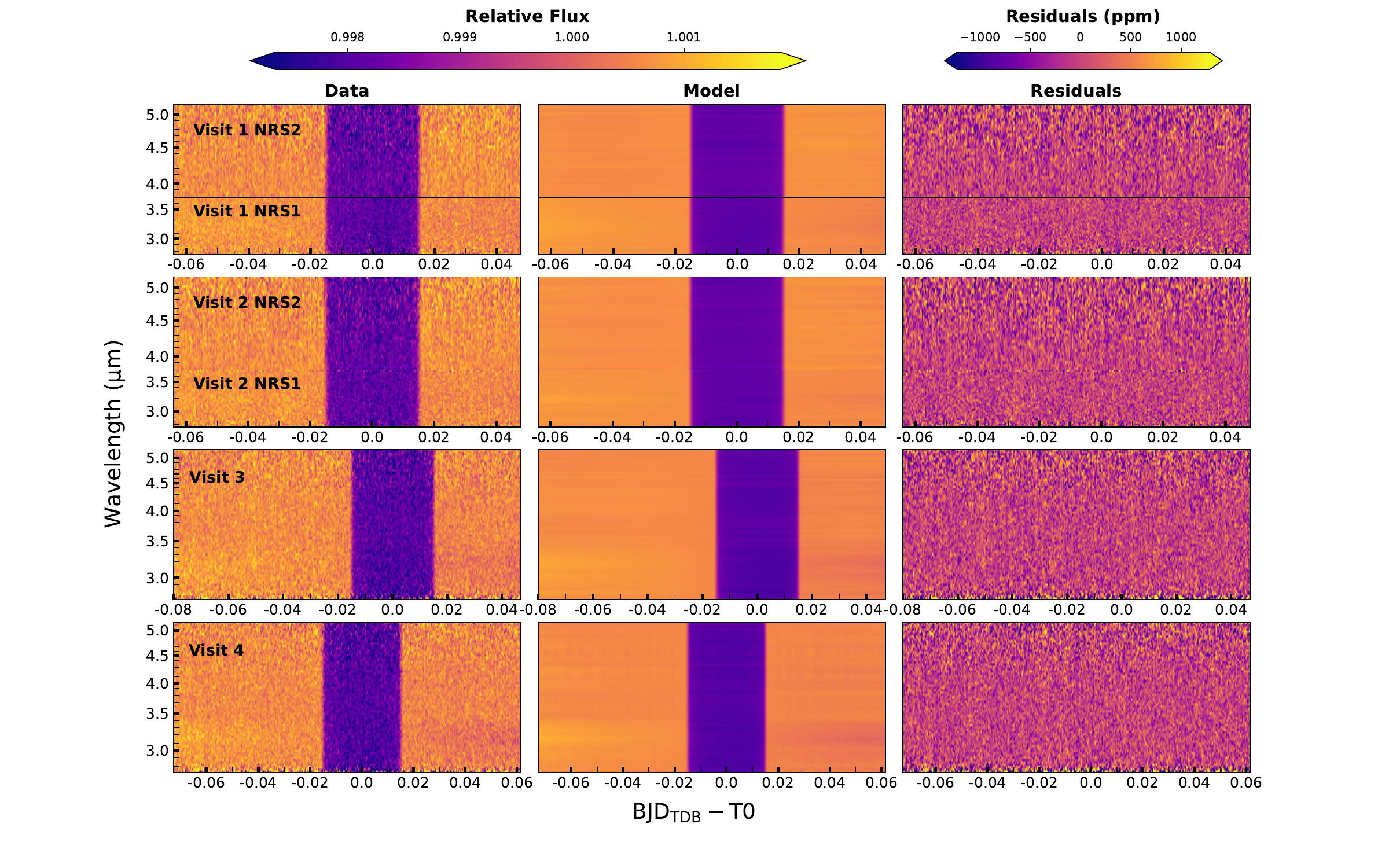}
    \caption{Spectroscopic light curve data, model, and residuals for all six light curves using the \firefly reduction. Each row reflects a different observation. The complete figure set (one figure per reduction, three images total) is available.}
    \label{fig:slcs}
\end{figure*}

For the spectroscopic fits, we use least-squares fitting with \texttt{lmfit} \citep{Newville2014}, as this is much more computationally efficient and has been shown to have consistent results with \texttt{emcee} fitting at the spectroscopic stage (\citetalias{May2023} \citeyear{May2023}). We apply a binning scheme of $R\sim100$, following \citetalias{May2023} (\citeyear{May2023}). We fix system parameters to the weighted-mean white light curve values and apply the same systematics model as in the white light curve fit. We fix the detrending parameter coefficients to their white light curve values, except for the linear term, which has a wavelength-dependency. We also extensively explore whether or not to fit or fix limb darkening in the spectroscopic fits. In past studies, \firefly has fixed limb darkening at this stage (\citetalias{May2023} \citeyear{May2023}, \citetalias{Moran2023} \citeyear{Moran2023}), since we are working at long wavelengths where in theory the limb darkening should not change that much. However, there was a slight wavelength-dependency (on the order of 10--20 ppm) when we compared the fit-limb-darkening versus fixed-limb-darkening transmission spectrum. Additionally, when plotting the weighted mean of the fitted $u_2$ values as a function of wavelength, there visually is a wavelength dependency shortward of $\rm \sim4\;\mu m$, whereas longward of $\rm \sim4\;\mu m$, $u_2$ is quite flat. This can be seen in each individual $u_2$ spectrum per visit as well, but is most apparent in the weighted mean spectrum. Thus, we adopt an ``empirical best-fit" approach in which we fit a linear slope to the fitted weighted mean $u_2$ values shortward of $\rm 4\;\mu m$ and fit a zero-order polynomial (i.e., just take the weighted mean) of the $u_2$ values longward of $\rm \sim4\;\mu m$. We then adopt these values as our fixed limb darkening values in the spectroscopic fits. Our spectroscopic light curves for \firefly  are illustrated in \autoref{fig:slcs}.

At both the white and spectroscopic light curve level, we rescale our error bars when needed so that $\chi^2_{\nu}=1$ between the data and model. While this allows us to more realistically describe the scatter of the data, it does not actually take into account correlated noise, the recurring demon of exoplanet transmission spectroscopy studies. Thus, in order to properly rescale our final transmission spectrum error bars, we use the binned rms of the residuals (hereafter referred to as a ``red noise diagnostic plot"; formerly referred to as an ``Allan variance plot", but see \citealt{Kipping2025}) to measure the correlated noise across different bin sizes in time \citep{Pont2006} for each spectroscopic channel. We then inflate the calculated error bar for that channel in quadrature with the median correlated noise across all bin sizes.

Fitted system parameters for each visit are shown in \autoref{tab:orbital_parameters}. We show each individual visit reduction for \firefly in \autoref{fig:spectrum_visit_compare}, and the final weighted mean spectrum (coadding all four visits) in \autoref{fig:spectrum_reduction_compare}.

\subsection{\eureka}

The \eureka\footnote{For this work, we use an untagged development version after v0.10 that addressed bug fixes prior to the large v1.0 release.} \citep{Bell2022} package begins with the \texttt{uncal} JWST files. Stages 1 and 2 are wrappers of the \texttt{jwst} pipeline \citep{Bushouse2024} (version 1.14.0 with crds context (pmap) 1253 for M visits and crds context (pmap) 1241 for H visits) with some modifications, primarily custom group-level background subtraction. This improves the precision of ramp fitting by removing the striping resulting from 1/f noise at the group level. It consists of identifying the center of light in each column and masking the values 6 pixels within each side of the center. For this study, we estimate the background as the median of all other pixels in a column with a 3$\sigma$ outlier threshold. Lastly, for Stage 1 we skip the jump step detection to avoid more noise being introduced in the light curves. For Stage 2, we skip both the flat-field and photom steps similar to \citetalias{May2023} \citeyear{May2023} since these steps give us absolute flux calculations which we do not need for this reduction. 

After running these stages of the \eureka pipeline, we continue running Stage 3, which first corrects the curvature of the trace. We identify the center of light in each column, then roll the pixels within that column so that the stellar spectrum aligns with the center of the detector. We then run an additional background subtraction, remove the median of all pixels more than 8 pixels away from the center of the trace with a 5$\sigma$ outlier threshold. We utilize optimal spectral extraction \citep{Horne1986} to extract the 1D stellar spectra time series. The median frame is assembled with a 5$\sigma$ outlier rejection threshold, smoothing over 20 frames. For Visit 3 with G395M, the 1D stellar spectra are extracted with a 20$\sigma$ outlier rejection threshold relative to the median frame and a 3 pixel half-width aperture size. For Visit 4, we use a 30$\sigma$ outlier rejection threshold and a 3 pixel half-width aperture size. We choose these thresholds and aperture sizes based on the combination of these parameters that result in the lowest median absolute deviation (MAD) in the white light curves.

We also reanalyze both G395H visits presented in \citetalias{May2023} (\citeyear{May2023}) following the same steps outlined within, while using updated NIRSpec reference files. For the G395H observations, white light curves are created by summing all light between 2.752 - 3.721 \micron\ for the NRS1 detector and 3.820 - 5.177 \micron\ for the NRS2 dector. For the two new G395M visits presented in this work, white light curves are created by summing all light between 2.726 - 5.181 \micron. Spectroscopic light curves for all visits are extracted on the same R$\sim$100 wavelength grid as \firefly.

We perform a joint white light curve fit with \texttt{emcee} \citep{emcee2013} of all six white light curves (two for each G395H visit, one for each G395M visit) using \texttt{batman} for the astrophysical signals and a linear temporal ramp for each light curve. For the temporal ramp we initially consider 0th - 3rd order polynomials in time, and any combination thereof (e.g. 3rd order with no quadratic term). We select the model with the lowest Bayesian Information Criteria (BIC), which corresponds to the standard linear ramp. In retrospect, we note that the BIC might not always be the best model selection tool to use, as there remains a higher level of red noise in the \eureka white light curves compared to those of \firefly, which uses the AIC to arrive at a more complex systematics model. The planetary orbital period, $T_0$, $a/R_s$, and orbital inclination ($i$) are required to be the same for all visits, while the transit depth is allowed to vary between visits and detectors. Quadratic limb darkening parameters are fixed independently to the MPS2 grid  \citep{Kostogryz2022} for each white light curve and calculated with \texttt{ExoTIC-LD} \citep{Grant2022, Grant2024} using the same stellar parameters as in \citetalias{May2023} (\citeyear{May2023}), namely T$_{\mathrm{eff}}$ = \startemp , $\log{g}$ = 5.02 \citep[both from][]{Stassun2019}, and a metallicity [Fe/H] = $-0.12$ \citep{Berta2015}. All chains are run to at least 50$\times$ the autocorrelation time to ensure convergence. Specifically, for these joint white light curve fits,  we use 25,000 steps, 250 walkers, with a burn-in of 5,000 steps. Best-fit orbital parameters are given in \autoref{tab:orbital_parameters}.

We then hold orbital parameters fixed at our best fit white light values for the spectroscopic light curves and only fit for the transit depth and temporal ramp parameters. Limb darkening is once again held to the values from the MPS2 grid with \texttt{ExoTIC-LD} for the given channel. Final error bars are inflated following the correlated-noise approach described in the \firefly section. Some type of error inflation like this is critical to do; 
though it only increases the error bars in the final weighted mean spectrum in select channels by $\sim5$ ppm, it changed the precise weighted mean spectrum enough to alter interpretation of the spectrum in our retrievals. Our final weighted mean spectrum is shown in \autoref{fig:spectrum_reduction_compare}.

\subsection{\exotic}

For our \exotic \citep{Alderson2022} reduction of the G395M data, we follow the same procedures as for the G395H data presented in \citetalias{May2023} (\citeyear{May2023}), with minor adjustments to account for the differences between the two modes. In summary, the \exotic reduction begins with a modified version of stage 1 of the \texttt{jwst} pipeline (v.1.14.0, \citealt{Bushouse2024}), using the custom bias correction available within \exotic \citep[e.g.,][]{Alderson2023, Alderson2024, Scarsdale2024}, before cleaning the 2D images and extracting 1D stellar spectra. The reduction of the G395M data uses the same reduction parameters as outlined in \citetalias{May2023} (\citeyear{May2023}) other than the full-width of the aperture, which for both Visit 3 and 4 is 3.5$\times$ the trace FWHM, equivalent to approximately 5 pixels. With the extracted 1D stellar spectra, we produce white light curves spanning the full G395M range (2.814--5.1\micron), as well as spectroscopic light curves at $R\sim100$. 

We additionally re-reduce the G395H data to take into account lessons learned from other \exotic analyses as well as updates to the \texttt{jwst} pipeline. This analysis also followed the same procedures as presented in \citetalias{May2023} (\citeyear{May2023}), however the \exotic custom bias subtraction was used as opposed to the default within the \jwst pipeline. With our re-reduced 1D stellar spectra, we produced new white light curves spanning the full G395H range (2.814--3.717\,$\mu$m for NRS1 and 3.824--5.111\,$\mu$m for NRS2) and spectroscopic light curves at $R\sim100$. Our pipeline does not extract flux blueward of 2.814 \,$\mu$m due to the limb darkening prescription used, as \texttt{ExoTiC-LD} \citep{Grant2024} is limited to the official throughput of the NIRSpec instrument.

To account for the full information provided by the four total observed transits, we fit the four individual G395H white light curves (one each for NRS1 and NRS2 per visit) along with the two G395M white light curves (one per visit) before taking the weighted average of the system parameters to ensure consistency when obtaining the final transmission spectra. In all cases, we fit the light curves using a least-squares optimizer with a \citet{Kreidberg2015} transit model (\texttt{batman}) and a systematic model and calculated limb darkening coefficients with \texttt{ExoTiC-LD} \citep{Grant2024} using the non-linear law \citep{Claret2000} and a Phoenix stellar model (\citealt{Husser2013}; T$_\mathrm{eff}$=3300\,K, log(g)=5.0, [Fe/H]=0.0). For all six sets of light curves (two for G395M and four for G395H), we used the same systematics model, $S(\lambda)$, as in \citetalias{May2023} \citeyear{May2023}, correcting for a linear trend in time, $t$, plus the change in x-position, $x_{s}$, multiplied by the absolute magnitude of the y-positional change, $|y_{s}|$, i.e., $S(\lambda) = s0 + (s1 \times x_{s}|y_{s}|) + (s2 \times t) \mathrm{,}$ (where $s0, s1, s2$ are coefficient terms). We also removed any data points that were greater than 4$\sigma$ outliers in the residuals, and rescaled the flux errors using the beta value \citep{Pont2006} to account for any remaining red noise. In the case of the G395H white light curves, we additionally removed the first 15 integrations from the time series to remove a settling ramp seen in the light curves.

For our initial white light curve fits, we fitted for $R_p/R_s$, $a/R_s$, $i$ and $T_0$, holding the period fixed to 1.628931\,days and the eccentricity fixed to 0 \citep{Bonfils2018}. We then calculated the weighted average of the $a/R_s$ and $i$ values from the two G395M and four G395H white light curves and refit them holding $a/R_s$ and $i$ fixed to these values (see \autoref{tab:orbital_parameters}). Finally, we fit the spectroscopic light curves for each visit, fitting only for $R_p/R_s$, fixing $T_0$ to the value obtained from each respective white light curve refit, and fixing all other parameters to the values as outlined in \autoref{tab:orbital_parameters}. The final weighted mean spectrum can be seen alongside the \firefly and \eureka spectra in \autoref{fig:spectrum_reduction_compare}.

\subsection{Intervisit and reduction agreement} \label{sec:reduction_intervisit_agreement}

\begin{figure*}
\figurenum{3}
    \centering
    \includegraphics[width=1\linewidth]{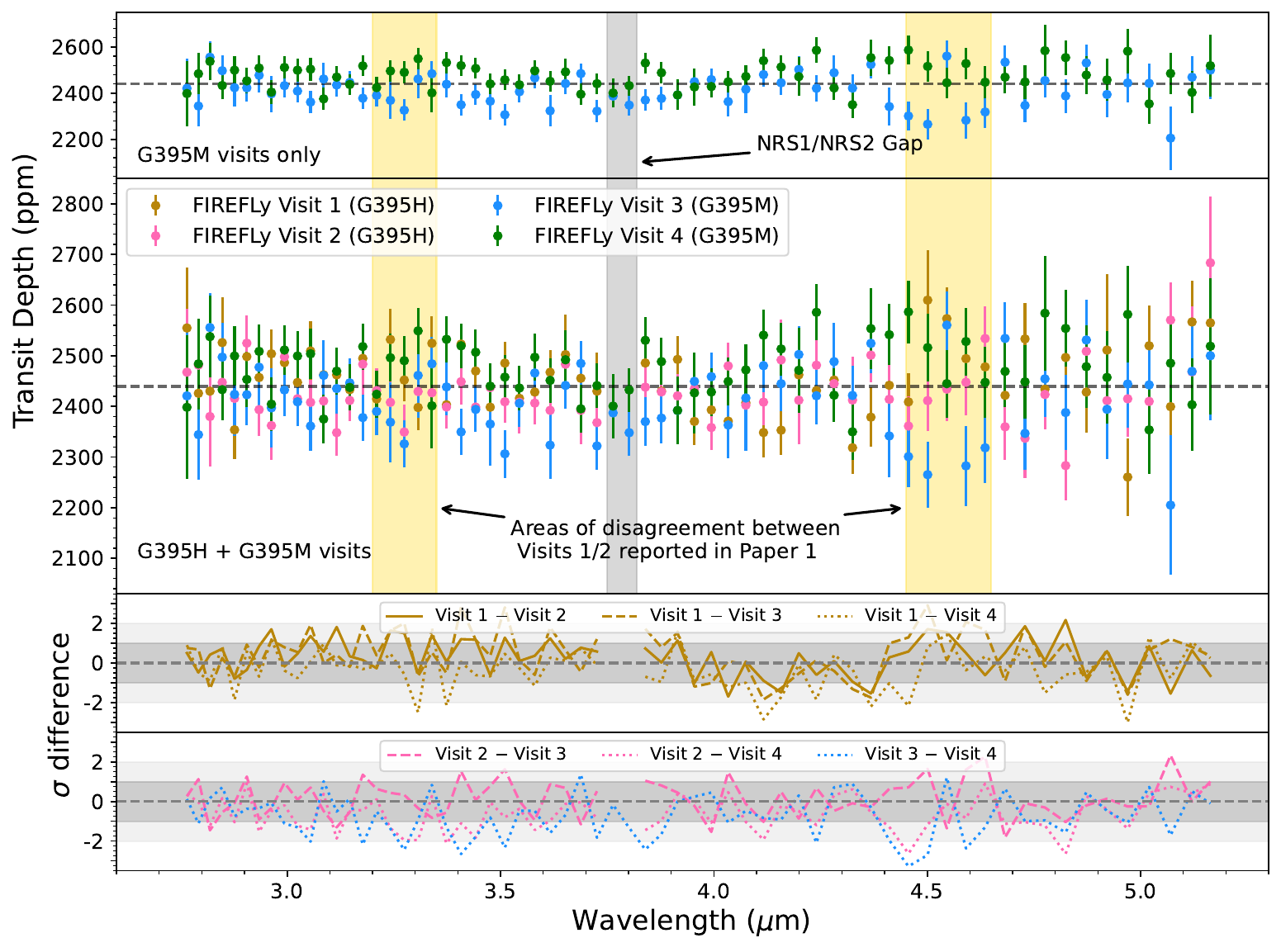}
    \caption{\textbf{Top:} Transmission spectra of \planetname shown individually for the two G395M visits (Visits 3 and 4) for the \firefly reduction. \textbf{Middle:} Same as top, but including the two G395H visits (Visits 1 and 2). For both panets, no offsets have been applied. Grey dashed horizontal line is the mean transit depth. Vertical gray region shows the gap between the NRS1 and NRS2 detectors, and the vertical yellow regions show the areas of the spectrum that drove differences in retrievals for the first two visits from \citetalias{May2023} \citeyear{May2023} (Paper 1). \textbf{Bottom}: $\sigma$ differences between different visits, with the $1\sigma$ and $2\sigma$ regions shaded in dark and light grey, respectively. There is strong intervisit and intermode agreement. Though there is scatter, all four visits largely show agreement to within $2\sigma$. The figure set for all three reductions is available.}
    \label{fig:spectrum_visit_compare}
\end{figure*}

\autoref{fig:spectrum_visit_compare} shows the \firefly transmission spectra for the two new G395M visits (we refer the reader to Figure 1 from \citetalias{May2023} (\citeyear{May2023}) for an illustration of the two G395H visits), as well as all four visits overplotted. Importantly, no offsets have been applied to any visits nor between the NRS1/NRS2 datasets in the G395H visits. The G395M data fill the gap between the NRS1/NRS2 detectors smoothly, suggesting that offsets between NRS1/NRS2 may not prevent useful data interpretation, at least for observations with a large number  of groups per integration ($\gtrsim7$). For more details, see Section \ref{sec:disc_g395m_vs_h}. 

In \autoref{fig:spectrum_visit_compare}, all four visits agree well, with the majority of points agreeing within $2\sigma$. Across the four visits, there is more visually scatter at the red end, but this is largely due to the decrease in stellar flux at these wavelengths, which increases the overall noise in the spectrum. To illustrate this, note that though the scatter increases visually in terms of ppm, the scatter as measured in $\sigma$ remains consistent across the spectrum. 

With the addition of two more visits, it appears by simple visual inspection that the disagreements between Visits 1 and 2 reported in \citetalias{May2023} (\citeyear{May2023}), highlighted in yellow in the figure, may be due to random noise. There is not inherently more scatter in the region around $\rm 3.3\;\mu m$ compared to elsewhere in the spectrum when looking across all four visits. We note, however that there does appear a disagreement between Visit 3 and the other visits in the region around $\rm 4.5\;\mu m$. The culprit here is unlikely to be planetary in nature, as the fluctuations are on the order of 200~ppm, too large to be explainable by atmospheric variation. Instrumental artifacts remain a possibility, though there is no evidence for increased red noise or scatter in these light curves. Intriguingly, this region aligns with the fundamental band of CO centered around $\rm 4.5\;\mu m$, suggesting that an astrophysical explanation is also worth speculating: the discrepancy could perhaps arise from spatially variable CO on the star itself. While the exoplanet community has been focused on stellar contamination driven by cooler water-dominated spots in the stellar atmosphere (e.g., \citealt{Rackham2018}, \citetalias{Moran2023} \citeyear{Moran2023}), stellar contamination by spatially variable CO has not been explored as deeply, though recent magnetohydrodynamic modeling efforts illustrate that excluding CO and other diatomic molecules can impact the stellar (specifically, the umbral and penumbral) spectra of early M-dwarfs longward of $\rm \sim4\;\mu m$ \citep{Smitha2025}. This suggests that stellar contamination at longer wavelengths cannot be ruled out, though of course random noise instances are also likely to play a role in the Visit 3 discrepancies. We will further discuss stellar variability in Section \ref{sec:stellar_spectra}.

\begin{figure*}
\figurenum{4}
    \centering
    \hspace*{-1.4cm}
    \includegraphics[width=1.18\linewidth]{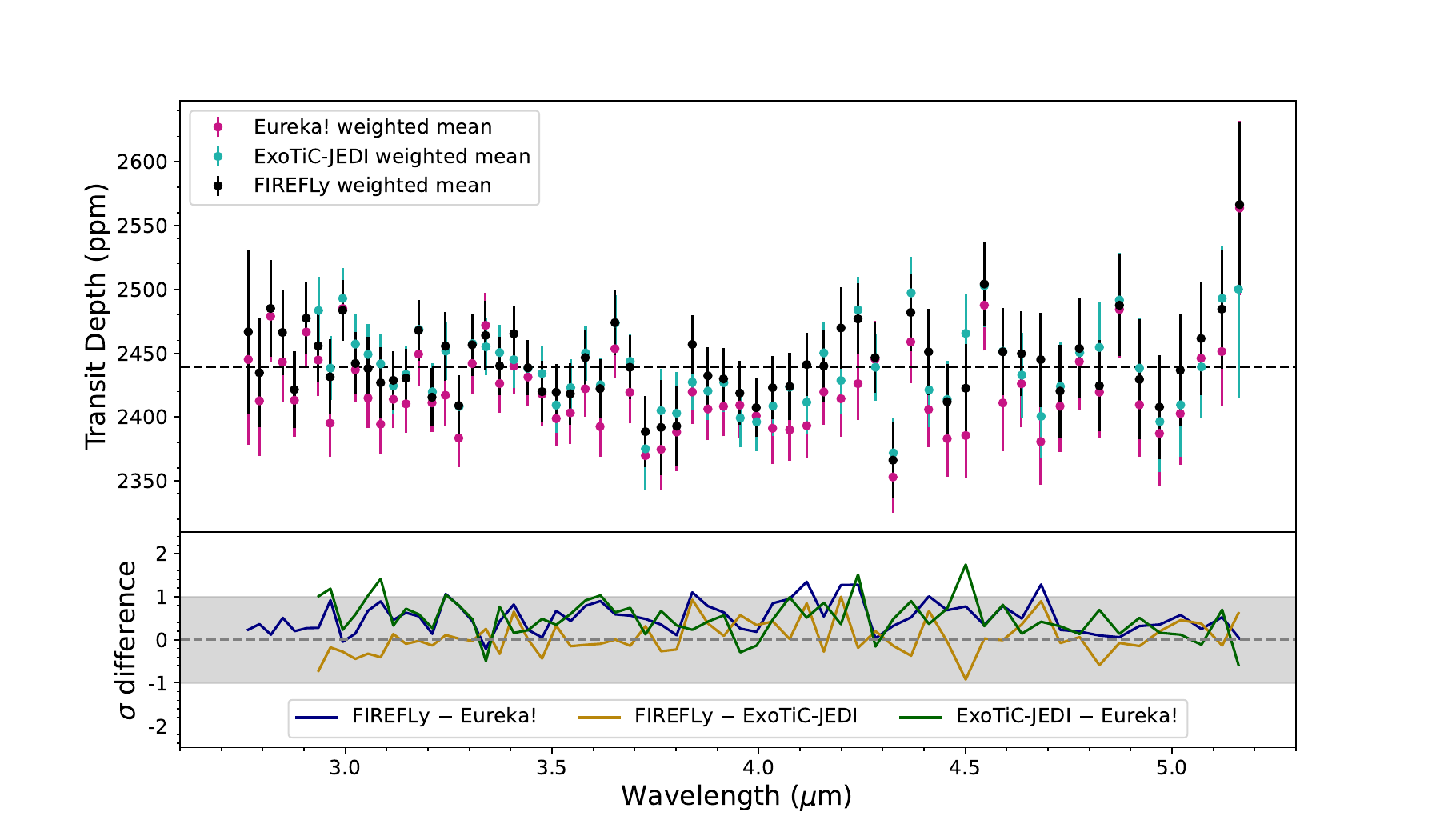}
    \caption{\textbf{Top:} Weighted mean of all four visits is shown for all three reductions, with the mean transit depth for \firefly shown as a horizontal dashed black line. No offsets have been applied to any dataset. \textbf{Bottom:} $\sigma$ difference between reductions is shown, with the $1\sigma$ difference region highlighted in gray. Reductions show exquisite agreement to within 1$\sigma$ for most channels.}
    \label{fig:spectrum_reduction_compare}
\end{figure*}

In addition to good intervisit agreement, we also have exquisite agreement between reductions, as shown in \autoref{fig:spectrum_reduction_compare}. This figure shows the weighted mean spectra from all three reductions, which agree to largely within $1\sigma$. Again, no offsets have been applied to any dataset.
The slightly lower transit depths for \eureka at the blue end are an artifact of limb darkening. When \eureka uses 3D stellar models from \cite{Magic2015} instead of the MPS2 grid, \eureka actually has slightly \textit{higher} transit depth values at the blue end compared to \firefly or \exotic. This demonstrates the importance of careful limb darkening selection/reduction comparison so as to not attribute limb darkening systematics to astrophysical phenomena. Finally, \autoref{tab:orbital_parameters} demonstrates the agreement in fitted orbital parameters across the three reductions, all of which agree to within $1.7\sigma$. 

\figsetstart
\figsetnum{3}
\figsettitle{Visit-to-Visit Transmission Spectra of GJ~1132~b}

\figsetgrpstart
\figsetgrpnum{3.1}
\figsetgrptitle{visit_compare_firefly}
\figsetplot{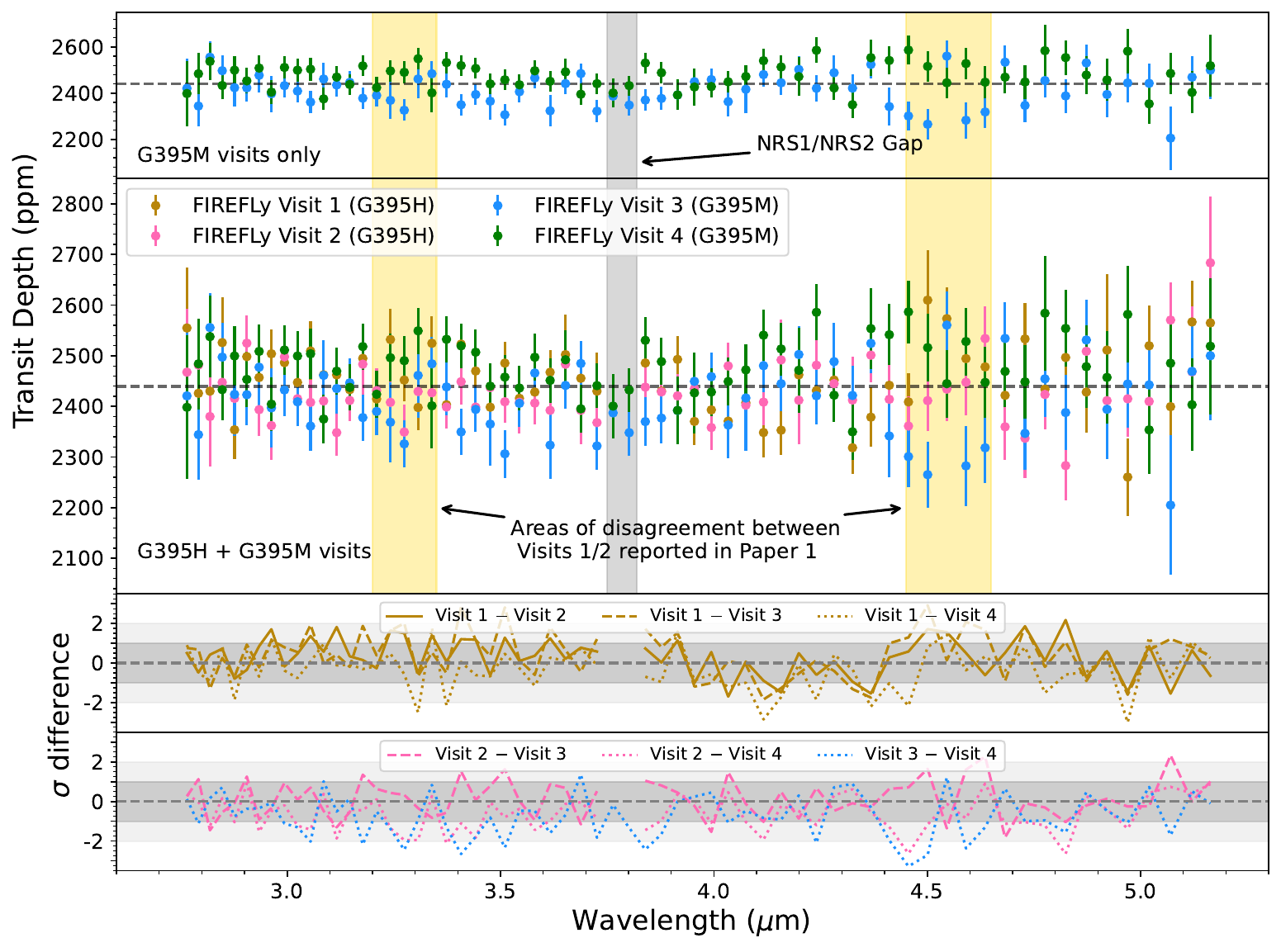}
\figsetgrpnote{\textbf{Top:} Transmission spectra of \planetname shown individually for the two G395M visits (Visits 3 and 4). \textbf{Middle:} Same as top, but including the two G395H visits (Visits 1 and 2). For both panets, no offsets have been applied. Grey dashed horizontal line is the mean transit depth. Vertical gray region shows the gap between the NRS1 and NRS2 detectors, and the vertical yellow regions show the areas of the spectrum that drove differences in retrievals for the first two visits from \citetalias{May2023} \citeyear{May2023} (Paper 1). \textbf{Bottom}: $\sigma$ differences between different visits, with the $1\sigma$ and $2\sigma$ regions shaded in dark and light grey, respectively. There is strong intervisit and intermode agreement. Though there is scatter, all four visits largely show agreement to within $2\sigma$.}
\figsetgrpend

\figsetgrpstart
\figsetgrpnum{3.2}
\figsetgrptitle{visit_compare_eureka}
\figsetplot{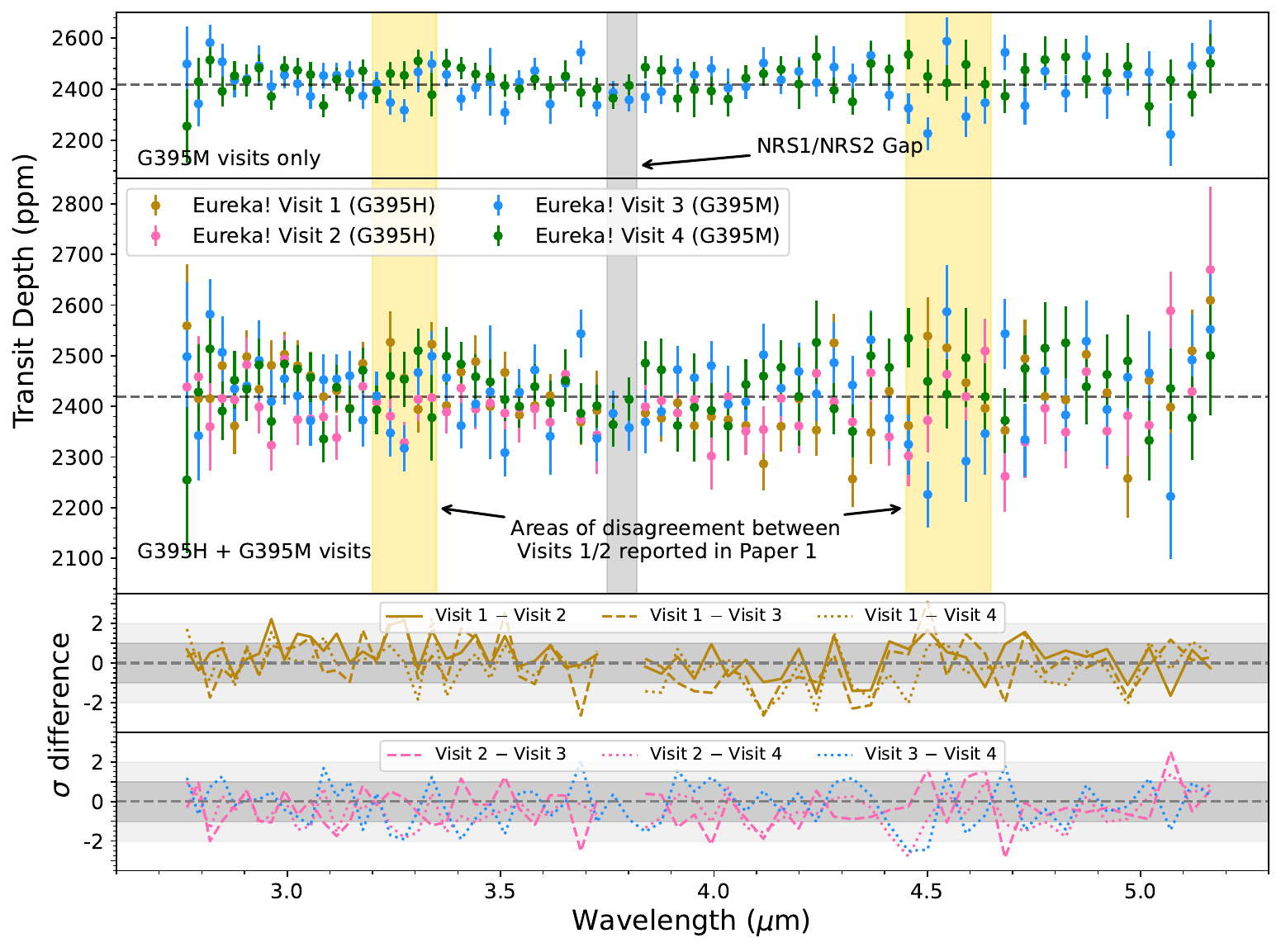}
\figsetgrpnote{\textbf{Top:} Transmission spectra of \planetname shown individually for the two G395M visits (Visits 3 and 4). \textbf{Middle:} Same as top, but including the two G395H visits (Visits 1 and 2). For both panets, no offsets have been applied. Grey dashed horizontal line is the mean transit depth. Vertical gray region shows the gap between the NRS1 and NRS2 detectors, and the vertical yellow regions show the areas of the spectrum that drove differences in retrievals for the first two visits from \citetalias{May2023} \citeyear{May2023} (Paper 1). \textbf{Bottom}: $\sigma$ differences between different visits, with the $1\sigma$ and $2\sigma$ regions shaded in dark and light grey, respectively. There is strong intervisit and intermode agreement. Though there is scatter, all four visits largely show agreement to within $2\sigma$.}
\figsetgrpend

\figsetgrpstart
\figsetgrpnum{3.3}
\figsetgrptitle{visit_compare_exotic}
\figsetplot{Gj1132b_visit_comparison_w_sigma_diff_exotic_g395m_separate.pdf}
\figsetgrpnote{\textbf{Top:} Transmission spectra of \planetname shown individually for the two G395M visits (Visits 3 and 4). \textbf{Middle:} Same as top, but including the two G395H visits (Visits 1 and 2). For both panets, no offsets have been applied. Grey dashed horizontal line is the mean transit depth. Vertical gray region shows the gap between the NRS1 and NRS2 detectors, and the vertical yellow regions show the areas of the spectrum that drove differences in retrievals for the first two visits from \citetalias{May2023} \citeyear{May2023} (Paper 1). \textbf{Bottom}: $\sigma$ differences between different visits, with the $1\sigma$ and $2\sigma$ regions shaded in dark and light grey, respectively. There is strong intervisit and intermode agreement. Though there is scatter, all four visits largely show agreement to within $2\sigma$.}
\figsetgrpend

\figsetend

\begin{deluxetable*}{rl|rlrlrl}
\tablewidth{0pt}
\tablecaption{Best fit orbital parameters from white light curve fitting. \label{tab:orbital_parameters}}
\tablehead{
    \multicolumn{2}{c}{Parameter} & 
    \multicolumn{2}{c}{\firefly} & 
    \multicolumn{2}{c}{\eureka} & 
    \multicolumn{2}{c}{\exotic}
}
\startdata
(R$_p$/R$_s$)$^2$ & (ppm) & 2428 & $\pm\;$ 8 & 2448 & $\pm\;$ 13 & 2439 & $\pm\;$ 8 \\
Visit 1 T$_0$ -- 2460000 & (BJD$_{TDB}$) & 0.97654 & $\pm\;$ 0.00002 & 0.97652 & $\pm\;$ 0.00001  & 0.97653 & $\pm\;$ 0.00002 \\
Visit 2 T$_0$ -- 2460009 & (BJD$_{TDB}$) & 0.12115 & $\pm\;$ 0.00002 & & -- & 0.12115 & $\pm\;$ 0.00002 \\
Visit 3 T$_0$ -- 2460349 & (BJD$_{TDB}$) & 0.56744 & $\pm\;$ 0.00002 & & -- & 0.56746 & $\pm\;$ 0.00002 \\
Visit 4 T$_0$ -- 2460466 & (BJD$_{TDB}$) & 0.85035 & $\pm\;$ 0.00002 & & -- & 0.85036 & $\pm\;$ 0.00002 \\
Period & (days) & 1.628931 & (fixed) & 1.62892962 & $\pm\;$ 7.15E-8 & 1.628931 & (fixed) \\
a/R$_s$ & (unitless) & 15.53 & $\pm\;$ 0.22 & 15.88 & $\pm\;$ 0.19 & 15.51  & $\pm\;$ 0.17  \\
b & (unitless) & 0.44 & $\pm\;$ 0.03 & 0.39 & $\pm\;$ 0.03  & 0.44 & $\pm\;$ 0.02 \\
i & (\textdegree) & 88.37 & $\pm\;$ 0.10  & 88.61 & $\pm\;$ 0.11 & 88.38 & $\pm\;$ 0.09   \\
\enddata
\tablecomments{\firefly and \exotic report (R$_p$/R$_s$)$^2$, $a/R_s$ and $b$ (or $i$) as the weighted mean from all six independent white light curves fits. \eureka reports values from a joint white light curve fit, and therefore only directly fits for the first center of transit time and the period. \eureka and \exotic use time units of BMJD$_{\rm TDB}$, which is converted to BJD$_{\rm TDB}$ here. $b$ is measured in the \firefly light curve fits and then converted to $i$ for this table, whereas the opposite is true for \exotic and \eureka. All fitted parameters agree to within $1.7\sigma$ across reductions.}
\vspace{-1cm}
\end{deluxetable*}

\section{Analysis} \label{sec:analysis}

\subsection{Gaussian Detection Test Favors a Flat Line} 

Following previous works (e.g., \citetalias{Moran2023} \citeyear{Moran2023}, \citetalias{May2023} \citeyear{May2023}, \citealp{Kirk2024}), we begin our analysis of the transmission spectrum by simply determining whether it rejects the null hypothesis. In this case, the null hypothesis is represented by a flat, featureless spectrum, indicative of no evidence for spectroscopic absorption features from an atmosphere. We perform two common statistical tests. 

In our first test, we assess the consistency of the data with the null hypothesis by computing the cumulative probability (CDF) of the observed chi-squared under the null model. We then convert this probability, $p$, into a Gaussian equivalent significance using $z=\sqrt{2} {\rm erf}^{-1}(p)$, representing the number of standard deviations by which the observation deviates from the null expectation. Results of this test for each data reduction are given in \autoref{tab:flat}. Small values of $z$ (e.g., $z<2$) indicate a high degree of consistency with the null hypothesis, while larger values (e.g., $z>3$) suggest increasing tension that may justify rejection of the null at high confidence. 

In our second test, we directly compare the null hypothesis model to a more complex, albeit still quite simple, alternative model with a Gaussian absorption feature added to the flat model. We elect to use a Gaussian model to approximate molecular absorption in a transmission spectrum \citep[e.g.,][]{Taylor2025}; we allow the feature to be as broad as the full G395H wavelength range and centered outside of this range so as to also approximate a spectral slope (e.g. from H$_2$O, as in \citetalias{Moran2023} \citeyear{Moran2023}) should that be preferred. Since the models are nested, they enable a comparison of their respective Bayesian evidence terms estimated using the \texttt{dynesty} Nested Sampling code \citep{Skilling2004, Speagle2020}. We calculate Bayes factors from the ratio of the two model evidences and then report equivalent sigma values in \autoref{tab:flat} for the detection significance of the Gaussian model over the null hypothesis \citep{Trotta2008}. We use a negative sign in this case to flip the interpretation and denote the detection of the flat line over the Gaussian feature.  

\autoref{tab:flat} reports results from both our null hypothesis and Gaussian feature detection tests for all three independent data reductions using the weighted mean of the four visits (\autoref{fig:spectrum_reduction_compare}). As previously discussed, the three reductions are in excellent agreement and this agreement holds for the statistical properties of our null hypothesis tests. All three reductions are consistent with a featureless spectrum to within approximately 1$\sigma$ of the expected range in $\chi^2_{\nu}$ values given the null hypothesis and no significant evidence is found to favor a Gaussian feature in the data. 

\begin{deluxetable}{l||c|c|c|c}
\tablewidth{0.98\textwidth}
\tabletypesize{\small}
\tablecaption{Is it Flat? \label{tab:flat}}
\tablehead{
\colhead{Reduction} & \colhead{$\chi^{2}_{\nu}$} & \colhead{$\nu$} & \colhead{Dev. from Null} & \colhead{Gaussian Detection}}
\startdata
\eureka   &   1.09 &  64 &    1.04$\sigma$ &   1.71$\sigma$ \\
\firefly  &   0.88 &  64 &    0.32$\sigma$ &  $-2.00\sigma$ \\
\exotic   &   1.13 &  58 &    1.20$\sigma$ &   1.49$\sigma$ \\
\enddata
\tablecomments{$\chi^{2}_{\nu}$ is the reduced chi-squared resulting from the best fitting featureless fit (null hypothesis) to the observed spectrum. ``Dev. from Null'' ($z$) refers to how many sigmas the $\chi^{2}_{\nu}$ deviates from expectation under the null. ``Gaussian Detection'' shows the equivalent detection significance of an agnostic Gaussian absorption feature in the spectrum over the flat null hypothesis from a Bayes factor model comparison, where the negative sign denotes preference instead for the flat model over the Gaussian.} 
\end{deluxetable}

\subsection{Models Hint At Evolution of Stellar Heterogeneity Between Visits} \label{sec:stellar_spectra}

To explore potential changes in stellar heterogeneity, we forward modeled the out-of-transit, flux-calibrated stellar spectra for all four visits. This analysis followed the same procedures presented in \citetalias{May2023} (\citeyear{May2023}), but using the \firefly reductions and adjusting each spectrum to match the lower resolution and wavelength sampling of the G395M visits (for consistency in determining inhomogeneities between visits). Additionally, we use a wider aperture for the flux-calibrated spectrum than the transmission spectrum, because we want to ensure we capture the entirety of the stellar flux. We include all flux within $9.7\sigma$ of the center of the column-level PSF (the maximum available width before exceeding the bounds of the detector), which equates to an aperture full-width of $13.35$ pixels for G395H NRS1, $15.15$ pixels for G395H NRS2, and and $14.12$ pixels for G395M. We confirm that the flux-calibrated spectra do not differ between reductions. Because the \jwst pipeline-estimated error bar size is underestimated \citep{Jakobsen2022}, we calculate empirical error bars. We estimate an initial error bar size based on the mission-required flux precision from JDox\footnote{\url{https://jwst-docs.stsci.edu/jwst-calibration-status/jwst-absolute-flux-calibration}}, then inflate/deflate this percentage until the $\chi^2_{\nu}$ between the data and best-fit model (see below) is 1. Assuming the best-fit model is true, we achieve a precision of $\sim2.2\%$ and $\sim1.3\%$ for the visits to G395H and G395M, respectively. This is consistent with the absolute flux calibration level ($\sim2\%$) of the NIRSpec instrument.

We utilize multi-component stellar models based on the \citet{Allard2012} \texttt{PHOENIX} models, aiming to detect any rotational or evolutionary modulation of surface features. Using a weighted combination of three \texttt{PHOENIX} models, we modeled the background photosphere, cooler regions representing starspots ($T_{\rm eff}$ $\leq$ $T_{\rm eff, \,photosphere}$ -- 100 K), and warmer regions representing faculae ($T_{\rm eff}$ $\geq$ $T_{\rm eff, \,photosphere}$ + 100 K). Each component shared the same log($g$) and metallicity (held constant at log($g$) = 4.5 cm/s$^2$ and Z=0; consistent with \citetalias{May2023} \citeyear{May2023}), and the coverage of the cooler/warmer regions was each allowed to represent up to 50\% of the surface. The model grid ranged from $T_{\rm eff}$ = 2500--4500~K.

For direct comparison with the observed spectra, we normalized the models to the observed flux at 3 $\mu$m and interpolated them to match the resolution and wavelength sampling of Visit 4 before computing $\chi_{\nu}^2$. Each $\chi_{\nu}^2$ was calculated across 1,273 wavelength points and five free parameters: $T_{\rm eff}$ of the photosphere, cool spots and warm spots, and coverage fractions of both feature types. 

\begin{figure*}[ht]
\figurenum{5}
    \centering
    \begin{minipage}{0.45\textwidth}
        \centering
        \includegraphics[width=\textwidth]{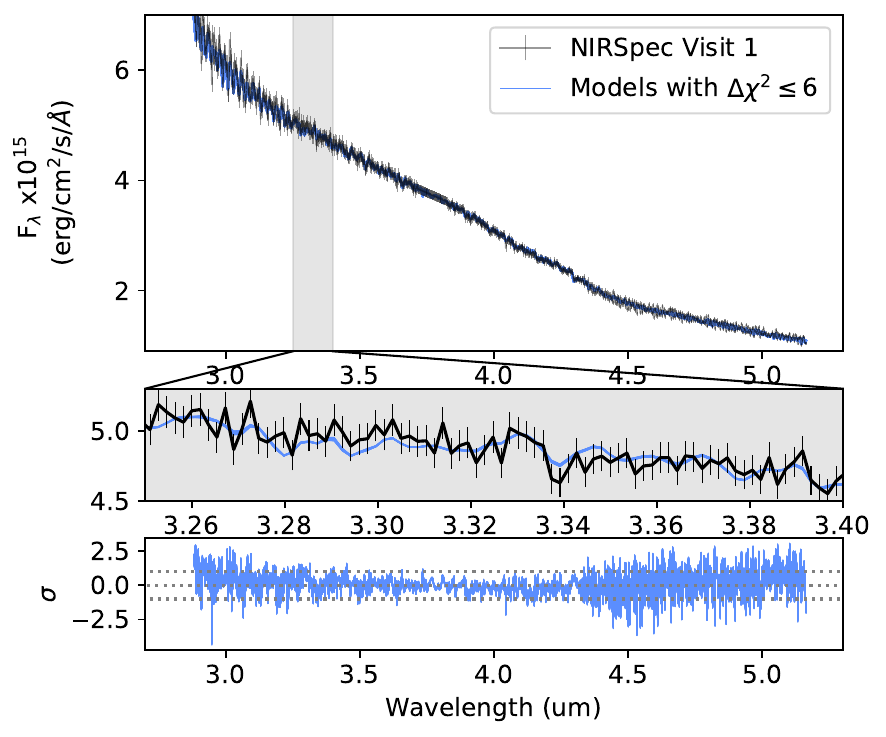}
    \end{minipage}
    \hspace{0.2cm}
    \begin{minipage}{0.45\textwidth}
        \centering
        \includegraphics[width=\textwidth]{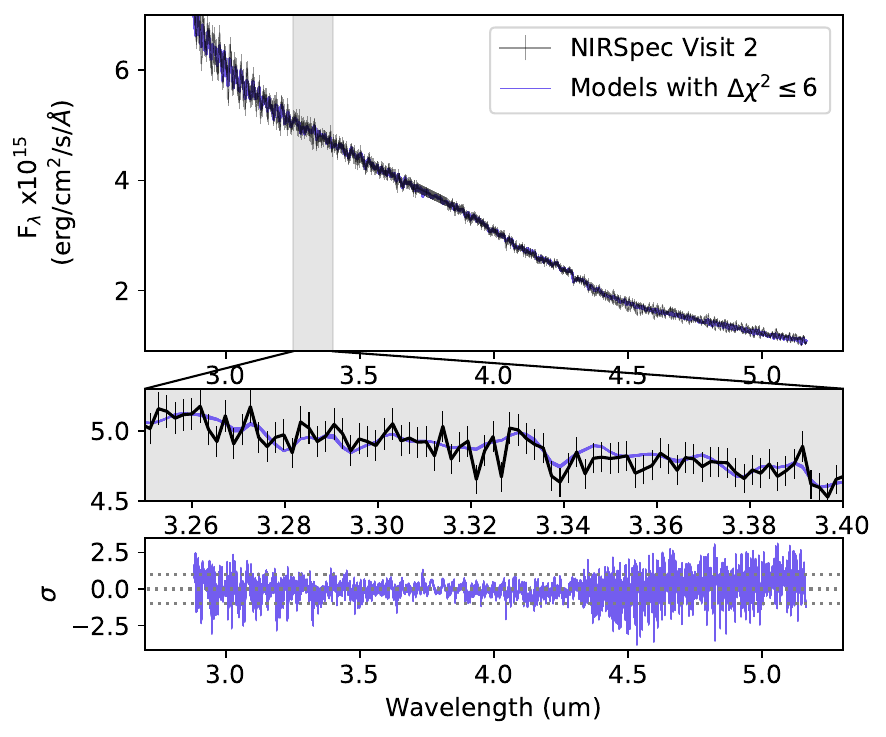}
    \end{minipage}

    \vspace{0.5cm} 

    \begin{minipage}{0.45\textwidth}
        \centering
        \includegraphics[width=\textwidth]{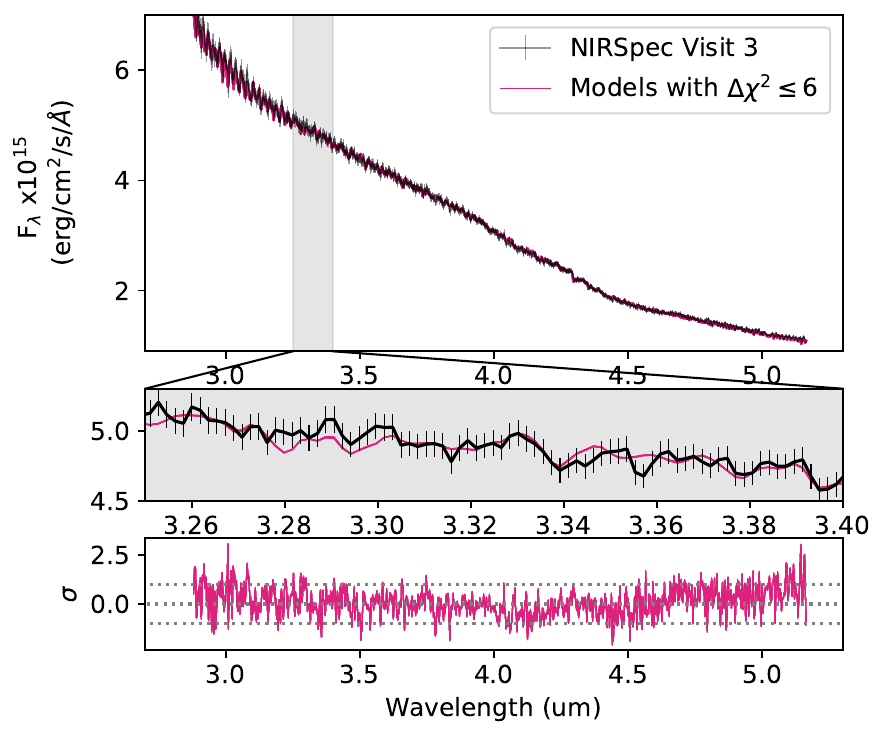}
    \end{minipage}
    \hspace{0.2cm}
    \begin{minipage}{0.45\textwidth}
        \centering
        \includegraphics[width=\textwidth]{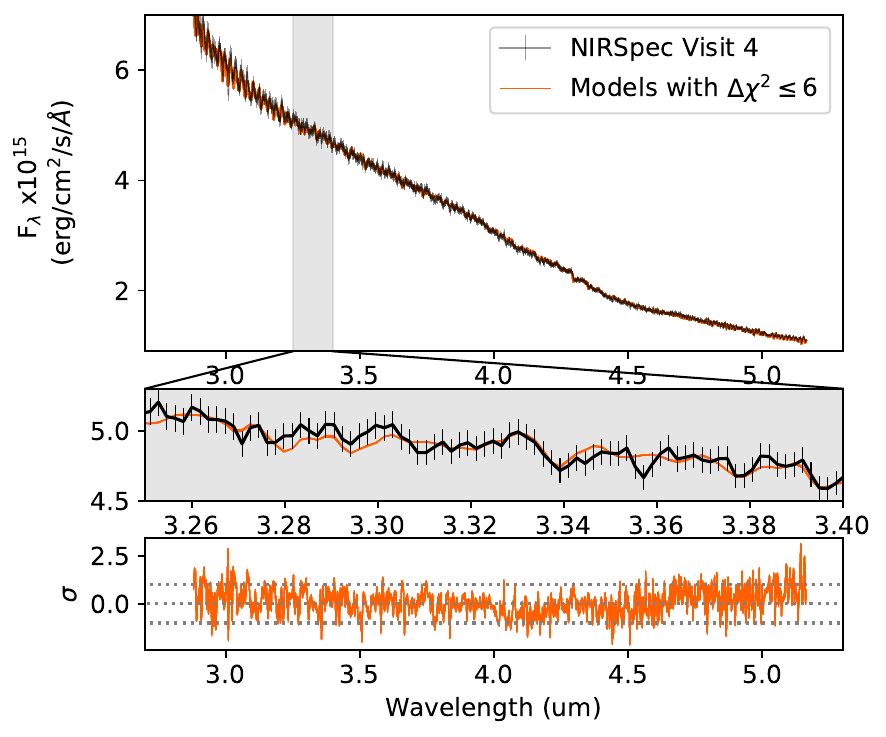}
    \end{minipage}

    \caption{Best-fit multi-component stellar models for each visit. Each of the four main panels corresponds to one visit, and contains three subpanels. In every subpanel, all models with $\Delta\chi^2 \leq 6$ from the best-fit are shown (25 models for Visit 1, 46 for Visit 2, 9 for Visit 3, and 10 for Visit 4). The models are nearly indistinguishable due to their close agreement. The observed spectra, smoothed to the resolution of G395M, are overplotted in black. The residuals (in units of $\sigma$) are shown in each bottom subpanel, with dotted lines indicating the $\pm1\sigma$ range.}
    \label{fig:stellarspectra}
\end{figure*}

All four visits favored a photosphere temperature of 3100 K, which is consistent within 2-$\sigma$ of the literature value of 3270 $\pm$140 K \citep{Bonfils2018}. They also all favor the same 2900 K for the cooler spots and 3300 K for warmer spots. The visits do differ, however, in the percent coverage, where Visit 1 is an outlier, suggesting $\sim$10\% more cool spots and $\sim$5\% less warm spots than the other three visits. The best-fit parameters and their uncertainties (computed from models with $\Delta\chi^2$ $\leq$ 6 above the minimum, which given the five model parameters yields a 1-$\sigma$ confidence interval) are given in \autoref{tab:spot_parameters}. All best-fit model spectra (including those captured in the uncertainties) are plotted in \autoref{fig:stellarspectra}.

We note that the results from Visit 1 and 2 presented here differ from those computed in \citetalias{May2023} (\citeyear{May2023}), where both visits were previously favored by a 3200 K photosphere and $\sim$40\% 2900 K spots and $\sim$25\% 3500 K faculae. This is due to myriad factors including different aperture sizes used in the reduction (which drive a 1\% difference in the spectra at native resolution), using a lower wavelength resolution in the current analysis, and allowing the cool and warm region coverage to now go up to 50\%, where before it was capped at 45\%. We made this change here to allow for the possibility of a two-component best-fit model.

The results presented here suggest that the visible surface of the star could have had a large starspot near the limb during Visit 1 that rotated off before Visit 2, 8 days later. Interestingly, in the \firefly reduction the fit spectroscopic limb-darkening coefficients are slightly different in Visit 1, compared to the rest of the visits.  Assuming a rotation period of the star of 122.31$^{+6.03}_{-5.04}$ days \citep{Cloutier2017}, there is a 7\% rotation between Visits 1 and 2 and a 95\% rotation between Visits 3 and 4. There are 340.43 days between the two sets of visits, corresponding to a 78\% rotation (when phase-folding). Therefore, we are looking at similar faces of the star for Visits 1 and 2 and for Visits 3 and 4, but between the two, we are seeing slightly different portions of the stellar disk. The total time period between Visit 1 and Visit 4 is 466 days, which allows for the additional possibility that the purported spot from Visit 1 evolved before Visits 3 and 4.

\begin{deluxetable}{l|cccc}
\tablewidth{0pt}
\tablecaption{Best fit multi-component stellar models to each visit. All observations were reduced to the resolution of G395M and the photospheric temperatures are held constant at 3100 K. Uncertainties are computed from models with $\Delta\chi^2\leq$6. \label{tab:spot_parameters}}
\tablehead{
    Parameter & 
    Visit 1&
    Visit 2&
    Visit 3&
    Visit 4
}
\startdata
    T$_{\rm photo}$ [K] & 3100 & 3100 & 3100 & 3100 \\
    $f_{\rm cooler}$ [\%] & $31 \pm 2$ & $21^{+3}_{-4}$ & $24 \pm 1$ & $22^{+2}_{-1}$\\
    T$_{\rm cooler}$ [K] & $2900 \pm 0$ & $2900^{+0}_{-100}$ & $2900 \pm 0$ & $2900 \pm 0$\\  
    $f_{\rm warmer}$ [\%] & $22^{+3}_{-4}$ &  $27^{+6}_{-3}$ & $26 \pm 2$ &  $28 \pm 2$\\
    T$_{\rm warmer}$ [K] & $3300 \pm 0$ & $3300 \pm 0$ & $3300 \pm 0$ & $3300 \pm 0$\\
    $\chi^{2}_{\nu, \text{min}}$ & 1.0 & 1.0 & 1.0 & 1.0 \\
    \textit{n}$^{\dagger}$ & 25 & 46 & 9 & 10\\
    $\delta$t$^{\dagger\dagger}$ [days] & 0 & 8.14 & 348.57 & 465.86 \\
    $\varphi^{\dagger\dagger\dagger}$ & 0 & $0.07 \pm 0.003$ & $0.85^{+0.13}_{-0.12}$ & $0.81^{+0.18}_{-0.16}$ \\
    \hline
\enddata
\tablecomments{$\dagger$: number of models with $\Delta\chi^2\leq$6.\\$\dagger\dagger$: time (in days) from the first observation\\ $\dagger\dagger\dagger$: phase-folded phase difference from the first observation assuming a rotation period of 122.31$^{+6.03}_{-5.04}$ days \citep{Cloutier2017}.}
\vspace{-1cm}
\end{deluxetable}

\subsection{Forward Models Favor Featureless Spectrum or Thin Steam Atmosphere} \label{sec:fwd_models}

\begin{figure*}[h]
\figurenum{6}
    \centering
    \includegraphics[width=0.8\linewidth]{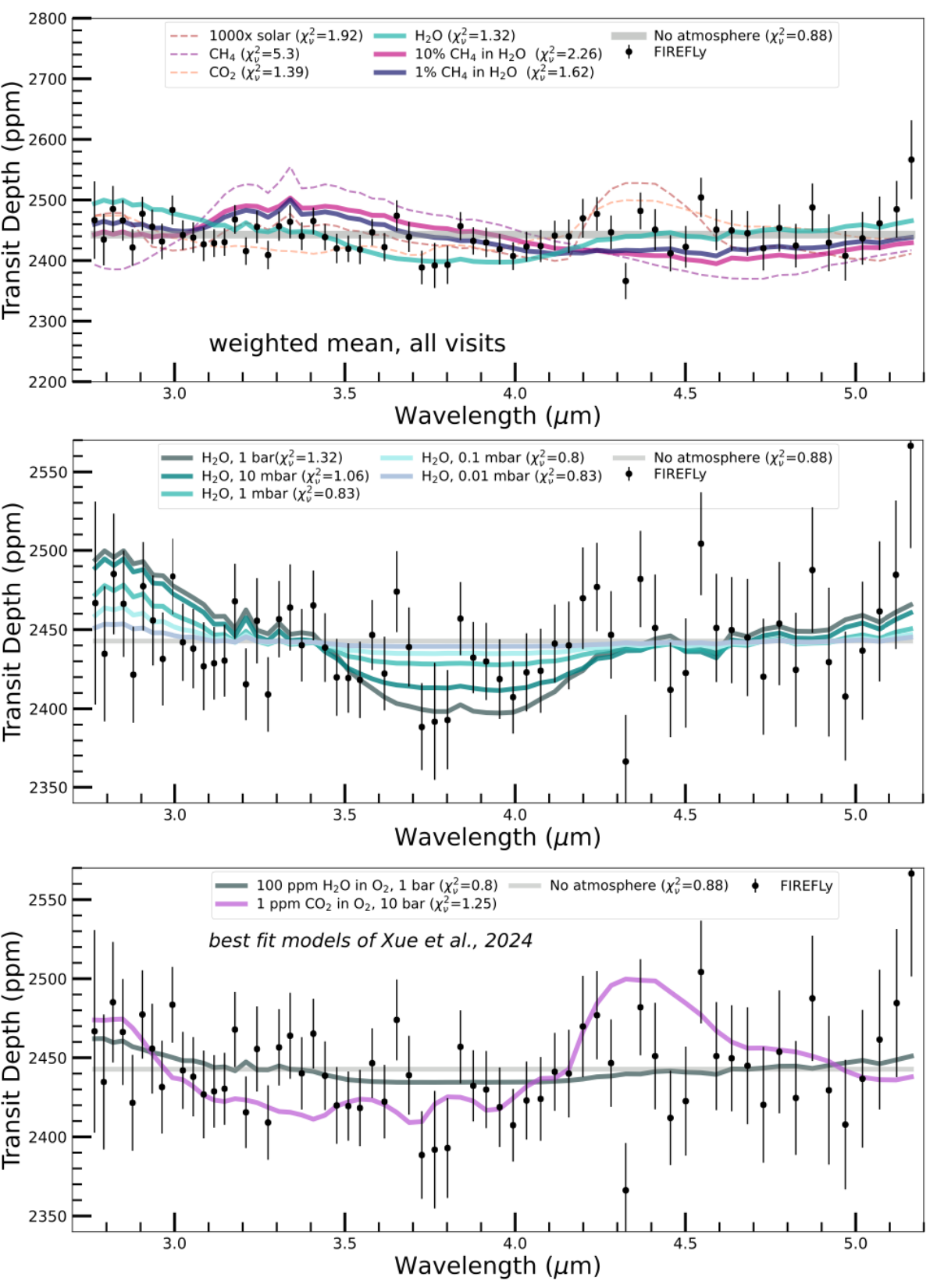}
    \caption{Forward model fits compared to the  weighted mean of all four visits are shown for the \firefly reduction. $\chi^2_{\nu}$ values, shown in the legend, are calculated assuming 64 degrees of freedom.  \textbf{Top:} Forward model fits for a variety of clear atmospheres, similar to those shown in \citetalias{May2023} (\citeyear{May2023}). The best-fit of those shown is a flat line model. \textbf{Middle:} Forward model fits for a pure water atmosphere with varying opaque pressure levels, which can represent either a surface pressure or that of a cloud-top. A water atmosphere approximately 1 mbar in pressure is more likely than a flat line model (see text for details). \textbf{Bottom:} The two best-fitting atmospheric forward models from the MIRI emission study of \citet{Xue2024} compared to our transmission spectra. We rule out their 1 ppm CO$_2$ in O$_2$ atmosphere to $\geq 3\sigma$ with our data while their 100 ppm H$_2$O case is consistent with our transmission spectrum. The complete figure set (one for each reduction) is available.}
    \label{fig:forward_models}
\end{figure*}

To further interpret the combined transmission spectrum, we follow our previous works (\citetalias{Lustig-Yaeger2023,Moran2023,May2023}, \citealt{Kirk2024}) and compare our observations of \planetname to a series of atmospheric forward models generated for \planetname's planetary mass, radius, and equilibrium temperature. We include the same series of models from our original \planetname study (\citetalias{May2023} \citeyear{May2023}), as summarized briefly below, as well as an expanded model set with additional atmospheric pressures given the increased precision enabled by our four visit combined spectrum. All models were produced using the \texttt{PICASO} \citep{Batalha2019} radiative transfer code, using molecular opacities derived from \texttt{PICASO}'s Zenodo v2 database of Resampled Opacities at R=60,000 \citep{Batalha2022}. We rebinned each model to the resolution of the combined transmission spectrum for each reduction to compute a goodness-of-fit in terms of the $\chi_\nu^2$ assuming 64 degrees of freedom ($\nu=64$). We show these results in \autoref{fig:forward_models}, where the best-fit results ($\chi^2_{\nu}<1.18$) have p-values corresponding to low ($<1\sigma$) significance levels for rejecting the models, indicating no significant disagreement with the data.

\figsetstart
\figsetnum{6}
\figsettitle{Forward Models for GJ~1132~b}

\figsetgrpstart
\figsetgrpnum{6.1}
\figsetgrptitle{fwd_models_firefly}
\figsetplot{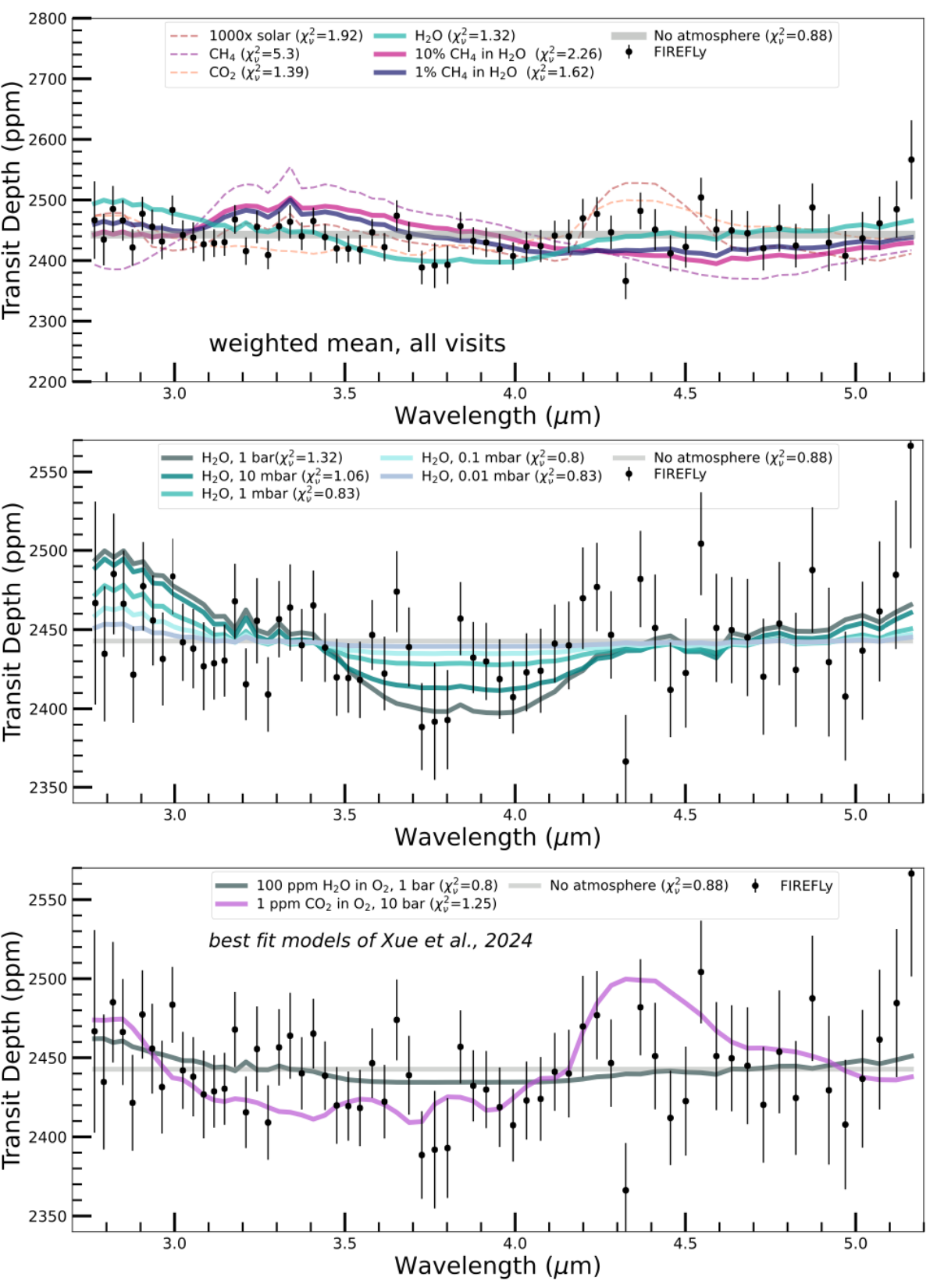}
\figsetgrpnote{Forward model fits compared to the  weighted mean of all four visits are shown. $\chi^2_{\nu}$ values, shown in the legend, are calculated assuming 64 degrees of freedom.  \textbf{Top:} Forward model fits for a variety of clear atmospheres, similar to those shown in \citetalias{May2023} (\citeyear{May2023}). The best-fit of those shown is a flat line model. \textbf{Middle:} Forward model fits for a pure water atmosphere with varying opaque pressure levels, which can represent either a surface pressure or that of a cloud-top. A water atmosphere approximately 1 mbar in pressure is more likely than a flat line model (see text for details). \textbf{Bottom:} The two best-fitting atmospheric forward models from the MIRI emission study of \citet{Xue2024} compared to our transmission spectra. We rule out their 1 ppm CO$_2$ in O$_2$ atmosphere to $\geq 3\sigma$ with our data while their 100 ppm H$_2$O case is consistent with our transmission spectrum.}
\figsetgrpend

\figsetgrpstart
\figsetgrpnum{6.2}
\figsetgrptitle{fwd_models_eureka}
\figsetplot{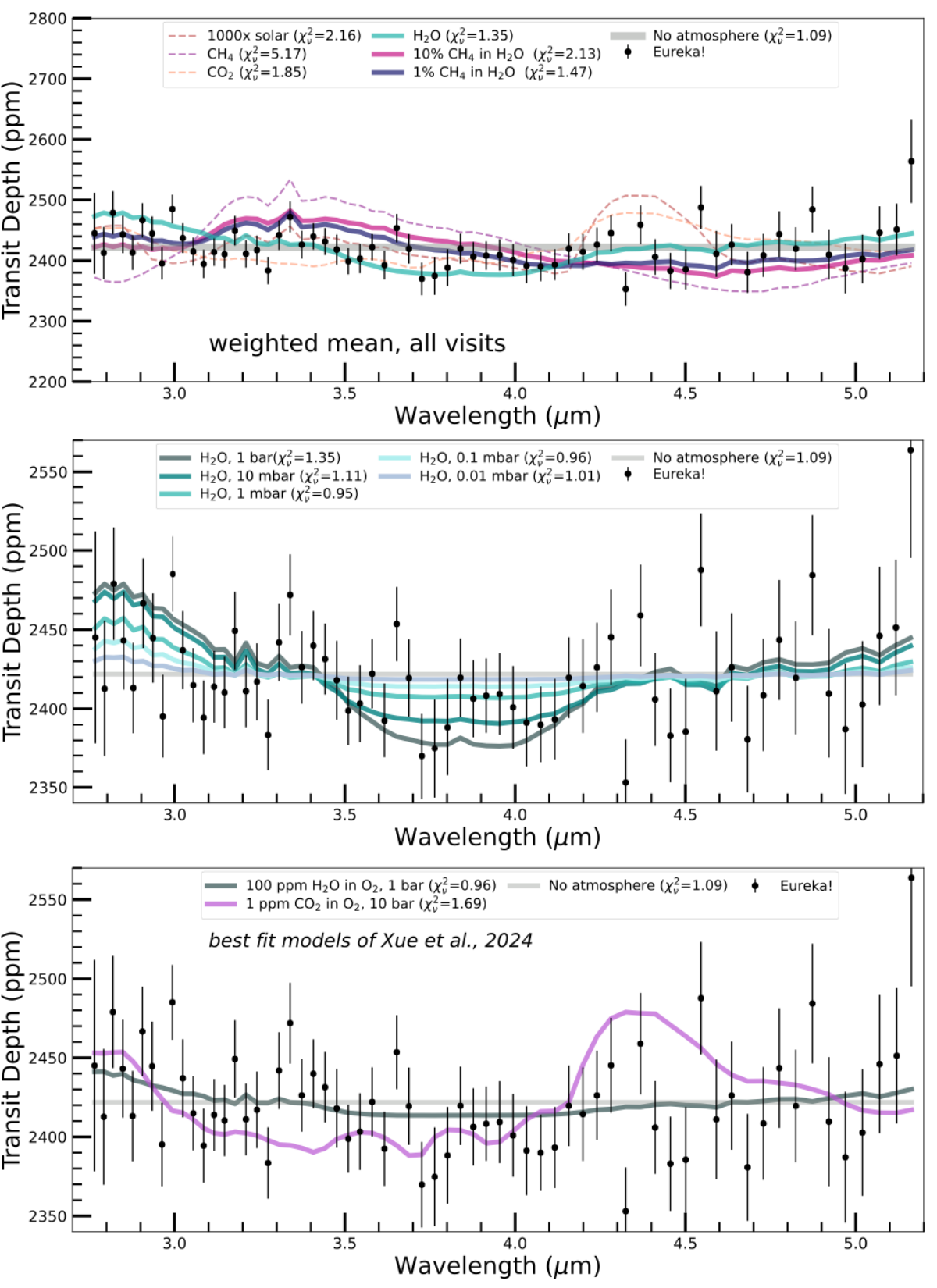}
\figsetgrpnote{Forward model fits compared to the  weighted mean of all four visits are shown. $\chi^2_{\nu}$ values, shown in the legend, are calculated assuming 64 degrees of freedom.  \textbf{Top:} Forward model fits for a variety of clear atmospheres, similar to those shown in \citetalias{May2023} (\citeyear{May2023}). The best-fit of those shown is a flat line model. \textbf{Middle:} Forward model fits for a pure water atmosphere with varying opaque pressure levels, which can represent either a surface pressure or that of a cloud-top. A water atmosphere approximately 1 mbar in pressure is more likely than a flat line model (see text for details). \textbf{Bottom:} The two best-fitting atmospheric forward models from the MIRI emission study of \citet{Xue2024} compared to our transmission spectra. We rule out their 1 ppm CO$_2$ in O$_2$ atmosphere to $\geq 3\sigma$ with our data while their 100 ppm H$_2$O case is consistent with our transmission spectrum.}
\figsetgrpend

\figsetgrpstart
\figsetgrpnum{6.3}
\figsetgrptitle{fwd_models_exotic}
\figsetplot{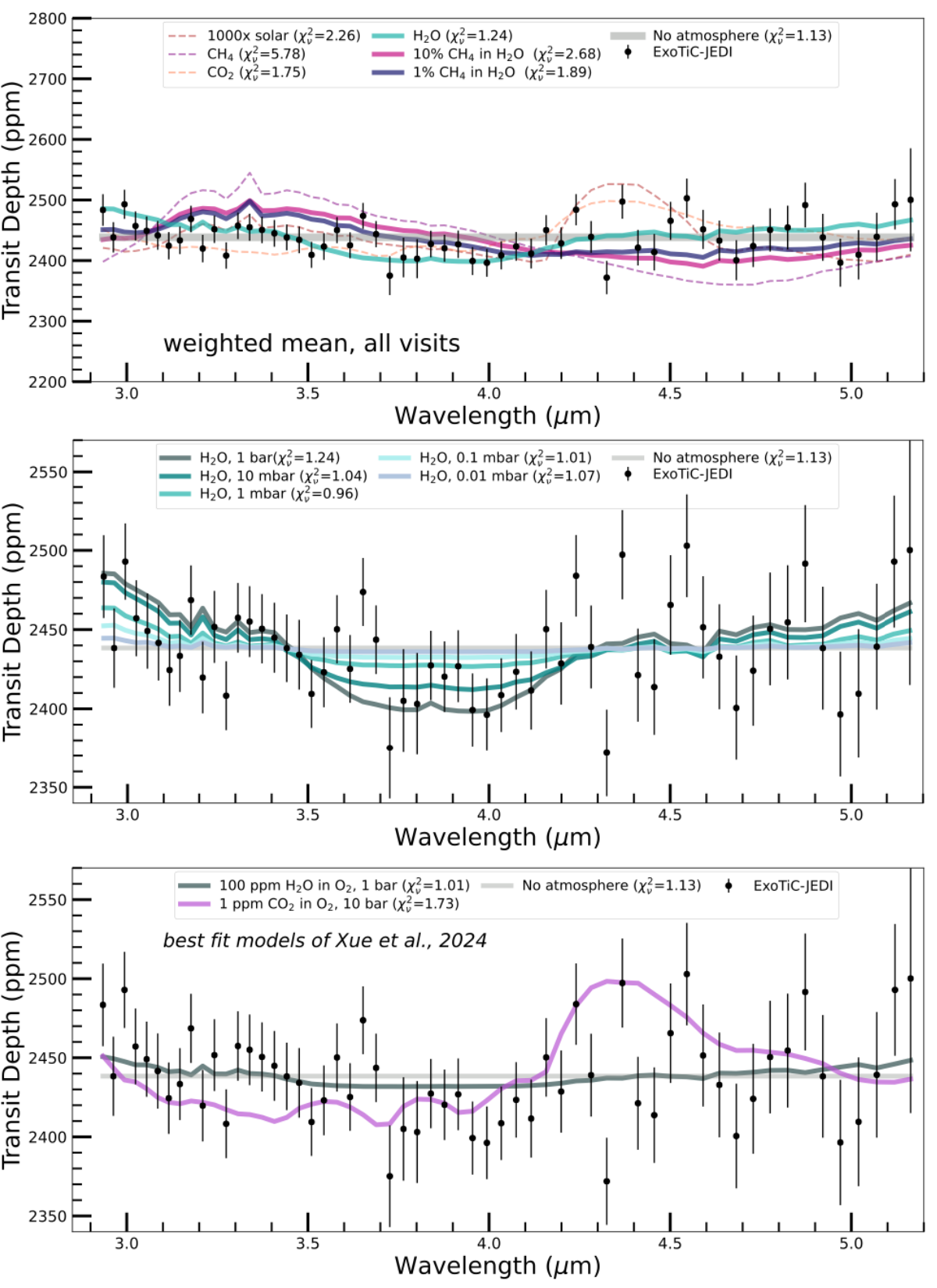}
\figsetgrpnote{Forward model fits compared to the  weighted mean of all four visits are shown. $\chi^2_{\nu}$ values, shown in the legend, are calculated assuming 64 degrees of freedom.  \textbf{Top:} Forward model fits for a variety of clear atmospheres, similar to those shown in \citetalias{May2023} (\citeyear{May2023}). The best-fit of those shown is a flat line model. \textbf{Middle:} Forward model fits for a pure water atmosphere with varying opaque pressure levels, which can represent either a surface pressure or that of a cloud-top. A water atmosphere approximately 1 mbar in pressure is more likely than a flat line model (see text for details). \textbf{Bottom:} The two best-fitting atmospheric forward models from the MIRI emission study of \citet{Xue2024} compared to our transmission spectra. We rule out their 1 ppm CO$_2$ in O$_2$ atmosphere to $\geq 3\sigma$ with our data while their 100 ppm H$_2$O case is consistent with our transmission spectrum.}
\figsetgrpend

\figsetend

We include one model using abundances of a 1000$\times$ solar metallicity, solar C/O ratio atmosphere generated using the \texttt{CHIMERA} \citep{line2013,line2014} thermochemical equilibrium model. This model includes opacities from $\rm H_2$ collision-induced absorption (CIA), H$_2$, He, H$_2$O, CH$_4$, CO, CO$_2$, NH$_3$, N$_2$, HCN, and H$_2$S with a \citet{guillot2010} temperature-pressure profile. Of the models explored in \citetalias{May2023} (\citeyear{May2023}), this 1000$\times$ solar metallicity model was not ruled out when considering Visits 1 and 2 independently. Now, when considering the weighted mean of all four visits, all three reductions robustly reject this atmospheric scenario by over 4$\sigma$.

We next compare to a series of either single or two-component atmospheric models at 1 bar in surface/cloud top pressure with an isothermal temperature set to \planetname's equilibrium temperature of $\sim$580~K. Two of these models -- either clear, pure CH$_4$ or CO$_2$ atmospheres -- are strongly rejected by each reduction at $>$10$\sigma$ and 3$\sigma$, respectively. These models can be rejected because the transmission spectrum lacks significant absorption in either the 3.35 $\mu$m or 4.3 $\mu$m regions where these molecules have spectral features. The CH$_4$ model was previously ruled out by our original two visits of \planetname (\citetalias{May2023}, \citeyear{May2023}), but we could not rule out the CO$_2$ model without stacking multiple transits. Finally, motivated by the potential water-rich, methane containing atmosphere we obtained from Visit 1 in \citetalias{May2023} (\citeyear{May2023}), we include water-rich atmospheres with varying amounts of CH$_4$. The pure 1-bar water atmosphere is a better fit ($\chi_\nu^2$ of $\sim$1.2--1.3 for each reduction) than that of either 10\% or 1\% CH$_4$ in water ($\chi_\nu^2$ of $\sim$1.5--2.7 across reductions). All of these clear atmospheric forward models are shown in the top panel of \autoref{fig:forward_models}. Finally, we also include a flat line model representative of either no atmosphere or a very high altitude cloud deck. Compared to the clear, 1 bar atmosphere models discussed above, the flat line model is the preferred forward model fit with the lowest $\chi_\nu^2$ for all three reductions ($\chi_\nu^2$ $\sim$0.9--1.1). 

Next, we examine how changing the pressure of our H$_2$O atmosphere affects our interpretation, generating models with an opaque pressure level from 1 bar to 0.01 mbar in five logarithmic steps. This opaque pressure level can be interpreted either as that of the pressure of an opaque cloud top or haze layer or the surface pressure.
The $\chi^2_{\nu}$ values do decrease with decreasing pressure levels, but all thin atmosphere water scenarios remain consistent with the data. Though it is not a significant difference, the water atmosphere models fit the data better than the flat line because they simultaneously maintain the slight visual slope from 2.8 to 3.5 $\mu$m while flattening the spectrum from 3.5 to 5.0 $\mu$m. In contrast, the higher pressure water atmosphere has more structure at these longer wavelengths. The results of these water atmosphere, pressure varying fits are shown in the middle panel of \autoref{fig:forward_models}.

Given a pure water atmosphere at the 580 K temperature of \planetname, there are few if any cloud species \citep[e.g.,][]{gao2020} likely to form to produce the low pressure cloud tops favored by our forward models. Hazes could potentially form in enough abundance at these pressures \citep[e.g.,][]{moran2018,he2024} to fully mute out water absorption at deeper layers, or the planet could simply have a very tenuous atmosphere dominated by water vapor. 

Finally, we examine whether we can refute or confirm the best-fit atmospheric models allowed by the emission study of \planetname reported in \citet{Xue2024}. Their best-fit atmosphere models (based on the lowest $\chi^2_{\nu}$ values reported in their Figure 5) included 1) a 1 bar O$_2$ atmosphere with 100 ppm H$_2$O and 2) a 10 bar O$_2$ atmosphere with 1 ppm CO$_2$. Both of these models fit their emission spectrum with a $\chi_\nu^2$ $<$1. We reproduce these conditions and compute transmission spectrum models (with an isothermal temperature-pressure profile set to 580 K) to compare to our transmission data. Their 100 ppm H$_2$O in an O$_2$ atmosphere is fully consistent with our data and similarly fit with $\chi_\nu^2$ $\leq$1 for each of our \eureka, \exotic, and \firefly reductions. However, we are able to rule out their 10 bar, 1 ppm CO$_2$ in O$_2$ atmosphere with moderate confidence, rejecting this possibility by over 3$\sigma$ from both the \eureka and \exotic reductions and to 1.6$\sigma$ from the \firefly reduction. These models are shown in the bottom panel of \autoref{fig:forward_models}. The combination of emission and transmission spectra together can clearly winnow down the allowed potential atmospheric scenarios beyond either set of observations individually \citep{WeinerMansfield2024}, with transmission observations more sensitive to more tenuous atmospheres. See Section \ref{sec:disc_miri_compare} for a more complete discussion of this.

\subsection{Atmospheric Retrievals Favor Thin or No Atmosphere and Reveal First Visit Dominates Interpretation} \label{sec:retrievals}

We also interpret \planetname's transmission spectrum via Bayesian atmospheric retrievals. We use the open source retrieval code \POSEIDON \citep{MacDonald2017,MacDonald2023}, following a similar approach to recent studies (\citetalias{Moran2023} \citeyear{Moran2023}; \citetalias{May2023} \citeyear{May2023}) whereby we explore planetary atmospheres, stellar contamination, and flat spectra as potential explanations. Given the high-precision of our multi-visit \planetname transmission spectrum, we additionally explore joint constraints on the mean molecular weight and (apparent) surface pressure for allowable atmospheres.

\subsubsection{Retrieval Configuration} \label{sec:retrieval_config}

We consider three retrieval model categories with \POSEIDON: (i) a planetary atmosphere with no stellar contamination; (ii) stellar contamination from unocculted stellar heterogeneities; and (iii) a flat line. We also consider a sub-model of the atmosphere category with a pure H$_2$O composition (i.e., a `steam world') motivated by recent studies (\citetalias{May2023} \citeyear{May2023}; \citealt{Piaulet-Ghorayeb2024}). We apply each retrieval model to the three data reductions described above and for each reduction we consider separate retrievals for the 4-visit combined dataset and 3-visit spectra with one visit excluded. The latter `leave one out' retrievals allow us to assess the robustness of results derived from all 4 visits combined. All our retrieval models use 10,000 \texttt{PyMultiNest} \citep{Feroz2009,Buchner2014} live points and calculate model transmission spectra at a resolving power of $R =$ 20,000 from 2.6--5.3\,$\micron$.

Our atmosphere models consider a variety of potential compositions. We consider 7 specific gases (H$_2$, H$_2$O, CO$_2$, CO, CH$_4$, N$_2$, and SO$_2$) alongside a `ghost' gas with a freely fit mean molecular weight. Any of these 8 gases can dominate the atmospheric composition, since we use the background-gas agnostic centered log-ratio prior (e.g. \citealt{Benneke2012}; \citetalias{Lustig-Yaeger2023} \citeyear{Lustig-Yaeger2023}). The molecular cross sections used in the retrieval forward model's radiative transfer uses the opacity database corresponding to \POSEIDON v1.2 \citep{Mullens2024}, which uses the following ExoMol \citep{Tennyson2018,Tennyson2024} line lists: H$_2$O \citep{Polyansky2018}, CO$_2$ \citep{Yurchenko2020}, CO \citep{Li2015}, CH$_4$ \citep{Yurchenko2024}, and SO$_2$ \citep{Underwood2016}. Collision-induced absorption from H$_2$-H$_2$ and pairs with N$_2$ is also included from HITRAN \citep{Karman2019}. Each model atmosphere spans $10^{-7}$--$100$\,bar, split into 100 layers spaced uniformly in log-pressure, with a freely-fit effective surface pressure (representing either a solid surface or cloud-top) and is assumed to be isothermal with pressure. We additionally fit for the atmospheric radius at the 10\,bar reference pressure. The priors for this 11-parameter model are listed in Table~\ref{tab:retrieval_priors}.

Our stellar contamination retrieval model uses a one-heterogeneity model \citep[e.g.,][]{Rathcke2021}. We initially explored a two-heterogeneity model (i.e. including both cold and hot active regions; e.g. \citealt{Fournier-Tondreau2024}), but found no need for a second heterogeneity population. Therefore, we focus on the one-heterogeneity model in what follows. This 4-parameter model is parametrized by the planetary radius, the stellar photosphere temperature, the heterogeneity covering fraction, and the stellar heterogeneity temperature (corresponding to spots if the heterogeneity is cooler than the photosphere, or faculae for the converse). The stellar spectra of the photosphere and heterogeneities are calculated by interpolating \texttt{PHOENIX} models \citep{Husser2013} using the \texttt{PyMSG} package \citep{Townsend2023}. We adopt the stellar radius and effective temperature from \citet{Bonfils2018} ($R_{*} = 0.2105\,R_{\odot}$; $T_{\rm{eff}}$ = 3270\,K), the stellar surface gravity from \citet{Southworth2017} ($\log_{10} g_{*}$ = 4.88) and the stellar metallicity from \citet{Berta2015} ([Fe/H] = $-0.12$). The priors are also summarized in Table~\ref{tab:retrieval_priors}.

\begin{deluxetable}{lcc}[t!]
\centering
\tablewidth{0pt}
\tablecaption{Atmospheric Retrieval Parameters and Priors}
\tablehead{
Parameter & \colhead{Prior}
}
\startdata
    \multicolumn{2}{l}{\textbf{Atmosphere Model}} \\[2pt]
    Reference Radius ($R_{\mathrm{p, \, ref}}$) & $\mathcal{U}(0.9605, 1.2995)\,R_{\Earth}$ \\
    Surface Pressure ($\log_{\rm{10}} (P_{\rm{surf}}$ / bar)) & $\mathcal{U}(-7, 2)$ \\
    `Ghost' Mean Molecular Weight ($\mu_{\rm{back}}$) & $\mathcal{U}(2.3, 100)$ \\
    Atmospheric Temperature ($T$) & $\mathcal{U}(200, 900)$\,K \\
    Molecule Volume Mixing Ratios ($\log_{\rm{10}} X_i$) & $\mathcal{CLR}(-12, 0)$ \\[2pt]
    \multicolumn{2}{l}{\textbf{Stellar Contamination Model}} \\[2pt]
    Planetary Radius ($R_{\mathrm{p}}$) & $\mathcal{U}(0.9605, 1.2995)\,R_{\Earth}$ \\
    Photosphere Temperature ($T_{\rm{phot}}$) & $\mathcal{N}(3270, 140^{2})$\,K \\
    Heterogeneity Temperature ($T_{\rm{het}}$) & $\mathcal{U}(2300, 3924)$\,K \\
    Heterogeneity Covering Fraction ($f_{\rm{het}}$) & $\mathcal{U}(0, 0.6)$ \\[2pt]
    \multicolumn{2}{l}{\textbf{Flat Line Model}} \\[2pt]
    Planetary Radius ($R_{\mathrm{p}}$) & $\mathcal{U}(0.9605, 1.2995)\,R_{\Earth}$ \\[2pt]
    \hline
\enddata
\tablecomments{The mixing ratios of H$_2$O, CH$_4$, CO$_2$, CO, SO$_2$, N$_2$, and H$_2$ are free parameters, with the mixing ratio of the `ghost' gas satisfying the summation to unity constraint. `CLR' refers to the centered log-ratio prior, which allows any gas to have equal probability \emph{a priori} to be the background gas \citep[e.g.][]{Benneke2012}. The Gaussian prior for $T_{\rm{phot}}$ is expressed as $\mathcal{N}(\mu, \sigma^2)$, where $\mu$ and $\sigma$ are the mean and standard deviation, respectively, for the stellar effective temperature from \citet{Bonfils2018}.}
\label{tab:retrieval_priors}
\end{deluxetable}

\begin{deluxetable}{lcccccc}[t!]
\centering
\tablewidth{0pt}
\tablecaption{Retrieval Model Statistics}
\tablehead{
Data + Model & $N_{\rm{param}}$ & $\ln \mathcal{Z}$ & $\chi^2$ & d.o.f. & $\chi^2_{\nu}$
}
\startdata
    \multicolumn{6}{l}{\textbf{\firefly}} \\
    \hspace{1pt} \textbf{4 Visits} & & & & \\
    \hspace{4pt} Atmosphere & 11 & 575.03 & 49.0 & 53 & 0.92 \\
    \hspace{4pt} Stellar Contam. & 4 & 575.44 & 50.8 & 60 & 0.85 \\
    \hspace{4pt} Thin Steam World & 3 & 576.36 & 50.6 & 61 & 0.83 \\
    \hspace{4pt} Flat & 1 & 575.45 & 56.1 & 63 & 0.89 \\
    \hspace{1pt} \textbf{3 Visits} & & & & \\
    \hspace{1pt} \textbf{(No Visit 1)} & & & & \\
    \hspace{4pt} Atmosphere & 11 & 563.59 & 53.0 & 53 & 1.00 \\
    \hspace{4pt} Stellar Contam. & 4 & 562.23 & 58.8 & 60 & 0.98 \\
    \hspace{4pt} Thin Steam World & 3 & 563.37 & 58.0 & 61 & 0.95 \\
    \hspace{4pt} Flat & 1 & 563.47 & 60.7 & 63 & 0.96 \\[2pt]
    \multicolumn{6}{l}{\textbf{\eureka}} \\
    \hspace{1pt} \textbf{4 Visits} & & & & \\
    \hspace{4pt} Atmosphere & 11 & 570.62 & 56.3 & 53 & 1.06 \\
    \hspace{4pt} Stellar Contam. & 4 & 571.68 & 61.4 & 60 & 1.02 \\
    \hspace{4pt} Thin Steam World & 3 & 572.50 & 60.4 & 61 & 0.99 \\
    \hspace{4pt} Flat & 1 & 569.83 & 69.5 & 63 & 1.10 \\
    \hspace{1pt} \textbf{3 Visits} & & & & \\
    \hspace{1pt} \textbf{(No Visit 1)} & & & & \\
    \hspace{4pt} Atmosphere & 11 & 561.76 & 59.5 & 53 & 1.12 \\
    \hspace{4pt} Stellar Contam. & 4 & 561.36 & 63.5 & 60 & 1.06 \\
    \hspace{4pt} Thin Steam World & 3 & 562.52 & 62.3 & 61 & 1.02 \\
    \hspace{4pt} Flat & 1 & 562.51 & 65.0 & 63 & 1.03 \\[2pt]
    \multicolumn{6}{l}{\textbf{\exotic}} \\
    \hspace{1pt} \textbf{4 Visits} & & & & \\
    \hspace{4pt} Atmosphere & 11 & 519.63 & 52.3 & 47 & 1.11 \\
    \hspace{4pt} Stellar Contam. & 4 & 521.89 & 52.3 & 54 & 0.99 \\
    \hspace{4pt} Thin Steam World & 3 & 521.74 & 55.3 & 55 & 1.01 \\
    \hspace{4pt} Flat & 1 & 518.59 & 65.7 & 57 & 1.15 \\
    \hspace{1pt} \textbf{3 Visits} & & & & \\
    \hspace{1pt} \textbf{(No Visit 1)} & & & & \\
    \hspace{4pt} Atmosphere & 11 & 514.76 & 48.8 & 47 & 1.04 \\
    \hspace{4pt} Stellar Contam. & 4 & 513.98 & 53.5 & 54 & 0.99 \\
    \hspace{4pt} Thin Steam World & 3 & 514.75 & 53.7 & 55 & 0.98 \\
    \hspace{4pt} Flat & 1 & 514.89 & 56.2 & 57 & 0.99 \\[2pt]
    \hline
\enddata
\tablecomments{The models are ordered by complexity ($N_{\rm{param}}$). $\mathcal{Z}$ is the Bayesian evidence, d.o.f. is the degrees of freedom ($N_{\rm{data}} - N_{\rm{param}}$), and $\chi^2_{\nu}$ is the reduced chi squared.}
\label{tab:retrieval_stats}
\end{deluxetable}

\subsubsection{Retrieval Results} \label{sec:retrieval_results}

Our \planetname retrieval analysis favors a thin H$_2$O-rich atmosphere when using all 4 visits, in agreement with our forward modeling analysis (Section \ref{sec:fwd_models}.) A thin ($\sim 10^{-4}$\,bar) pure H$_2$O atmosphere maximizes the Bayesian evidence and is preferred over both a flat line and stellar contamination for all three data reductions (\autoref{tab:retrieval_stats}). The Bayes factors for the simple 3-parameter pure H$_2$O model compared to a flat line are 2.5 (\firefly), 14 (\eureka), and 23 (\exotic).
The lower significance from \firefly arises from the slightly larger error bars in certain channels in this reduction, which also manifests in the smaller reduced-$\chi^2$ values compared to \eureka and \exotic (Table~\ref{tab:retrieval_stats}). The improvement in the Bayesian evidence between the pure H$_2$O model and the stellar contamination model is not statistically significant, so these NIRSpec G395 data cannot distinguish between stellar contamination and an atmosphere (as in previous studies, e.g. \citetalias{Moran2023} \citeyear{Moran2023}).

\begin{figure*}[h]
\figurenum{7}
    \centering
    \vspace{-0.3cm}
    \includegraphics[width=0.88\textwidth]{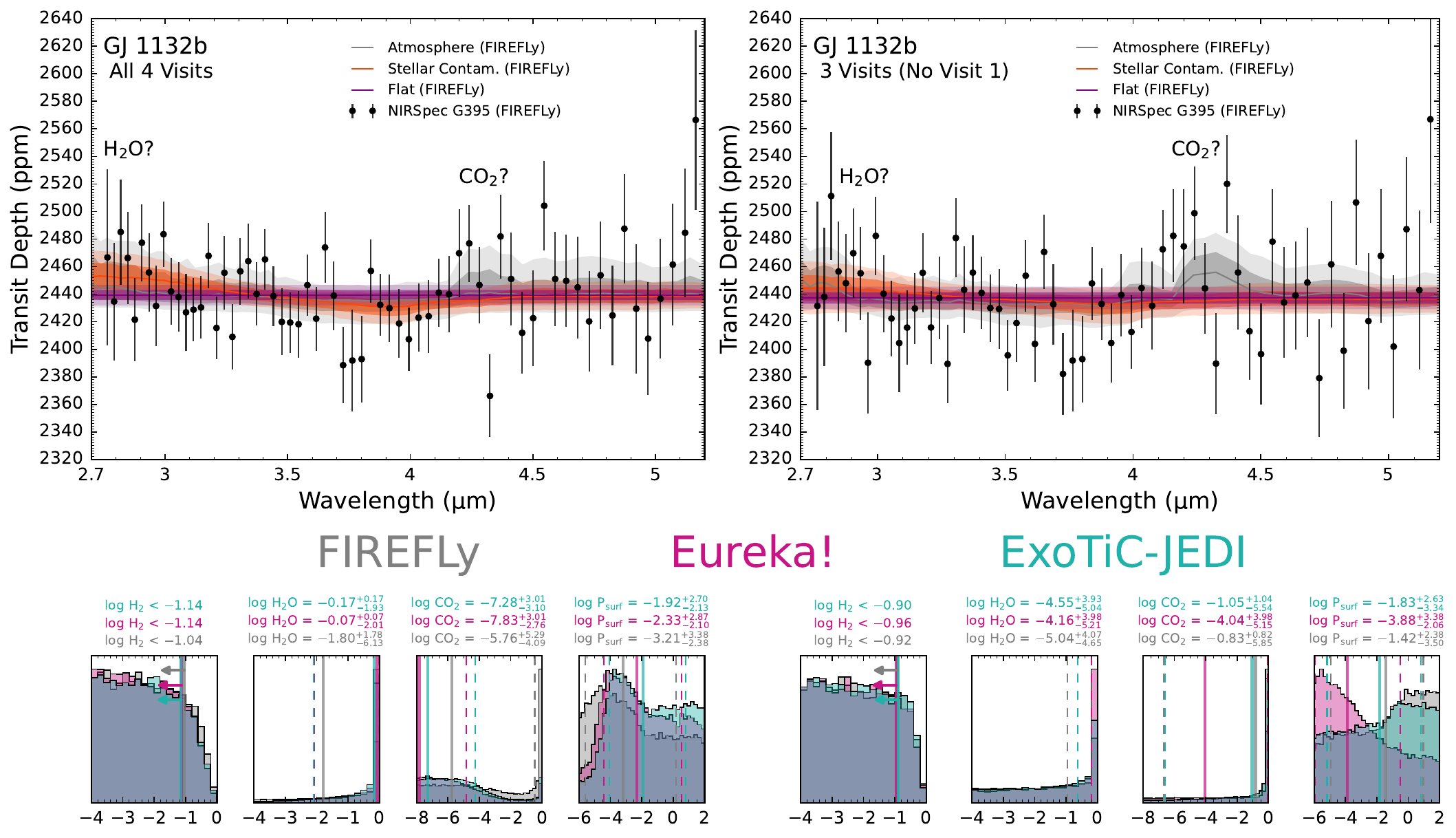}
    \caption{Retrieval results for \planetname. Top panels: comparison between the retrieved transmission spectra for the \firefly reduction with all four visits (left) and for three visits with Visit 1 excluded (right). Three \POSEIDON retrieval models are shown: (i) an atmosphere (grey), (ii) stellar contamination (orange), and (iii) a flat line (purple). The corresponding median retrieved spectra (solid lines) and 1$\sigma$ and 2$\sigma$ model confidence intervals (dark and light contours) are overlaid. Labels indicate H$_2$O and CO$_2$ absorption bands (neither are detected, but CO$_2$ is allowed by \firefly). Bottom panels: atmosphere posterior histograms for the three data reductions: \firefly (grey), \eureka (pink), and \exotic (cyan). The 4-visit combined analysis weakly favors a thin H$_2$O-rich atmosphere (driven by a short-wavelength slope), but this solution is not favored when Visit 1 is excluded.
    }
    \label{fig:retrieved_spectra}
\end{figure*}

\autoref{fig:retrieved_spectra} shows the retrieved spectra fits to the \firefly data and the posterior distributions for the atmosphere model for all three data reductions. We show the results from the `full' atmosphere model, rather than the simple pure H$_2$O model, since the full model encompasses the best-fitting pure H$_2$O solution while also marginalizing over other potential atmosphere components. While our retrievals identify H$_2$O as the most likely background gas for all three data reductions, we see that the H$_2$O abundance posterior displays a long tail to lower abundances without a clear lower limit. Compared to the results from \citetalias{May2023} (\citeyear{May2023}) (for Visit 1 only), the weaker H$_2$O abundance constraints here are due to our 4-visit average spectrum being flatter, which requires a low apparent surface pressure peaking at $\sim 10^{-4}$\,bar. Besides the preference for H$_2$O, the \firefly data tentatively allow for the presence of CO$_2$ (see \autoref{fig:retrieved_spectra}), but the \eureka and \exotic data do not indicate any evidence for CO$_2$ --- this likely arises from small differences in the data scatter near 4.3\,$\micron$. Our data also rules out H$_2$-rich atmospheres, with 2$\sigma$ upper limits of $\approx$ 10\% H$_2$ (\autoref{fig:retrieved_spectra}), even after marginalizing over surface pressure, for all three data reductions.

Given that our 4-visit retrieval analysis finds weaker evidence for an atmosphere than the Visit 1 only results from \citetalias{May2023} (\citeyear{May2023}), we repeated our retrievals with 3-visit average spectra excluding one visit in turn. This leave-one-out test allows us to assess whether a single visit drives our inferences. We find generally similar results when excluding Visits 2, 3, or 4 to the results described above, so we do not discuss these further. However, the exclusion of Visit 1 significantly alters the retrieval results. Therefore, in what follows we present the retrieval results for both the 4-visit and 3-visit (with Visit 1 excluded) datasets on equal grounds.

\begin{figure*}[h]
\figurenum{8}
    \centering
    \includegraphics[width=0.88\textwidth]{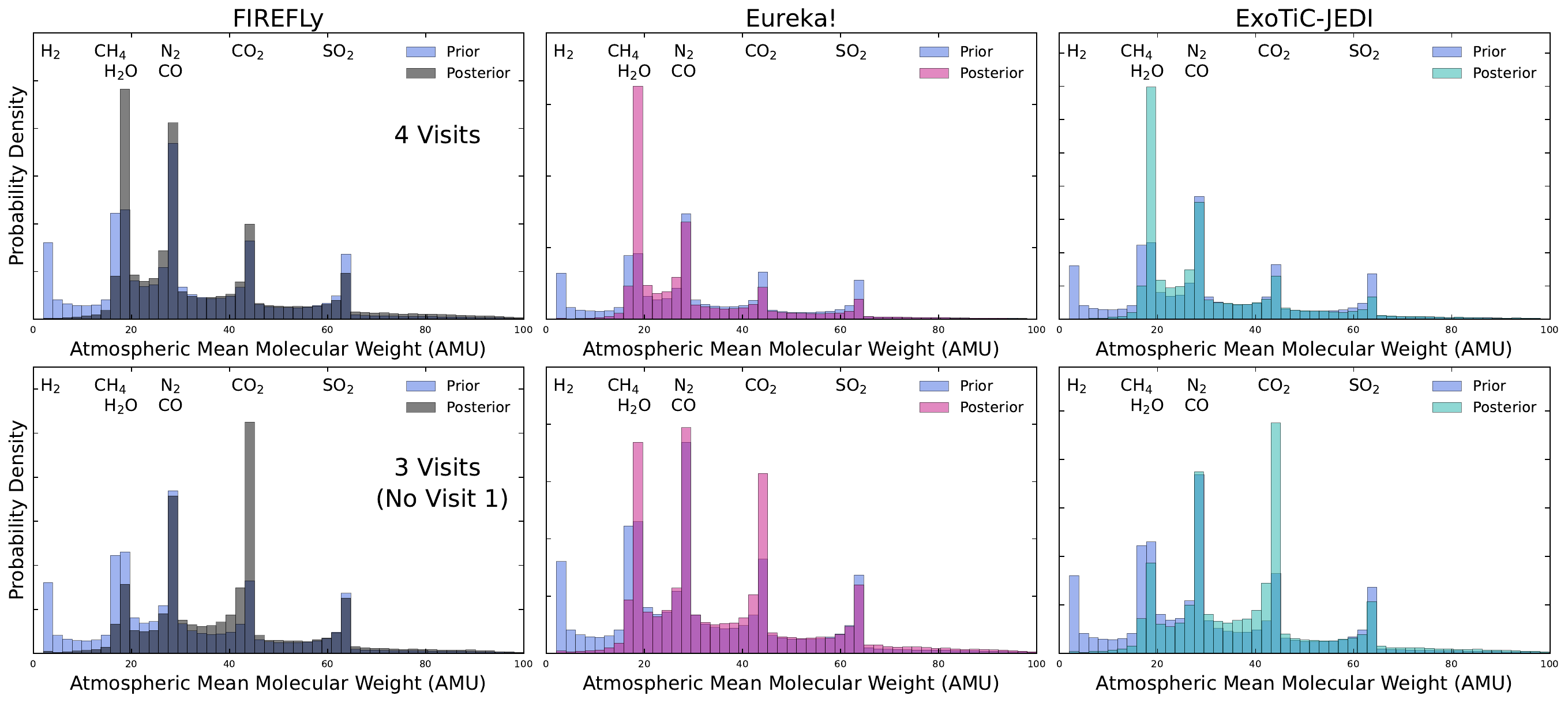}
    \caption{Mean molecular weight constraints for \planetname's atmosphere. Top panels: constraints from all 4 visits combined, comparing the prior distribution (blue) with the posteriors from \firefly (grey; left), \eureka (pink; middle), and \exotic (cyan; right). Bottom panels: constraints from 3 visits with Visit 1 excluded. The peaks in the prior at $\sim$ 100\% abundances of each considered molecule (annotated) arise from the use of the centered log-ratio prior \citep[e.g.][]{Benneke2012}, which formally encodes that all gases are equally likely to be the background gas. Higher relative prior amplitudes arise from multiple molecules overlapping in a given mean molecular weight bin (e.g. N$_2$ and CO both have $\mu = 28$). The 4-visit results favor a H$_2$O-dominated composition for all three reductions, but without Visit 1 this solution is disfavored relative to higher mean molecular weights (e.g. CO$_2$-dominated atmospheres). H$_2$-dominated atmospheres are excluded in all cases.
    }
    \label{fig:retrieved_mu}
\end{figure*}

With Visit 1 excluded, we find that \planetname's transmission spectrum is statistically flat with no evidence for atmospheric absorption. \autoref{tab:retrieval_stats} demonstrates that the thin pure H$_2$O model that was preferred for the 4-visit spectrum has a similar or slightly lower Bayesian evidence than the flat model for all three data reductions, indicating that a flat model sufficiently describes the data (see \autoref{fig:retrieved_spectra}). Should \planetname possess an atmosphere, the transmission spectrum can be rendered flat by a very low surface pressure (\eureka), or a high mean molecular weight dominated by CO$_2$ (\firefly and \exotic). Nevertheless, atmospheres rich in H$_2$ remain ruled out with a 2$\sigma$ upper limit of $\approx$ 10\% H$_2$ for the 3 visits without Visit 1 (see \autoref{fig:retrieved_spectra}).

Our atmospheric mean molecular weight constraints for \planetname are shown in \autoref{fig:retrieved_mu}. The prior probability distribution for the mean molecular weight displays distinct `spikes' at the mean molecular weight (given in AMU, atomic mass units) of specific molecules (H$_2$ = 2, CH$_4$ = 17, H$_2$O = 18, N$_2$ = CO = 28, CO$_2$ = 44, SO$_2$ = 64), corresponding to atmospheres dominated by one of the considered gases. Atmospheres dominated by the molecular weight of a single gas are a manifestation of the centered log-ratio prior, which allows every gas an \emph{a priori} equal prior probability of being the dominant background gas. Since at least one gas must have a significant abundance in order for the mixing ratios to sum to unity, the prior probability for a mean molecular weight corresponding to one of the considered gases is higher than intermediate mean molecular weights produced by a combination of different gases. We note that the prior distribution for $\mu_{\rm{atm}} > 64$ arises from the `ghost' gas in our retrieval model, which we allow up to 100. While the ghost gas also follows the CLR prior, since its mean molecular weight is also a free parameter it produces a diffuse prior density across the parameter space. The prior density falls towards the upper limit ($\mu_{\rm{atm}} = 100$) because for high values of $\mu_{\rm{atm}}$ there are fewer samples where the ghost gas is both the dominant gas and has a high randomly sampled mean molecular weight. In summary, the prior distribution in \autoref{fig:retrieved_mu} is a reflection of our prior knowledge that not all mean molecular weights are equally likely \emph{a priori}, since there are characteristic molecular weights corresponding to the specific molecules that can dominate an atmosphere.

We find robust lower limits on \planetname's atmospheric mean molecular weight. When considering all 4 visits, we find 99.7\% credible interval lower limits of $\mu_{\rm{atm}} >$ 7.5 (\firefly), 7.6 (\eureka), and 8.8 (\exotic). The equivalent limits excluding Visit 1 are  $\mu_{\rm{atm}} >$ 5.9 (\firefly), 6.2 (\eureka), and 6.5 (\exotic). Therefore, a H$_2$-rich atmosphere is ruled out to high significance across all three data reductions, regardless of the inclusion of Visit 1. While a specific atmospheric composition is not uniquely identified for our atmospheric retrieval model considering multiple gases, \autoref{fig:retrieved_mu} demonstrates that the most likely composition from the 4-visit spectrum is an H$_2$O-dominated atmosphere. The \eureka and \exotic reductions yield the highest posterior probability for a H$_2$O-dominated atmosphere, but the \firefly data allows a significant contribution from N$_2$. However, the 3-visit spectrum without Visit 1 produces significantly different solutions. \autoref{fig:retrieved_mu} shows that the relative posterior probability shifts from the H$_2$O-rich peak to atmospheres dominated by CO$_2$ (\firefly and \exotic) or N$_2$ (\eureka). In essence, the flatter transmission spectrum without Visit 1 (\autoref{fig:retrieved_spectra}) requires a higher mean molecular weight than even a pure H$_2$O atmosphere can provide, resulting in a probability density shift towards heavier background gases (in particular, CO$_2$).

\begin{figure}[ht!]
\figurenum{9}
    \centering
    \includegraphics[width=\columnwidth]{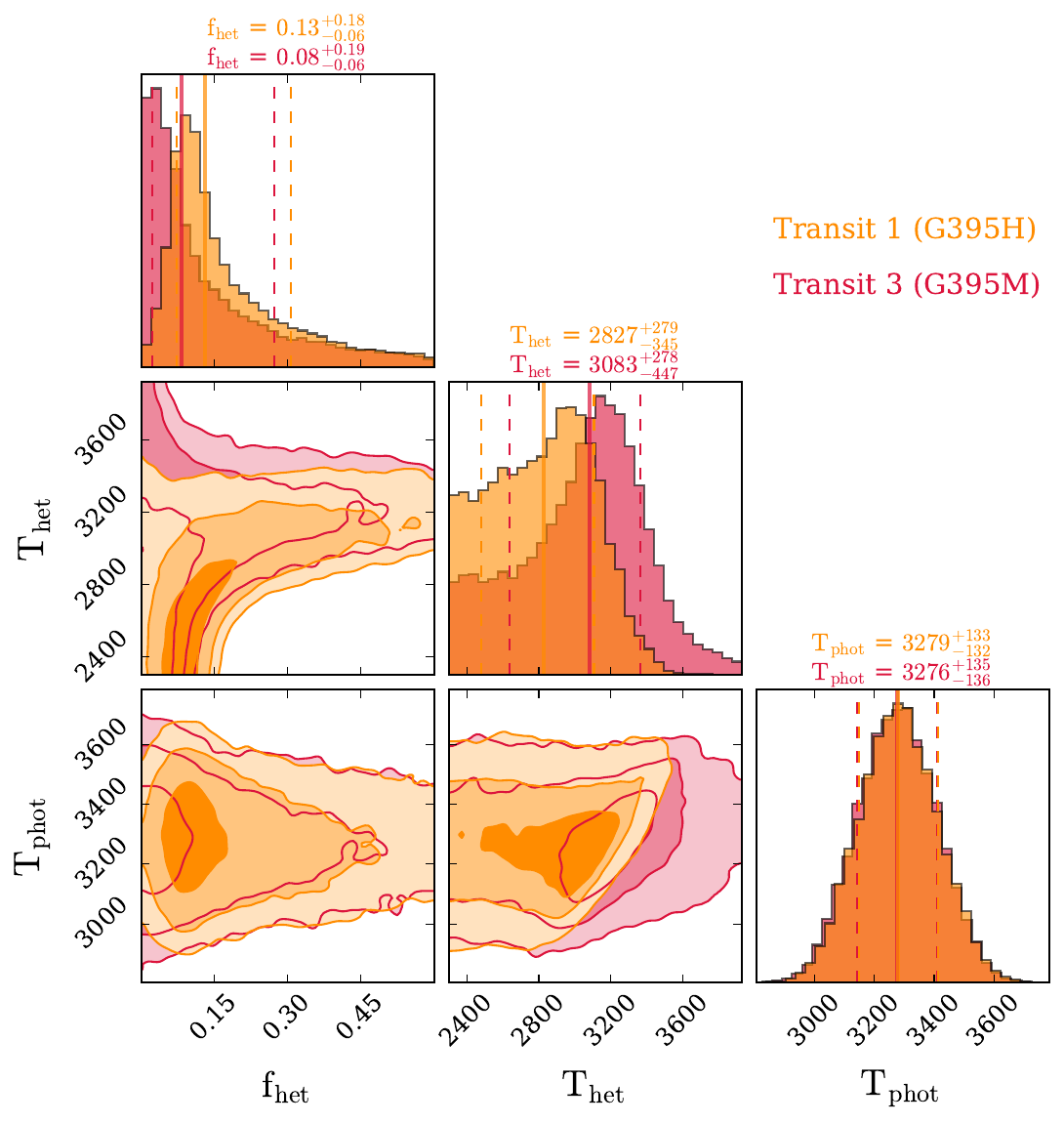}
    \caption{Retrieved stellar properties from \planetname's \firefly transmission spectra from Visit 1 (NIRSpec G395H) and Visit 3 (NIRSpec G395M). These stellar contamination retrievals assume all spectral features arise from unocculted heterogeneities (i.e. no planetary absorption). The slope seen in Visit 1 (\citetalias{May2023}, \citeyear{May2023}) can be explained with a moderate starspot covering fraction ($f_{\rm{het}} \sim 10\%$), but the relatively flat Visit 3 spectrum does not require unocculted starspots ($f_{\rm{het}} \rightarrow 0$).
    }
    \label{fig:retrieved_stellar_contam}
\end{figure}

Given the significant sensitivity of \planetname's transmission spectrum and retrieval results to the inclusion of Visit 1, we briefly consider whether time-variable stellar contamination could offer an explanation, as in Section \ref{sec:stellar_spectra}. We compare the flux-calibrated stellar spectra comparison approach to the retrieval approach to determine if the retrieval approach can also tease out meaningful information on the cooler/warmer spot coverage fractions and temperatures on a visit-to-visit basis. We would also like to understand if the constraints in the retrieval approach differ between the G395H and G395M modes. For this, we test a single G395H visit (Visit 1) and a single G395M visit (Visit 3). \autoref{fig:retrieved_stellar_contam} shows retrieval results only using a stellar contamination retrieval model. We see that the spectral slope exhibited in the Visit 1 transmission spectrum requires $13^{+18}_{-6}$\% starspot coverage to be explained by stellar contamination, but Visit 3 does not require stellar contamination ($f_{\rm{het}} \rightarrow 0$), since this visit is statistically flat. However, the two posterior distributions are formally consistent within $1\sigma$, given the strong $f_{\rm{het}}$--$T_{\rm{het}}$ degeneracy, so the presence of time-variable stellar contamination cannot be established from the individual visit spectra alone using this technique, illustrating that analyzing the flux-calibrated stellar spectrum (Section \ref{sec:stellar_spectra}) may be more information-rich.

\begin{figure*}[ht!]
\figurenum{10}
    \centering
    \vspace{0.3cm}
    \includegraphics[width=\textwidth]{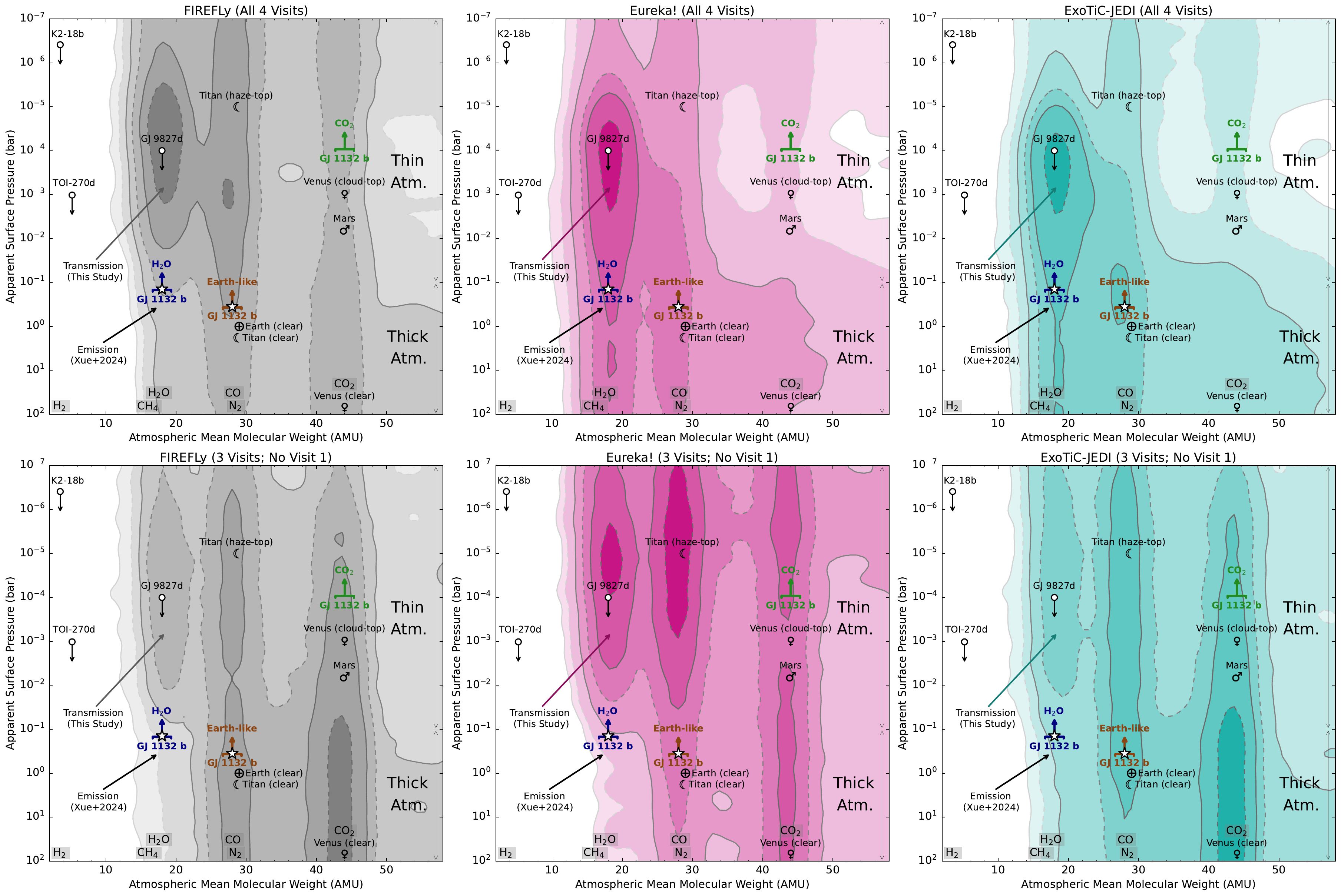}
    \caption{Surface pressure vs. mean molecular weight constraints for \planetname. \textbf{Top:} posterior probability density (colored contours) from 4 NIRSpec G395H/M transits combined from \firefly (grey; left), \eureka (pink; middle), and \exotic (cyan; right). \textbf{Bottom:} Same as top panel, but excluding Visit 1. The 1$\sigma$, 2$\sigma$, and 3$\sigma$ credible regions have solid line boundaries, while the 0.5$\sigma$, 1.5$\sigma$, and 2.5$\sigma$ credible regions have dashed boundaries. Pressure upper limits from emission spectroscopy of \planetname (2$\sigma$ limits from \citealt{Xue2024}) are annotated for 100\% H$_2$O (blue arrow), an Earth-like composition (brown arrow), and 100\% CO$_2$ (green arrow). Three exoplanets with detected atmospheres and well-constrained mean molecular weights heavier than pure H$_2$+He (TOI-270d: \citealt{Benneke2024}; GJ~9827d: \citealt{Piaulet-Ghorayeb2024}; and K2-18b: \citealt{Schmidt2025}) are annotated (circles with arrows, located at the 2$\sigma$ lower limit on the surface / cloud pressure), alongside terrestrial atmospheres in the solar system. \planetname's transmission spectrum favors a thin $\sim 10^{-4}$\,bar H$_2$O-dominated atmosphere when using all 4 visits, but the exclusion of Visit 1 allows a wider range of heavier and/or thin atmospheres (or no atmosphere) as viable ways to produce a featureless 3-visit transmission spectrum (see \autoref{fig:retrieved_spectra}).
    }
    \label{fig:retrieved_mu_Psurf}
\end{figure*}

Finally, since different surface pressures can affect the viability of atmospheric compositions differently, we present the joint posteriors for surface pressure and mean molecular weight in \autoref{fig:retrieved_mu_Psurf}. This visual allows a direct comparison between our \planetname retrieval results and the solar system terrestrial atmospheres (Venus, Mars, Earth, and Titan), three exoplanets with well-measured atmospheric mean molecular weights heavier than H$_2$+He \citep{Piaulet-Ghorayeb2024,Benneke2024,Schmidt2025}, and previous atmospheric constraints for \planetname from JWST emission spectroscopy \citep{Xue2024}. In \autoref{fig:retrieved_mu_Psurf}, the CLR prior spikes seen in  \autoref{fig:retrieved_mu} translate to columns in the $P_{\rm{surf}}$--$\mu_{\rm{atm}}$ parameter space.

\autoref{fig:retrieved_mu_Psurf} demonstrates that the H$_2$O-rich atmospheric solution for the 4-visit combined spectrum is localized in a thin pressure region centered near $10^{-4}$\,bar. This solution is consistent with the \planetname surface pressure upper limit from \citet{Xue2024} for a pure H$_2$O composition ($P_{\rm{surf}} \lesssim 10^{-1}$\,bar) and with the `steam world' atmosphere detected for GJ~9827~d \citep{Piaulet-Ghorayeb2024}. However, the removal of Visit 1 shifts the solution space to favor heavier atmospheres. For 3 visits, the \eureka reduction favors a thin ($P_{\rm{surf}} < 10^{-3}$\,bar) N$_2$-rich atmosphere, while the \firefly and \exotic reductions favor thick ($P_{\rm{surf}} > 10^{-1}$\,bar) CO$_2$-rich atmospheres. Given the featureless spectrum without Visit 1, small differences between the reductions alters the relative probability between the thin N$_2$-rich and thick CO$_2$-rich solutions. We note that the \planetname MIRI LRS emission spectrum from \citet{Xue2024} strongly disfavors a thick CO$_2$-rich atmosphere, so our transmission spectra results should not be interpreted as evidence for such an atmosphere, despite the low significance possibility of 4.3\,$\micron$ CO$_2$ absorption shown in \autoref{fig:retrieved_spectra}. Our most robust result is that \planetname cannot have an atmosphere with a mean molecular weight $<$ 6, for any surface pressure $> 10^{-7}$\,bar, which rules out both H$_2$-rich atmospheres and a miscible supercritical envelope similar to that favored for TOI-270d by \citet{Benneke2024} ($\mu_{\rm{atm}} \approx 5$).

The extreme sensitivity of our retrieval results to the inclusion of Visit 1 suggests that this Visit is a statistical outlier, and therefore that the most reliable constraints on \planetname's atmosphere should be drawn from the 3-visit transmission spectrum with Visit 1 excluded. For this dataset, we find a non-detection of atmospheric absorption, in particular a lack of H$_2$O absorption, which can result from \planetname being either a bare rock or possessing a thin atmosphere dominated by heavy molecules such as N$_2$ or CO$_2$. 

\section{Discussion} \label{sec:discussion}

In this section, we discuss both the technical implications of utilizing the NIRSpec/G395M mode and stacking multiple transits as well as the astrophysical implications for \planetname.

\subsection{Mind the Gap: NIRSpec G395M versus G395H} \label{sec:disc_g395m_vs_h}

For transmission spectroscopy with \jwst targeting the 3--5$\rm \;\mu m$ range, the exoplanet community by and large has utilized NIRSpec/G395H (or NIRSpec/PRISM for fainter targets). To date, only two other studies have observed with G395M \citep{Mayo2025, Davenport2025}. (There are multiple ongoing/upcoming programs intending to use this mode, but none have been published as of this writing.)

Importantly, our observations are the first time both the G395H and G395M mode have been utilized for a single target, offering a critical opportunity to compare the two modes. G395M is an attractive alternative to the higher resolution mode primarily as a way to avoid the notorious detector gap with G395H, which splits the incoming light between the NRS1 and NRS2 detectors with a gap between $\rm \sim 3.75-3.82\;\mu m$ (see \autoref{fig:spectrum_visit_compare}). This allows for greater spectral resolution, but there has not actually yet been evidence that suggests the higher resolution afforded by G395H improves abundance constraints for detected molecules compared to G395M \citep{Guzman2020}. Additionally, the use of G395H comes at the price of needing to characterize any potential offsets between the datasets due to this gap, which can confound results (e.g., \citealt{Madhu2023}). With the G395M mode, on the other hand, light falls solely on the NRS1 detector while still covering the full $\rm \sim2.9-5.1\;\mu m$ wavelength range, allowing us to avoid the gap. 

In \autoref{fig:spectrum_visit_compare}, we see that the data for Visits 3 and 4 (the G395M visits) bridge the detector gap and ``fill in the blank" very smoothly. Visually, there is no apparent need for an offset between NRS1/NRS2 in the first two visits (the G395H visits). Though large offsets have been seen in G395H data ($\sim10$s of ppm), this is frequently seen in datasets observing bright targets with few ($\leq3$) groups per integration (\citetalias{Moran2023} \citeyear{Moran2023}, \citealt{Alderson2024, Wallack2024, BelloArufe2025}). Our G395H visits, on the other hand, use 14 groups per integration. This hints that this detector offset is driven by poor constraints on the up-the-ramp fit due to this low number of groups. This could subsequently lead to increased systematics in the light curves, necessitating the need for more complex models and thus inducing offsets. However, offsets are also seen in observations with as many as six or seven groups per integration \citep[e.g.,][]{Gressier2024, Alderson2025}. It could be, therefore, that 1) approximately $>7$ groups for G395H data are required in order to mitigate detector offsets, or 2) the offsets are unrelated to the number of groups and instead are a result of some other underlying instrumental effect. Overall, coupling the G395M mode provides an important check on whether an offset is needed in G395H data.

Beyond detector gaps, we can evaluate the usefulness of G395H versus G395M based on noise levels seen in the two modes. As far as the light curve noise levels, we see comparable levels of photon noise reached between G395M and G395H for both the white and spectroscopic light curves. Additionally, the standard deviation and mean absolute deviation in the white light curve fits are comparable between modes, as illustrated in \autoref{tab:g395m_vs_g395h_stats}, which illustrates these statistics for Visits 1 (G395H) and 3 (G395M). Comparable values are seen in the other visits for each mode. We note that this may be somewhat reduction dependent, as \firefly and \eureka have marginally larger SD/MAD values for G395M relative to G395H/NRS1 (but marginally smaller values compared to G395H/NRS2), but \exotic has G395M values slightly worse than both NRS1 and NRS2. Neverless, there are no glaring differences in white light noise levels between the observing modes. 
\begin{deluxetable}{c|cccc}
\tablewidth{0pt}
\tablecaption{White light curve and transmission spectrum statistics for G395H versus G395M. \label{tab:g395m_vs_g395h_stats}}
\tablehead{ & 
    Photon &
    SD &
    MAD & 
    Spectroscopic\\
    & Noise & (ppm) & (ppm) & Error (ppm)} 
\startdata
    G395H (NRS1) & 1.4$\times$ & 131 & 88 & 48\\
    G395H (NRS2)& 1.6$\times$ & 177 & 123 & 58\\
    G395M & 1.4$\times$ & 154 & 103 & 60\\
\enddata
\tablecomments{Visit 1 (G395H) and Visit 3 (G395M) are shown for the \firefly reduction. First three columns refer to the white light curve fits. Level of photon noise is calculated by dividing the observed scatter by the predicted pipeline scatter, SD is standard deviation, and MAD is mean absolute deviation. Median spectroscopic error is shown in the final column. Statistics are comparable between G395H and G395M.}
\vspace{-1cm}
\end{deluxetable}

\begin{figure}
\figurenum{11}
    \centering
    \includegraphics[width=1\linewidth]{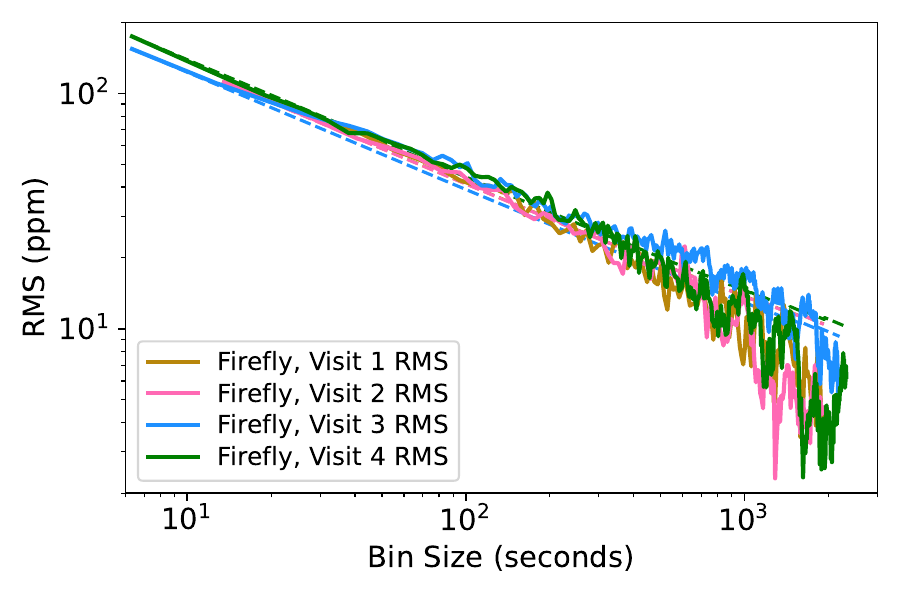}
    \includegraphics[width=1\linewidth]{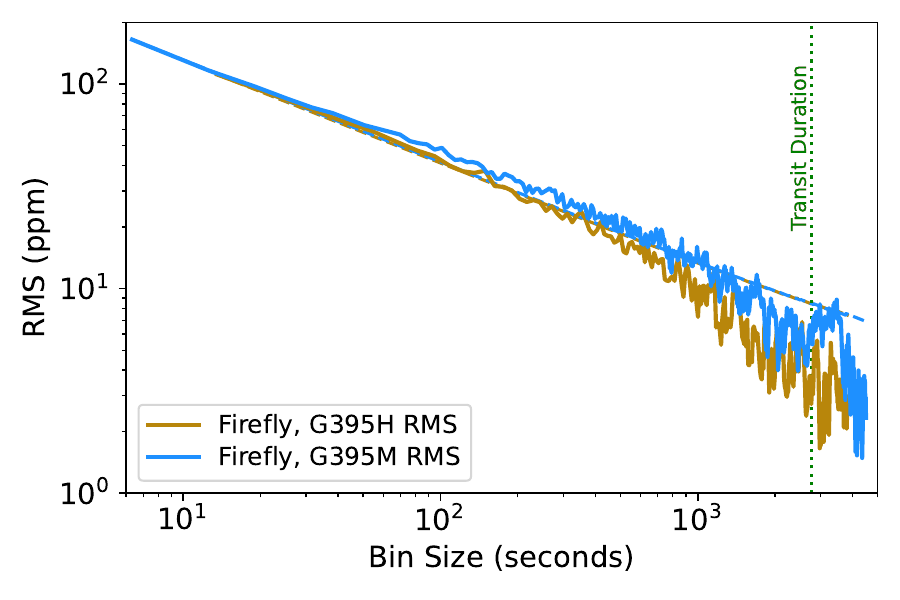}
    \caption{Red noise diagnostic plot (depicting the rms of the residuals as a function of bin size in time) for the residuals from the four \firefly visits (\textbf{top}) and each observing mode with residuals from visits concatenated (\textbf{bottom}). RMS as a function of bin size is shown, with the dotted lines representing the expected decrease in noise with the square root of bin size ($\sqrt{N}$). Bin size is in seconds, not number of integrations, in order to appropriately compare G395H and G395M. There is no apparent difference in correlated noise between the G395H (Visits 1 and 2) and G395M (Visits 3 and 4) observations, nor is there an indication that an instrumental noise floor is being reached.}
    \label{fig:wlc_variance}
\end{figure}

Additionally, we observe no meaningful differences in correlated noise between the detectors, as visualized by the white light red noise diagnostic plots shown in the top panel of \autoref{fig:wlc_variance}. All four visits bin down close to photon noise. Likewise, the spectroscopic red noise diagnostic plots are similar, with a similar number of wavelength channels requiring error inflation between G395H and G395M, as described in Section \ref{sec:firefly_reduction}. We note, however, that this is also somewhat reduction-dependent. \firefly sees a similar number of channels requiring error inflation, but \eureka requires more channels for error inflation in G395M, although this is likely due to the simpler systematics model employed by \eureka.  

Still, at the end of the day, this error inflation technique does not meaningfully impact the average error bars between the visits, which is also comparable across the observing modes, regardless of reduction. Median error bar sizes for Visits 1 and 3 (binned to the same resolution wavelength grid, as in \autoref{fig:spectrum_visit_compare}) are shown in \autoref{tab:g395m_vs_g395h_stats}. G395M sees a very small increase in median (and mean) error bar size compared to G395H - on the order of 5 ppm per-visit for all reductions - but this is well within the per-visit error ($\sim$50--60 ppm). 

One positive benefit of utilizing G395H is the possibility of doing isotopologue investigations using cross-correlation with the pixel-resolution data (e.g., \citealt{EsparzaBorges2023}). However, this is a new strategy that is likely only applicable to targets with large signals with \jwst, such as brown dwarfs and gas giants. Rocky planets have such small and noisy transit depths in the pixel-resolution data that the signal is insufficient for this type of science. 

Our last remaining point in evaluating the two modes is that G395H does allow for more information-rich flux-calibrated stellar spectra. As shown in Section \ref{sec:stellar_spectra}, examining the flux-calibrated spectra is a powerful tool to assess underlying stellar heterogeneity that is missed in the transmission spectrum. The precision of spot coverage/temperature constraints is comparable between the modes. 
However, the increased spectral detail for G395H allows for greater flexibility in model matching. Higher resolution provides more data points and finer structure in the observed spectrum, which can accommodate a broader range of models that capture different physical conditions. This effectively reduces degeneracies and makes it easier for multiple models to achieve a statistically acceptable fit. In contrast, the G395M spectra contain fewer constraints, making it harder for many models to simultaneously match the observed features.
Whether or not the stellar models are optimally extracting this possible source of extra information from the higher resolution spectra is a topic for a separate investigation.   

Overall, the decision to utilize G395M versus G395H should be made on a case-by-case basis. A balance must be struck between concerns about systematics and concerns of underlying stellar heterogeneity when stacking multiple transits. If 
observing sources with at least $\sim$8 groups per integration, and stacking multiple visits, it may be worth utilizing G395H, as it is less likely that NRS1/NRS2 offsets will appear in the data and more likely to need robust stellar heterogeneity measurements to ensure minimal visit-to-visit stellar heterogeneity. On the other hand, if transits are not being stacked and the source can be observed with G395M without saturating, G395M will suffice. We note that in practice, any target that requires few (less than $\sim8$) groups with G395H will likely saturate (or come close to saturating) G395M, so one may still be constrained to a given observing mode by the saturation limits of NIRSpec. 

\subsection{Stacking Multiple Transits: A Treasure or a Curse?} \label{sec:disc_stacking}

This study marks the second time as many as four transits for a single planet have been co-added together using \jwst data, after L98-59b \citep{BelloArufe2025}.
The hope of the rocky exoplanet community for many years has been to use the power of stacking to bring down transmission spectra error bar size and reach the regime where tiny, 10--20 ppm signals are discernible from a secondary atmosphere composed of high mean molecular weight gases (e.g., \citealt{Kaltenegger2009, Barstow2016, Morley2017, Lustig-Yaeger2019}). 
Additionally, coadding may be better suited for small planets because even if the atmosphere shows some variability (e.g. changes in average cloud coverage), the amplitude of the changes in the spectra introduced by atmospheric variability are smaller than the error bars on individual transits. The same may not be true for giant planets.

Our study demonstrates that reaching these small signals is technically possible. The median error for all four individual transits is 55 ppm, which bins down to a median error of 28 ppm in the weighted mean. We also find that red noise in individual wavelength channels becomes less important when stacking visits, as those channels with inflated error bars due to red noise simply become weighted less. 

Additionally, co-adding transits still shows promise because there is still no indication that we are hitting an instrumental noise floor. We test this by concatenating the residuals from each observing mode and computing a red noise diagnostic plot for each, as shown in the bottom panel of \autoref{fig:wlc_variance}. We do not concatenate the G395M and G395H residuals, because for every integration of G395H (with 14 groups), there are 2.3 integrations of G395M (with 6 groups), so we cannot simply bin an integer number of G395M residuals to match G395H. \autoref{fig:wlc_variance} illustrates that the residuals are still binning down with increasing bin size down to the duration of the transit itself, which tells us no noise floor is hit down to 8 ppm. Again, this shows promise for the potential of stacking transits. 

However, the community is also now coming to terms with the reality of stacking transits in the face of stellar contamination that may impact the transmission spectrum visit-to-visit \citep[e.g.,][]{Lim2023, Radica2025}. Typically, the presence of stellar contamination is discerned by unocculted cold or hot spots that present themselves as unexpected slopes and/or water features in the spectrum itself (this was first postulated by \citealt{Pont2008} and popularized as the ``transit light source effect" by \citealt{Rackham2018}). Attempts to mitigate this include incorporating stellar contamination into retrievals (e.g., \citealt{Cadieux2024}) and using a transmission spectrum from another (presumably bare rock) planet in the same system as an empirical correction factor \citep[e.g.,][]{Rathcke2025}. 

However, if there is no obvious sign of stellar variability from the weighted mean transmission spectrum, as we demonstrate in our study, there is no clear reason to suspect that there might be a problem stacking multiple transmission spectra together. However, as we see in Sections \ref{sec:stellar_spectra} and \ref{sec:retrievals}, Visit 1 is an outlier in terms of its stellar heterogeneity and outsized influence on retrieval results. By simply looking at the visit-by-visit transmission spectrum (see Section \ref{sec:reduction_intervisit_agreement} and \autoref{fig:spectrum_visit_compare}), it appears that the apparent discrepancies between Visits 1 and 2 discussed in \citetalias{May2023} (\citeyear{May2023}) are likely random noise instances based on the scatter of the data seen in all four visits, with the notable exception of the curious disagreement at $\rm 4.5\;\mu m$ between Visit 3 and the other visits (see Section \ref{sec:reduction_intervisit_agreement}). However, further investigation revealed that the random noise interpretation is not favored. Rather, the star may be changing in more subtle ways than previously assumed.

This exercise demonstrates that single visits can still dominate interpretation of planetary spectra when stacking multiple visits, even in the apparent face of little-to-no obvious stellar variability in the transmission spectrum. It is unclear if a single visit could have such a dominating influence if, say, 10 transits were stacked instead of four. There should be a limit in which a single transit, no matter how contaminated, becomes ``diluted" in the face of many transits. In any case, moving forward, we recommend a ``leave-one-out" approach in atmospheric retrievals and forward models to ensure that one visit is not dominating the results.

\subsection{Comparison with MIRI/LRS Findings} \label{sec:disc_miri_compare}

We examine the transmission and emission constraints together by combining our retrieval results (using the three visit case, excluding Visit 1) with the atmospheric constraints from \citet{Xue2024}. We compare our results to their emission spectroscopy constraints, which are more stringent compared to their atmospheric constraints from their white light eclipse depth.

In \autoref{fig:retrieved_mu_Psurf}, we overplot the \citet{Xue2024} $2\sigma$ lower limit (see their Figure 5) on atmospheric pressures allowable for atmospheres of three end-member compositions: pure $\rm H_2O$, Earth-like, and pure $\rm CO_2$. With transmission alone (considering the case where we exclude Visit 1), our data are equally consistent with thin ($<$100 mbar) atmospheres that are 14 AMU or greater, or thick ($>$100 mbar) atmospheres that are 25 AMU or greater. The latter scenario, however, is rejected by the emission data. 

The emission $2\sigma$ lower limit of either a $\sim$0.4 bar Earth-like atmosphere or a $\sim$0.1~mbar $\rm CO_2$ atmosphere agrees with our findings across reduction pipelines. These limits help contextualize our findings that, for example, our transmission data are consistent with a thick pure $\rm CO_2$ atmosphere, but given the emission lower limit of 0.1~mbar $\rm CO_2$, we can exclude this as a possibility. Thus, the emission limits for atmospheres $>$25 AMU generally place more stringent constraints on lower pressures than we can with transmission alone. (This is illustrated in the figure by the fact that the emission lower limit lies within or above our $1\sigma$ contour.) The only exception to this is in the case of the \eureka reduction, which favors a very thin $\rm N_2$-dominated Earth-like atmosphere. However, as discussed in Section \ref{sec:retrievals}, this demonstrates that different noise properties across reductions lead to slightly different preferences in retrieval results in the case of a featureless spectrum (as also seen in GJ~341b, \citealt{Kirk2024}), as opposed to implying that an $\rm N_2$-dominated atmosphere is more likely than a $\rm CO_2$ atmosphere.

In the case of the pure $\rm H_2O$ atmosphere, however, results from transmission spectroscopy are more constraining than those from emission (2$\sigma$ limit of $\sim0.2$ bar), as our results suggest only a very tenuous steam atmosphere is possible. 
To further test the possibility of a steam atmosphere in conjunction with emission data, we estimate the nightside temperature from the white light eclipse MIRI results \citep{Xue2024} to determine if a steam atmosphere is possible or if it would condense out on the nightside. Assuming $A_b=0$ and the $1\sigma$ upper and lower limits from \cite{Xue2024} on the heat redistribution efficiency, $\epsilon$, we use Equation 5 from \cite{Cowan2011}, 
\begin{equation}
    T_n = T_{\rm eff} \left(\frac{R_s}{a}\right)^{1/2} (1-A_b)^{1/4} \left(\frac{\epsilon}{4}\right)^{1/4},
\end{equation}

\noindent to calculate the $1\sigma$ limits of the maximum nightside temperature, $T_n$. \cite{Xue2024} report a $1\sigma$ upper limits of $\epsilon=0.52$ and $1\sigma$ lower limit of $\epsilon=0.06$ (see their Figure 4), giving us $287<T_n<492$ K. At these temperatures, therefore, it is still possible from an energy transport perspective to have water vapor in the upper atmosphere. 

This exercise demonstrates the utility of combining results from transmission and emission to provide a complementary and constraining picture of any possible atmosphere on a rocky exoplanet. Together, our results show that any atmosphere on \planetname is tenuous at best. 

\subsection{On the Nature of \planetname}

Combining the emission and transmission spectroscopy data together tell us that \planetname has a thin atmosphere, or no atmosphere at all. Which of these scenarios is more likely? Is it possible for \planetname to retain a thin high mean molecular weight atmosphere? 

Early attempts to predict the atmosphere composition of \planetname by and large found that the most likely outcome, given that the planet is interior to the runaway greenhouse limit \citep{Ingersoll1969} and assuming that it started with some amount of water, was that \planetname has a tenuous atmosphere with at most a few bars of $\rm O_2$ \citep{Schaefer2016}. This is a result of dissociation of $\rm H_2O$ and subsequent preferential loss of $\rm H_2$. A steam atmosphere on this (at least) 5~Gyr \citep{Berta2015}, 1.6-day orbital period planet would require that a magma ocean still be present on the planet's surface (as a source for the outgassed steam) \citep{Schaefer2016}, a scenario that is highly implausible based on predicted solidification timescales of magma oceans. The only mechanism by which a magma ocean could have endured is if the initial water content of the planet was $>10\%$ by weight. \cite{Schaefer2016} could not rule out this high level of initial water content due to the imprecise mass from the discovery paper ($\rm 1.62\pm0.55\;M_{\oplus}$; \citealt{Berta2015}). However, using the updated radius from \cite{Dittman2017} ($\rm 1.13\pm0.056\;R_{\oplus}$) and updated mass from \cite{Bonfils2018} ($\rm 1.66\pm0.23\;M_{\oplus}$), we use the mass-radius interior composition curves from \cite{Zeng2016} to argue that a $10\%$ $\rm H_2O$ by weight composition can be rejected at ${\sim}2.8\sigma$, and that the composition is most consistent (within $1\sigma$) with an Earth-like interior. Therefore, it does not seem feasible for current-day \planetname to host a surface magma ocean and have any significant interior water fraction capable of maintaining a steam atmosphere. This 
supports our conclusion that the tentative hints of a steam atmosphere, driven largely by Visit 1, are erroneous. 

We acknowledge, however, that the other scenario postulated by \cite{Schaefer2016} - an $\rm O_2$ atmosphere leftover from preferential atmospheric escape of $\rm H_2$ - cannot so easily be ruled out. A 1 bar $\rm O_2$-dominated atmosphere with 100 ppm $\rm H_2O$ (see \autoref{fig:forward_models}) is consistent with both the emission and transmission data. However, we are able to rule out the 10 bar $\rm O_2$ atmosphere with 100 ppm of $\rm CO_2$ that is consistent with the emission data (see the lower panel of \autoref{fig:forward_models}), as our data should be sensitive enough to detect the $\rm CO_2$ feature at this atmospheric pressure. We note however, that thinner $\rm O_2$ atmospheres with some $\rm CO_2$ would likely be more consistent with our data, as the lower pressures would decrease the expected feature size, analogous to our lower pressure $\rm H_2O$ scenario illustrated in the middle panel of \autoref{fig:forward_models}. 

In theory, the lack of an $\rm O_3$ detection at 4.7 \micron{} could be used to rule out an oxygen-dominated atmosphere, as $\rm O_3$ would readily be produced by photochemistry \citep{Lincowski2018}. However, additional retrieval tests with $\rm O_3$ included fail to place an upper limit on the species, suggesting that the precision of our transmission spectrum is insufficient to help rule out any meaningful amounts of $\rm O_3$ in an oxygen-dominated atmospheres within the NIRSpec wavelength range.  We conclude, therefore, that though we cannot entirely rule out a $\rm O_2$ atmosphere, we find that maintenance of a remnant steam atmosphere that has become $\rm O_2$-dominated to be unlikely.

What of the other two scenarios highlighted in Section \ref{sec:retrievals} and \autoref{fig:retrieved_mu_Psurf}? Is it plausible for \planetname to host a thin ``Earth-like" ($<0.4$ bar) or $\rm CO_2$ ($<0.1$ mbar) atmosphere? To answer this, we must consider atmosphere generation (via volcanic outgassing), stability (considering the radiative timescale and collapse), and stripping (via irradiation and stellar wind). In terms of atmospheric generation, volcanism wanes over geologic time, especially for smaller planets that deplete their mantle volatiles more quickly \citep{Dorn2018}. Assuming GJ~1132 is $\sim$5 Gyr old places it right at the cusp of whether or not volcanism could be possible \citep{Kite2009}. We note this is a conservative estimate; \planetname is likely \textit{at least} 5~Gyr old; if it is much older than this, the possibility of present-day volcanism is drastically lower. If volcanism is present, many theoretical studies predict atmospheres much thicker and/or with clouds than what our analysis suggests \citep{Kite2009, Noack2017, Dorn2018}, assuming the planet is in a stagnant-lid regime. For example, one study predicted that a volcanically outgassed $\rm CO_2$ atmosphere for a planet of \planetname's mass (1.66~$\rm M_{\oplus}$) and temperature (588~K) should have surface pressures between that of Mars ($\sim$7~mbar) and Venus ($\sim$92~bar) (see \citealt{Dorn2018}, their Figure 17). Our lower limit is clearly less than this. 

Atmospheric stability must also be considered. We can first consider the radiative versus convective timescale, which tells us how fast the wind speeds must be to propagate the incident stellar flux from day to night before the heat is radiated away. Using a back-of-the-envelope calculation (equations 10.1 and 10.2 from \citealt{2010eapp.book.....S}), we estimate the radiative timescale for a very thin (0.1 mbar) $\rm CO_2$ atmosphere to be on the order of 10--20 seconds, which translates to wind speeds on the order of hundreds of m/s. This is an order of magnitude greater than typical wind speeds measured on our own terrestrial planets. This suggests that in the case of a very thin atmosphere, heat is radiated away before it can be propagated to the nightside.

This also has important implications for atmospheric stability, as atmospheric collapse is a cause for concern on tidally-locked rocky exoplanets when day-night heat transport is inefficient \citep{Wordsworth2015}, as the nightside can condense and cold-trap atmospheric species \citep{Wordsworth2022}. \cite{Wordsworth2015} estimates that close-in tidally-locked exoplanets collapse at $\lesssim0.03-0.1$ bar (see their Figure 12), although they do not extrapolate their findings to planets as irradiated as \planetname. 

Lastly, we must also consider the fact that \planetname lies only 0.0153 au from its star \citep{Bonfils2018}, and that atmospheric stripping may occur. \cite{Kite2020} predict that M-dwarf super-Earths that started out with primordial $\rm H_2$ atmospheres are unlikely to retain secondary atmospheres if they are $T_{\rm eq}>500$~K. This is because high mmw species get lost to space alongside $\rm H_2$ before the underlying magma ocean (which can absorb and protect the volatile high mmw species from loss) crystallizes. This would leave only a tenuous atmosphere that would be difficult to replenish via volcanic outgassing. Hot planets such as \planetname could theoretically have atmospheres if they formed without a $\rm H_2$ envelope or if they began with a high volatile content. However, the latter, as previously discussed, is unlikely for \planetname. Finally, stellar winds may play a significant role in stripping thin atmospheres for planets interior to the habitable zone (e.g., \citealt{Dong2017, Dong2018}).

Taken together, our results demonstrate that though both the transmission and emission data still allow several thin atmospheres on \planetname, this scenario is disfavored from a first-principles modeling perspective. We conclude, therefore, that our featureless spectrum is more likely a reflection of a true bare rock planet, the simplest explanation that is consistent with the existing data. However, further and more detailed planetary evolution modeling is warranted to fully understand (1) the possible states for this planet considering the many remaining uncertainties (e.g., the interior composition, escape history, etc.) and (2) how the stringent constraints from the emission and transmission measurements in turn help constrain these uncertain unobservable properties of \planetname.  

\section{Conclusions} \label{sec:conclusions}

In this study, we follow up on the two apparently discrepant NIRSpec/G395H transits of \planetname reported in \citetalias{May2023} (\citeyear{May2023}). We observe two additional transits using NIRSpec/G395M in this case, and use all four transits and the recent MIRI/LRS emission spectroscopy results \citep{Xue2024} to place strict limits on any atmosphere around \planetname.

This is the first time G395H and G395M observations have studied a single target, offering a critical opportunity for comparison between modes. NIRSpec/G395M functions very well for rocky exoplanet transmission spectroscopy, with no meaningful differences in scatter, red noise, or final error bar size between the two modes. However, utilizing the G395M mode does lose the opportunity to gain high-resolution flux-calibrated stellar spectra, which can be critical for uncovering hidden stellar heterogeneity that is not obvious in the transmission spectrum. Characterizing this possible hidden heterogeneity is critical in the case of co-adding multiple transits. In the case of rocky M-dwarf studies, we therefore recommend that G395M be utilized only if not stacking multiple visits.

For \planetname, we find good intervisit agreement in the transmission spectra, with the G395M data cleanly filling in the NRS1/NRS2 gap from the G395H data with no offsets required. Simply by looking at the transmission spectrum, it would appear that the differences in spectra reported in \citetalias{May2023} (\citeyear{May2023}) are simply due to random noise. However, upon further investigation, there is evidence that underlying stellar heterogeneity in Visit 1 (relative to the other visits) is driving these differences. Using the flux-calibrated stellar spectra, we find that Visit 1 likely has a higher coverage fraction of cool spots (and lower coverage fraction of warm spots). Because Visit 1 and 2 were taken only eight days apart, this suggests (along with the slight different in fit limb darkening values for Visit 1) that spots on the limb may have rotated completely or partially out of view in this timeframe. 

The fact that Visit 1 is an outlier is strongly corroborated by the retrieval analysis, which finds a steam atmosphere is preferred in the weighted mean of all visits. However, a featureless spectrum is preferred if Visit 1 is excluded. This illustrates the challenge of stacking multiple transits and attempting to interpret small deviation from flatness. In the future, we strongly recommend the rocky exoplanet community adopt this ``leave-one-out" approach to test if a single visit is dominating one's interpretation.

What of the planet's atmosphere? Forward models prefer a featureless spectrum over most 1 bar atmospheres, whereas a thinner ($<10$~mbar) steam atmosphere is equally consistent with the data. However, this preference for a steam atmosphere disappears with the exclusion of Visit 1. We further explore possibilities with retrievals that use an agnostic ``ghost gas" to more broadly explore the surface pressure vs. mmw parameter space allowed by the data. These show that our data confidently exclude atmospheres with mmw $<6$ AMU (this includes $\rm H_2$ atmospheres, as well as ``miscible envelope" atmospheres as in \citealt{Benneke2024}) at all pressures, but are consistent with (within $1\sigma$ of) thin, $<$100 mbar atmospheres that are at least 14 AMU or any atmosphere that is 25 AMU or greater. 

We combine these results with the MIRI/LRS emission spectroscopy results \citep{Xue2024}, which are more sensitive to thicker atmospheres, which tell us that a thick atmosphere $>$25 AMU is unlikely. For a thinner, steam atmosphere (18 AMU), transmission results are more stringent; thin ($\lesssim10$ mbar) steam atmospheres remain consistent with the data. We further note that certain thin $\rm O_2$ atmospheres remain consistent with both data sets. Due to \planetname's age and proximity to its host star, however, it is unlikely to be able to retain such a thin atmosphere, regardless of composition. At this point, the simplest explanation is in fact that \planetname is a bare rock. 

If this is true, this would push our endeavor to locate the M-dwarf cosmic shoreline toward larger ($\rm >1.13\;R_{\oplus}$) and/or cooler ($T_{\rm eq}<580$~K) rocky planets. Importantly, we note that at face value, GJ~1132 was always a promising M-dwarf target for transmission spectroscopy, with its slow rotation period and old age. If \planetname is indeed a bare rock, this demonstrates that even optimal targets from an observability standpoint may not make optimal targets for atmospheres. Importantly, \planetname is in a very similar XUV instellation/escape velocity parameter space as the first two targets (LTT~1445Ac and GJ~3929b) selected for the STScI 500 hour Rocky World Director's Discretionary Time program\footnote{\url{https://outerspace.stsci.edu/pages/viewpage.action?pageId=257035126}}. If none of these planets show hints of atmospheres, we as a community may want to strive toward targets that are at lower instellations and therefore more challenging observationally, but are more likely to host atmospheres. 

The quest to determine which M-dwarf planets have atmospheres is no where near over. What's more, the finding of a bare rock is just as critical a finding as an atmosphere, as it helps us refine our understanding of the cosmic shoreline as well as opens up a world of possible exogeologic studies \citep{Hu2012, Whittaker2022, First2024}. Regardless of the ubiquity or lack thereof of nearby M-dwarf rocky exoplanet atmospheres, we are thus taking the first observational steps toward mapping our rocky neighbors and placing the Solar System rocky planets in their true Galactic context.

\section*{Acknowledgments}
The authors thank the reviewer for their insightful and helpful feedback on this work. K.B. thanks N. Allen for helpful conversations about stellar contamination. This work was done based on observations using the NASA/ESA/CSA JWST. The data were obtained via the Mikulski Archive for Space Telescopes (MAST) at the Space Telescope Science Institute, which is operated by the Association of Universities for Research in Astronomy, Inc. (AURA). Support for this work was provided by NASA grants from the JWST Cycle 1 GO Program 1981. S.P. acknowledges support from NASA under award number  80GSFC24M0006. 
H.R.W was funded by UK Research and Innovation (UKRI) framework under the UK government’s Horizon Europe funding guarantee for an ERC Starter Grant [grant number EP/Y006313/1]. S.E.M. is supported by NASA through the NASA Hubble Fellowship grant HST-HF2-51563 awarded by the Space Telescope Science Institute, which is operated by the Association of Universities for Research in Astronomy, Inc., for NASA, under contract NAS5-26555.
L.A. acknowledges funding from UKRI STFC Consolidated Grant ST/V000454/1 and is supported by the Klarman Fellowship. J.K. acknowledges financial support from Imperial College London through an Imperial College Research Fellowship grant.


\vspace{5mm}
\facilities{JWST (NIRSpec)}

\software{\\ astropy \citep{astropy:2013, astropy:2018, astropy:2022}, \\ 
\texttt{batman} \citep{Kreidberg2015}, \\
\texttt{CHIMERA} \citep{line2013,line2014}, \\
\texttt{emcee} \citep{emcee2013}, \\
\eureka \citep{Bell2022}, \\
\exotic \citep{Wakeford2016, laginja2020}, \\
\texttt{ExoTIC-LD} \citep{Grant2022, Grant2024}, \\
\firefly \citep{Rustamkulov2022, Rustamkulov2023}, \\
\texttt{jwst} \citep{JWST_pipeline}, \\
\texttt{lacosmic} \citep{vanDokkum2001}, \\
\texttt{lmfit} \citep{Newville2014}, \\
matplotlib \citep{Hunter2007}, \\
numpy \citep{Harris2020}, \\
pandas \citep{McKinney2010}, \\
\texttt{PHOENIX} \citep{Allard2012, Husser2013}, \\
\texttt{PICASO} \citep{Batalha2019}, \\
\texttt{POSEIDON} \citep{MacDonald2017, MacDonald2023}, \\
\texttt{PyMSG} \citep{Townsend2023}, \\
\texttt{PyMultiNest} \citep{Feroz2009, Buchner2014}, \\
scipy \citep{Virtanen2020}}

\section*{Data Availability}

All data presented in this article were obtained from the Mikulski Archive for Space Telescopes (MAST) at the Space Telescope Science Institute (STScI). The four transit observations can be accessed via \dataset[doi:10.17909/e97r-g212]{https://doi.org/10.17909/e97r-g212}.\\
Data products are available on Zenodo: \url{https://zenodo.org/doi/10.5281/zenodo.15318916}. 


\bibliography{main_revised_arxiv}{}
\bibliographystyle{aasjournal}

\end{document}